\documentclass[a4paper,11pt]{article}
\pdfoutput=1
\usepackage{jcappub}
\usepackage[T1]{fontenc}
\usepackage{amssymb}
\usepackage{graphicx}
\usepackage{amsmath}
\usepackage{hyperref}
\usepackage{tensor}

\title{\boldmath The gravitational waves from the first-order phase transition with a dimension-six operator}

\author[1,2,3]{Rong-Gen Cai}
\author[2,4]{Misao Sasaki}
\author[1,2,3]{Shao-Jiang Wang}

\affiliation[1]{CAS Key Laboratory of Theoretical Physics, Institute of Theoretical Physics, Chinese Academy of Sciences, No.55 Zhong Guan Cun East Road, Beijing 100190, China}
\affiliation[2]{Center for Gravitational Physics, Yukawa Institute for Theoretical Physics, Kyoto University, Kyoto 606-8502, Japan}
\affiliation[3]{School of Physical Sciences, University of Chinese Academy of Sciences, No.19A Yuquan Road, Beijing 100049, China}
\affiliation[4]{International Research Unit of Advanced Future Studies, Kyoto University, Kyoto 606-8502, Japan}

\emailAdd{cairg@itp.ac.cn}
\emailAdd{misao@yukawa.kyoto-u.ac.jp}
\emailAdd{schwang@itp.ac.cn}

\abstract{We investigate in details the gravitational wave (GW) from the first-order phase transition (PT) in the extended standard model of particle physics with a dimension-six operator, which is capable of exhibiting the recently discovered slow first-order PT in addition to the usually studied fast first-order PT. To simplify the discussion, it is sufficient to work with an example of a toy model with the sextic term, and we propose an unified description for both slow and fast first-order PTs. We next study the full one-loop effective potential of the model with fixed/running renormalization-group (RG) scales. Compared to the prediction of GW energy density spectrum from the fixed RG scale, we find that the presence of running RG scale could amplify the peak amplitude by amount of one order of magnitude while shift the peak frequency to the lower frequency regime, and the promising regime of detection within the sensitivity ranges of various space-based GW detectors shrinks down to a lower cut-off value of the sextic term rather than the previous expectation.
\begin{flushleft}
  YITP-17-67
\end{flushleft}}

\begin{document}
\maketitle
\flushbottom

\section{Introduction}\label{sec:introduction}

The direct detections of gravitational-waves (GWs) from binary-black-holes (BBHs) merging events~\cite{Abbott:2016blz,Abbott:2016nmj,Abbott:2017vtc} open up an unprecedented era to probe the fundamental physics~\cite{Cai:2017cbj}, such as the gravitational theories alternative to general relativity (GR), the cosmological models alternative to Lambda-Cold-Dark-Matter ($\Lambda$CDM), and the Beyond-Standard-Model (BSM) of particle physics. One of the approaches of utilizing GWs to constrain the BSM of particle physics is the relic stochastic background of GWs from various first-order phase transitions (PTs) happened during the evolution of the universe. See earlier~\cite{Binetruy:2012ze,Caprini:2015zlo} and recent~\cite{Cai:2017cbj,Weir:2017wfa} reviews on GWs from first-order PTs.

The generation of GWs from first-order PTs follows from three processes and consists of three sources, which are summarized in the recent review~\cite{Cai:2017cbj} and references therein. The first-order PT begins with the nucleation of true vacuum bubbles within the false vacuum background, and then the nucleated bubbles expand until they collide with other bubbles, finally the first-order PT is completed once there is significant percolation of bubbles. The GWs are generated through the collisions of the uncollided envelop of the overlapping bubbles, the magneto-hydrodynamics (MHD) turbulent flows, and the acoustic waves of bulk fluid motion. Due to the stochastic nature of the relic GWs of first-order PT, only the energy density spectrums can be fitted with power-law forms according to the simulation results. See~\cite{Weir:2017wfa} for a summary adopting the most recent simulation~\cite{Hindmarsh:2017gnf} based on the analytic model~\cite{Jinno:2016vai} for bubble collisions and analytic model~\cite{Hindmarsh:2016lnk} for sound waves. See also~\cite{Jinno:2017fby} for recent improved analytic model even without the use of envelope approximation in the estimation of GWs from bubble collisions and sound waves.

One of the simplest extension to the SM with a first-order PT is to add a dimension-six operator~\cite{Grojean:2004xa,Bodeker:2004ws} to the standard model, which is expected from some new strong dynamics at electroweak (EW) scale or simply integrating out the heavy field to have an effective-field-theory (EFT) description. The preliminary study in~\cite{Huber:2007vva} reveals the possibility of detecting the relic GWs from the first-order PT with the sextic term. The reanalysis made in~\cite{Delaunay:2007wb} of the dimension-six operator uses the full finite temperature effective potential at one-loop order with more complete scope, however, without taking into account the renormalization-group (RG) improvement. The revisit in~\cite{Huber:2013kj} focuses on the hydrodynamical description for the scalar-fluid system with friction coupling, where the bubble wall velocity was computed in the stationary regime with an example of dimension-six operator. Later in~\cite{Leitao:2015fmj} the calculations on the bubble wall velocity was generalized into the runaway regime besides the detonation and deflagration regimes with several examples including the dimension-six operator. See also~\cite{Huang:2016odd} for an example of the sextic term that the future space-based interferometers could detect the parameter space beyond current particle collider and provide alternative cross-check for future particle collider.

We study two issues in this paper on the GWs from the first-order PTs: The first one is to find an unified description for both slow and fast first-order PTs. The numerical  simulations  have told us that the energy density spectrum of GWs from bubble collisions can be characterized by a few phenomenological parameters, among which the strength factor and the characteristic length scale of bubble collisions evaluated at some reference temperature are used to determine the peak position and peak amplitude of the energy density spectrum. For the usually studied fast first-order PT, the reference temperature is defined at the equality between the bubble nucleation rate and the Hubble expansion rate, the strength factor is defined by the ratio of the released latent heat with respect to radiation background, and the characteristic length scale of bubble collisions is defined by the short time-duration of PT. For the slow first-order PT identified recently in~\cite{Kobakhidze:2017mru} in the model~\cite{Kobakhidze:2016mch}, the reference temperature is defined at the moment of percolation when 30\% of space has transited into bubbles, the strength factor is defined by the ratio of the kinetic energy density of bubble walls with respect to radiation background, and the characteristic length scale of bubble collisions is defined by the radius of bubbles of majority at percolation temperature. However, both two prescriptions have their limitations, and we propose an unified prescription for both slow and fast first-order PTs : the reference temperature is chosen at the percolation temperature, the strength factor comes from the total released vacuum energy density from all kinds of bubbles at percolation with different sizes and number densities, and the characteristic length scale of bubble collisions is defined by the mean size of bubbles at percolation. We illustrate our unified prescription in an example of a toy model, where the specific values should not be taken literally.

The second issue is to show the effect of RG improvement on the energy density spectrum of GWs from first-order PTs. The primary motivation we harbor for considering a running RG scale is that, those coupling constants that conspire to give rise to a potential barrier needed for the first-order PTs could flip signs when considering a running RG scale during PT. Therefore it is interesting to see how large a running RG scale could affect the predictions compared to a fixed RG scale. We exemplify the effect of RG improvement in the context of the dimension-six term with full one-loop effective potential. It turns out as a surprise that, compared to the prediction of GW energy density spectrum from the fixed RG scale, the presence of running RG scale could amplify the peak amplitude by amount of one order of magnitude while slightly shift the peak frequency into the lower frequency regime. Therefore, the effect from the RG improvement cannot be simply neglected in future for the precise prediction of energy density spectrum of GWs from first-order PT. We also find that the promising regime of detection within the sensitivity ranges of various space-based GW detectors shrinks down to a lower cut-off value associated with the sextic term rather than the previous expectation~\cite{Leitao:2015fmj}.

The outline for this paper is as follows. In section~\ref{sec:slowfast}, the unified approach is proposed in both regimes of slow and fast first-order PTs, of which the phenomenological parameters from the effective potential are calculated in details in subsection~\ref{subsec:model}, while the predictions of the relic GWs from both slow and fast first-order PTs are presented in details in subsection~\ref{subsec:simulation}. In section~\ref{sec:fixrun}, the full one-loop effective potential is given in subsection~\ref{subsec:looppotential}, of which the predictions of the relic GWs from first-order PTs are presented in~\ref{subsec:loopGW} with and without RG improvement. The section~\ref{sec:conclusion} is devoted to the conclusions.

\section{An unified description of slow/fast first-order phase transition}\label{sec:slowfast}

In this section, we will give an unified description for both slow and fast first-order PTs. The energy density spectrum of the GWs from first-order PT can be characterized by a few phenomenological parameters as suggested from numerical simulations of bubble collisions. In subsection~\ref{subsec:model}, these phenomenological parameters are introduced with appropriate account for both regimes of slow and fast first-order PTs. After obtaining the bounce action evaluated at the bounce solution of the bounce equation in~\ref{subsubsec:potential}, the characteristic temperature at which the GW energy density spectrum manifests a peak frequency is evaluated in~\ref{subsubsec:temperature}, the energy budget of released vacuum energy into the bubble wall and bulk fluid is described in~\ref{subsubsec:budget}, and the characteristic length scale of the GW energy density spectrum at peak frequency is estimated in~\ref{subsubsec:length}. The predictions of GWs from slow/fast first-order PT are presented in subsection~\ref{subsec:simulation}.

\subsection{Phenomenological parameters from the effective potential}\label{subsec:model}

Since the potential barrier is already presented at the tree-level for the sextic term, it should be sufficient for us to start with a toy model with tree-level potential and thermal corrections under high-temperature approximation. It is worth noting that, the toy model presented below only serves as a pedagogical example to illustrate all the main characteristic features of GWs from slow/fast first-order PT. Therefore, the specific values involved in this section should not be taken literally.

\subsubsection{A toy model}\label{subsubsec:potential}

We start with the SM Higgs potential $V(H)=m^2(H^\dagger H)+\lambda(H^\dagger H)^2$, where the Higgs doublet
$H=\frac{U(x)}{\sqrt{2}}\left(\begin{array}{c}0\\h(x)\\\end{array}\right)$ leaves only one physical degree of freedom (d.o.f.) under unitary gauge $U(x)=1$. The resulting physical Higgs potential $V(h)=\frac12m^2h^2+\frac14\lambda h^4$ manifests a spontaneous sysmtry breaking when $m^2<0$ at $V'(h=v)=0$, where the vacuum-expectation-value (vev) $v^2=-\frac{m^2}{\lambda}$. When Taylor expanded around the true vacuum $h^2=v^2$, the physical Higgs potential becomes $V(h)=-\frac14\lambda v^4+\frac14\lambda(h^2-v^2)^2$.
When further expanded around the real physical fluctuation around the true vacuum $h=v+\sigma$, the physical Higgs potential $V(\sigma)=-\frac14\lambda v^4+\lambda v^2\sigma^2+\lambda v\sigma^3+\frac14\lambda\sigma^4$ gives the physical Higgs mass $m_h^2=2\lambda v^2$. The essential point here is that one should use the physical observables $v\simeq246\,\mathrm{GeV}$ and $m_h\simeq125\,\mathrm{GeV}$ to express the Lagrangian parameters $m_{\mathrm{SM}}^2=-\frac12m_h^2\simeq-(88\,\mathrm{GeV})^2$ and $\lambda_{\mathrm{SM}}=\frac{m_h^2}{2v^2}\simeq0.13$ by requiring the normalization condition:
\begin{align}
\begin{split}\label{eq:normalization}
V'(h=v)&=0,\\
V''(h=v)&=m_h^2
\end{split}
\end{align}
at the true vacuum. Since it is known that~\cite{Kajantie:1996mn} the true vacuum $h=v$ is relaxed by a cross-over type transition from the unbroken phase $h=0$ within the SM, one needs to go beyond the SM in order to have a first-order PT that would generate the relic GWs constrained by the GW detectors. The second motivation driving us to go beyond SM is that the Higgs mechanism in SM is merely a phenomenological description~\cite{deFlorian:2016spz}, where the current collider experiments only provide us the local shape of Higgs potential around the true vaccum. The above normalization condition~\eqref{eq:normalization} is just another way of saying that the position and local shape of true vacuum remain unchanged for different BSM models.

Among many non-renormalizable BSM extensions within the EFT frame, the extra dimension-six term
\begin{align}
V(H)=m^2(H^\dagger H)+\lambda(H^\dagger H)^2+\frac{\kappa}{\Lambda^2}(H^\dagger H)^3
\end{align}
is the most simplest construction from the gauge invariant combination $H^\dagger H$ alone. Here $\kappa$ will be restricted to be positive in order to have a first-order PT and further restricted to be unity due to its degeneration with the cut-off scale $\Lambda$. The other choices of dimension-six terms~\cite{Elias-Miro:2013mua,Jenkins:2013zja,Jenkins:2013wua,Alonso:2013hga}, including the gravitational-induced dimension-six term $(H^\dagger H)R$ and kinetic-induced dimension-six term $(\partial_\mu H^\dagger H)(\partial^\mu H^\dagger H)$, will be reserved for future works. Similar to the SM case, the unitary gauge leaves only one physical d.o.f., and the physical Higgs potential reads
\begin{align}\label{eq:tree}
V(h)=\frac12m^2h^2+\frac14\lambda h^4+\frac18\frac{\kappa}{\Lambda^2}h^6,
\end{align}
The normalization condition~\eqref{eq:normalization} around the true vacuum allows us to reexpress the BSM Lagrangian parameters in terms of the SM Lagrangian parameters,
\begin{align}
\begin{split}\label{eq:reparameterization}
m^2&=m_{\mathrm{SM}}^2+\frac{3\kappa v^4}{4\Lambda^2}\\
\lambda&=\lambda_{\mathrm{SM}}-\frac{3\kappa v^2}{2\Lambda^2}.
\end{split}
\end{align}
Here the SM Lagrangian parameters $m_{\mathrm{SM}}^2$ and $\lambda_\mathrm{SM}$ are directly related to the physical observables $v$ and $m_h^2$, therefore~\eqref{eq:reparameterization} leaves us only one BSM Lagrangian parameter, namely the cutoff scale $\Lambda$. Having the tree-level potential~\eqref{eq:tree} in mind, it is easy to have a potential barrier separating the false and true vacuums by choosing positive $m^2$ and $\kappa/\Lambda^2$ and negative $\lambda$. A toy model is adopted throughout this section by simply supplying the tree-level potential~\eqref{eq:tree} with a thermal correction under the high-temperature approximation~\cite{Grojean:2004xa} (See also~\cite{Huber:2013kj} for an improved hight-temperature approximation),
\begin{align}\label{eq:treehighT}
V(h)=\frac12m^2h^2+\frac14\lambda h^4+\frac18\frac{\kappa}{\Lambda^2}h^6+\frac12c T^2h^2
\end{align}
where $c=\frac{1}{16}\left(-\frac{12v^2}{\Lambda^2}+g'^2+3g^2+4y_t^2+\frac{4m_h^2}{v^2}\right)$. The exact form where above high-temperature approximation is deduced will be presented in~\ref{subsubsec:loopthermal}, along with the initial conditions $v=246.21971 \mathrm{GeV}$, $m_h=125.15 \mathrm{GeV}$, $g'=0.35830$, $g=0.64779$, $y_t=0.93690$ detailed in~\ref{subsubsec:RGcondition}.

\begin{figure}
  \includegraphics[width=0.5\textwidth]{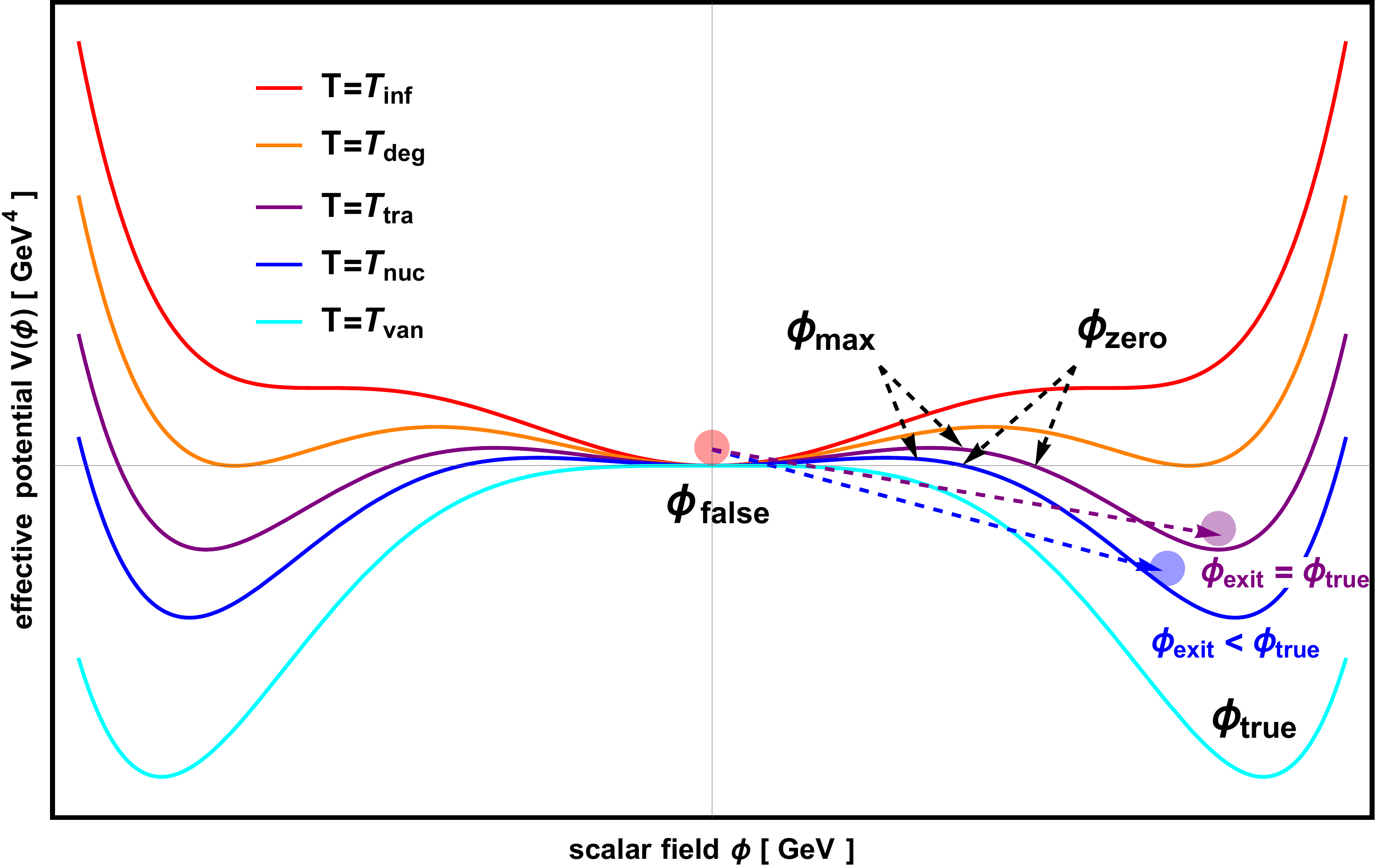}
  \includegraphics[width=0.5\textwidth]{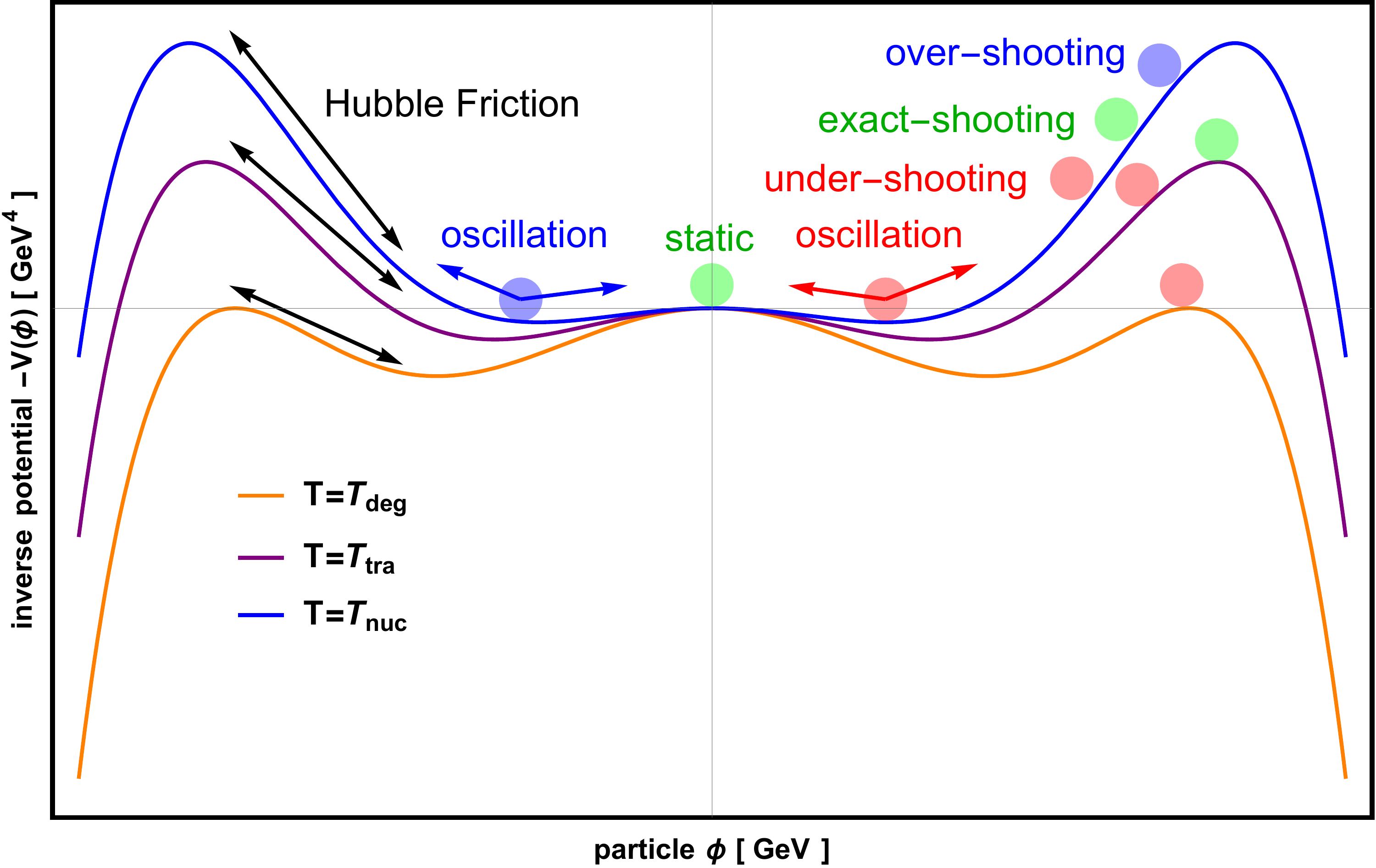}\\
  \includegraphics[width=0.5\textwidth]{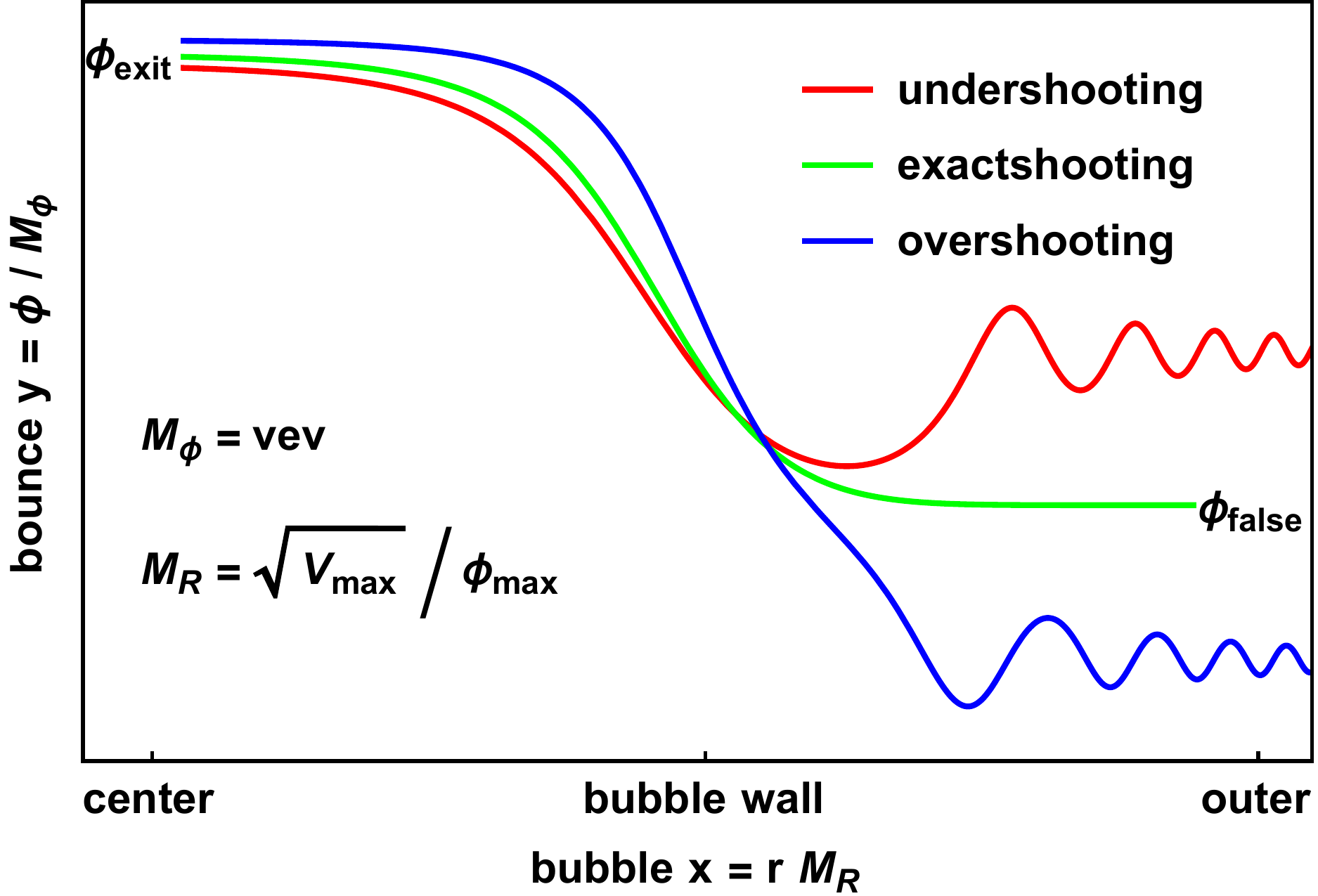}
  \includegraphics[width=0.5\textwidth]{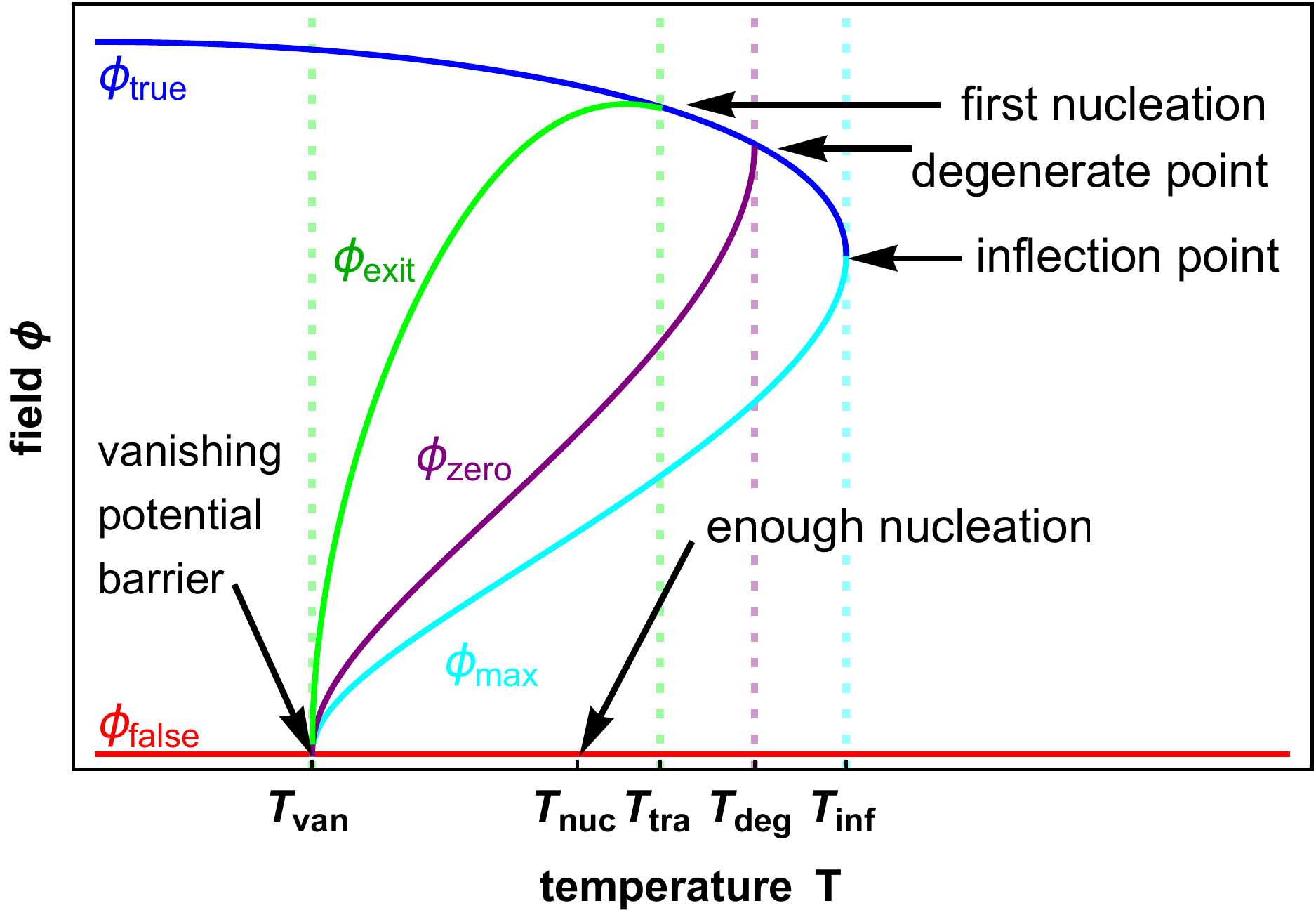}\\
  \caption{The pedagogical introduction of the bounce equation and bounce solution for the first-order PT. The upper left panel presents a schematic illustration of the effective potential at some characteristic temperatures, for example, the inflection temperature $T_\mathrm{inf}$ when a second minimum is about to appear, the degeneration temperature $T_\mathrm{deg}$ when the second minimum is degenerated with the first one, the transition temperature $T_\mathrm{tra}$ when the exit point of bounce solution is exactly sitting at the true vacuum, the nucleation temperature $T_\mathrm{nuc}$ when there are enough nucleated bubbles for the unbroken phase to be transited to the broken phase, and the vanishing temperature $T_\mathrm{van}$ when the potential barrier separating the two vacuums disappeares. The upper right panel presents a schematic illustration of an equivalent particle moving in the inverse of field potential $-V(\phi,T)$ with Hubble friction, where its position is labeled by the field value and its time is labeled by the radial coordinate of bounce solution. The bottom left panel presents a schematic illustration of the shooting algorithm, namely the particle is released from a finely adjusted exit point of bounce solution, above/below which the particle overshoots/undershoots and oscillates around the inverse potential barrier. Only when the appropriate exit point of bounce solution is found could the particle stand still at the origin. The bottom right panel presents a schematic illustration of various field values evolving with the decreasing temperature, for example, the field value $\phi_\mathrm{false}/\phi_\mathrm{true}$ where the false/true vacuum sits, the field value $\phi_\mathrm{zero}$ where the potential crosses zero, the field value $\phi_\mathrm{max}$ where the potential barrier lies, and the field value $\phi_\mathrm{exit}$ where the field penetrates from the other side of potential barrier. The first three panels are also used in~\cite{Cai:2017cbj}.}\label{fig:shooting}
\end{figure}

\begin{figure}
  \includegraphics[width=0.5\textwidth]{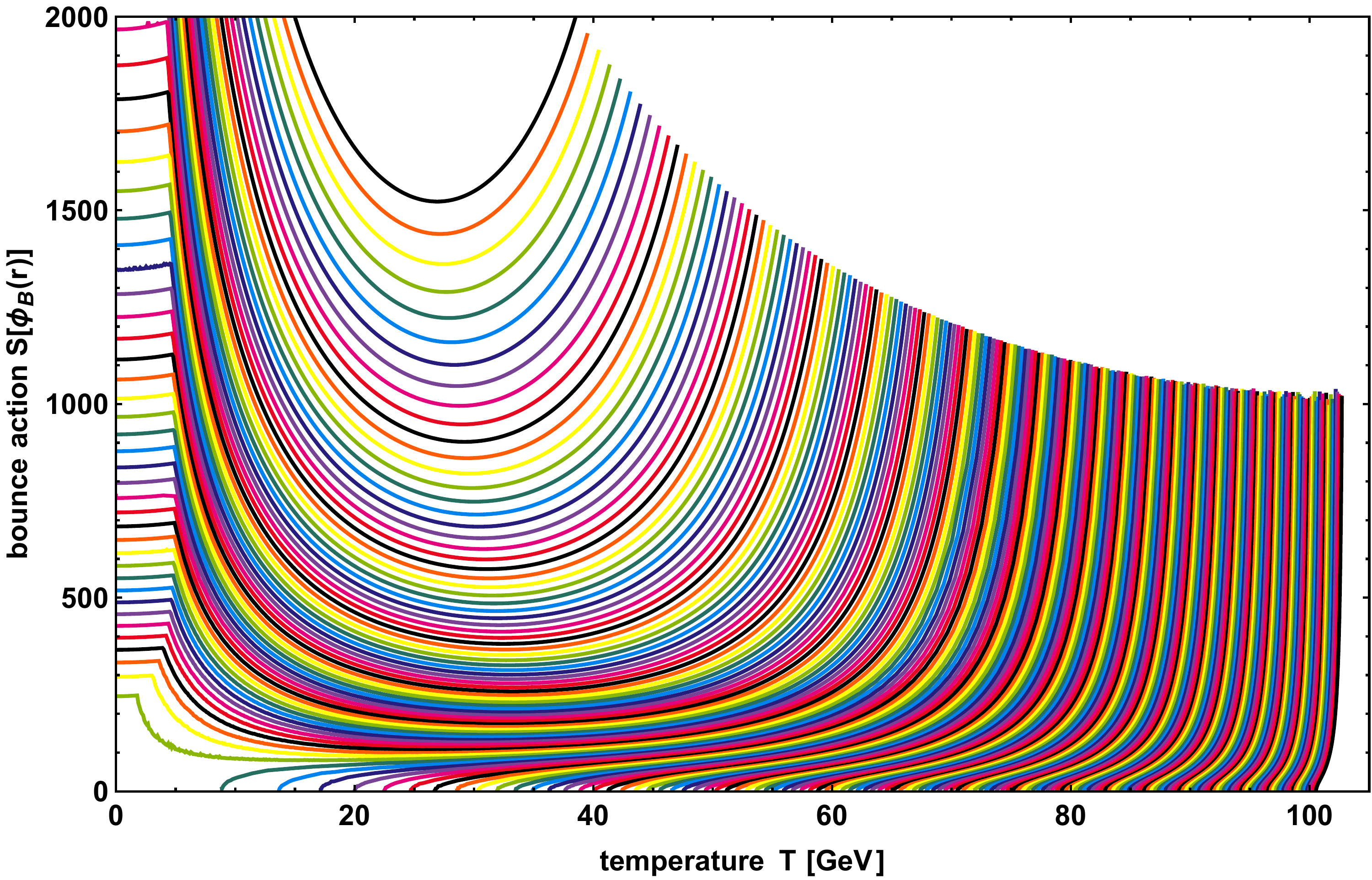}
  \includegraphics[width=0.5\textwidth]{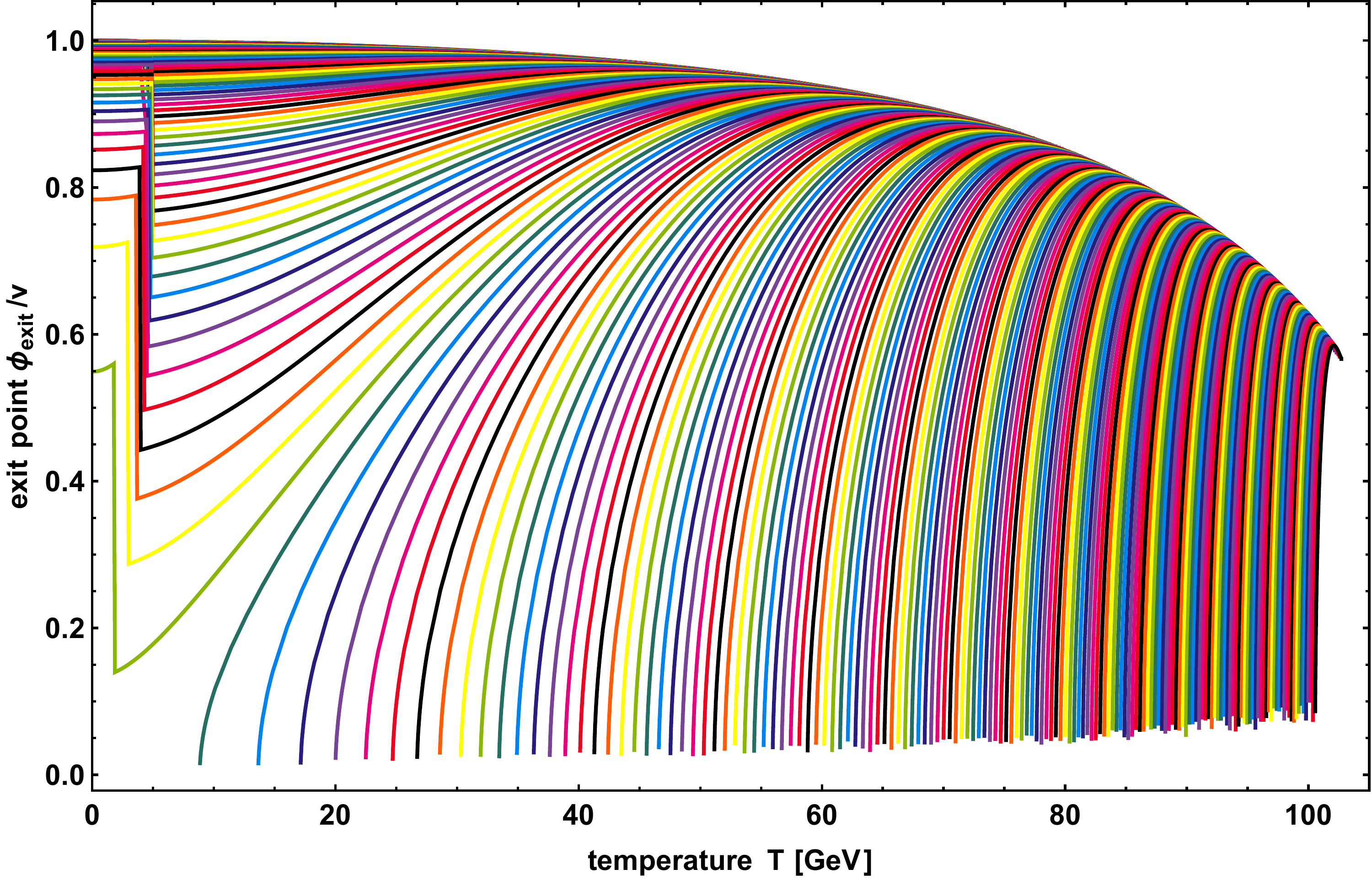}\\
  \caption{Left: Each curve presents the approximated bounce action $S(T)\approx\min[S_4(T),S_3(T)/T]$ at different temperatures $T$ and cut-off scale $\Lambda$. Right: Each curve presents the exit point $\phi_\mathrm{exit}$ of bounce solution at different temperature $T$ and cut-off scale $\Lambda$. Both panels use the cut-off scales between $530\,\mathrm{GeV}$ and $750\,\mathrm{GeV}$ with interval $1\,\mathrm{GeV}$ from top-left to bottom-right. With decreasing cut-off scale in the left panel, the first non-monotonic curve has cut-off scale $\Lambda=593\,\mathrm{GeV}$, below which there is a near-flat platform for each choice of $\Lambda\leq593\,\mathrm{GeV}$ below some characteristic temperature $T<1/R_0$.}\label{fig:BounceActionExitPoint}
\end{figure}

We then move to solve the equation-of-motion (EOM) with a field configuration that connects the false and true vacuum, namely the bounce solution or simply the true vacuum bubble. The nucleation rate per unit volume per unit time for such a bubble~\cite{Coleman:1977py,Callan:1977pt} scales as $\Gamma(T)=A(T)\mathrm{e}^{-S(T)}(1+\mathcal{O}(\hbar))$, where the prefactor $A(T)$ will be given later, and the exponential factor is given by the bounce action evaluated between the bounce solution and the false vacuum configuration, $S(T)=S[\phi_B,T]-S[0,T]\equiv S[\phi_B,T]$. For a generality, we will switch to $\phi$ in replace of $h$ to denote the Higgs field. With the full generality~\cite{Salvio:2016mvj}, the bounce action should be in the form of
\begin{align}
S[\phi,T]=4\pi\int_0^{1/T}\mathrm{d}\tau\int_0^\infty\mathrm{d}r r^2\left[\frac12\left(\frac{\partial\phi}{\partial\tau}\right)^2+\frac12\left(\frac{\partial\phi}{\partial r}\right)^2+V(\phi,T)\right]
\end{align}
where the Euclidean time $\tau=it$ and the inverse temperature $1/T$. The EOM
\begin{align}\label{eq:EOM}
\frac{\partial^2\phi}{\partial\tau^2}+\frac{\partial^2\phi}{\partial r^2}+\frac{2}{r}\frac{\partial\phi}{\partial r}=\frac{\partial V}{\partial\phi}
\end{align}
along with its initial/boundary conditions
\begin{align}\label{eq:action}
\left.\frac{\partial\phi}{\partial\tau}\right|_{\tau=0,\pm1/2T}=0,\quad\left.\frac{\partial\phi}{\partial r}\right|_{r=0}=0,\quad\lim\limits_{r\rightarrow\infty}\phi(r)=\phi_\mathrm{false}\equiv0
\end{align}
admits a partial-derivative-equation (PDE), of which the full numerical solution requires time-consuming lattice calculation. Fortunately it was found in~\cite{Linde:1980tt,Linde:1981zj} that the bounce action~\eqref{eq:action} from the full solution of~\eqref{eq:EOM} can be approximated as $S(T)\approx\min[S_4(T),S_3(T)/T]$ from the bounce action $S_4(T)$ evaluated at $O(4)$-bounce solution at low-temperature and the bounce action $S_3(T)/T$ evaluated at $O(3)$-bounce action at high-temperature. The benefit of such an approximation is that the EOM
\begin{align}\label{eq:EOMO4}
\frac{\mathrm{d}^2\phi}{\mathrm{d}\rho^2}+\frac{3}{\rho}\frac{\mathrm{d}\phi}{\mathrm{d}\rho}=\frac{\mathrm{d}V}{\mathrm{d}\phi},\quad\phi'(0)=0,\quad \phi(\infty)=0
\end{align}
of $O(4)$-bounce solution $\phi(\rho=\sqrt{\tau^2+\mathbf{r}^2})$ with $O(4)$-symmetric metric $\mathrm{d}s^2=\mathrm{d}\rho^2+\rho^2\mathrm{d}\Omega_3^2$ from the bounce action
\begin{align}\label{eq:S4}
S_4=2\pi^2\int_0^\infty\mathrm{d}\rho\rho^3\left[\frac12\left(\frac{\mathrm{d}\phi}{\mathrm{d}\rho}\right)^2+V\right]
\end{align}
and the EOM
\begin{align}\label{eq:EOMO3}
\frac{\mathrm{d}^2\phi}{\mathrm{d}r^2}+\frac{2}{r}\frac{\mathrm{d}\phi}{\mathrm{d}r}=\frac{\partial V}{\partial \phi},\quad\phi'(0)=0,\quad\phi(\infty)=0
\end{align}
of $O(3)$-bounce solution $\phi(r=\sqrt{\mathbf{r}^2})$ with $O(3)$-symmetric metric $\mathrm{d}s^2=\mathrm{d}\tau^2+\mathrm{d}r^2+r^2\mathrm{d}\Omega_2^2$ from the bounce action
\begin{align}\label{eq:S3T}
\frac{S_3}{T}=\frac{4\pi}{T}\int_0^\infty\mathrm{d}r r^2\left[\frac12\left(\frac{\mathrm{d}\phi}{\mathrm{d}r}\right)^2+V\right]
\end{align}
are ordinary-derivative-equations (ODEs), which can be easily solved using shooting algorithm as shown in Fig.~\ref{fig:shooting}. The results are summarized in Fig.~\ref{fig:BounceActionExitPoint}.

The upper left panel of Fig.~\ref{fig:shooting} illustrates some characteristic temperatures from the effective potential~\eqref{eq:treehighT}, for example, the inflection temperature $T_\mathrm{inf}$ when a second minimum is about to appear, the degeneration temperature $T_\mathrm{deg}$ when the second minimum is degenerated with the first one, the transition temperature $T_\mathrm{tra}$ when the exit point of bounce solution is exactly sitting at the true vacuum, the nucleation temperature $T_\mathrm{nuc}$ when there are enough nucleated bubbles for the unbroken phase to be transited to the broken phase, and the vanishing temperature $T_\mathrm{van}$ when the potential barrier separating the two vacuums is disappeared. The vanishing temperature could be zero if the potential barrier never vanishes even lowering the temperature down to zero.
The upper right panel and bottom left panel illustrate the shooting algorithm of solving both~\eqref{eq:EOMO4} and~\eqref{eq:EOMO3}, which can be regarded as the EOM of a particle moving in the inverse potential $-V(\phi,T)$ with Hubble friction, and its position is labeled by the field value and its time is labeled by the radial coordinate of bounce solution. The exact bounce solution (green) is found by finely adjusting the initial released point (corresponding to the exit point $\phi_\mathrm{exit}$ in field case at the center of bubble) so that it would stand still at the origin (corresponding to the false vacuum value in field case at spatial infinity of bubble) instead of oscillating around the inverse potential barrier after overshooting (blue) or undershooting (red).
The exit points of bounce solutions at different temperatures are drawn in the bottom right panel of Fig.~\ref{fig:shooting}, where the exit point (green curve) is degenerated with the true vacuum (blue curve) at a temperature called transition temperature $T_\mathrm{tra}$ just below the degeneration temperature $T_\mathrm{deg}$. Due to the Hubble frictions in the second terms of EOM~\eqref{eq:EOMO4} and~\eqref{eq:EOMO3}, the particle will always be undershooting even released from the true vacuum at temperature above transition temperature. Therefore the bubble can only possibly be nucleated below the transition temperature, which will be regarded as the starting point of first-order PT in the following.

In the left panel of Fig.~\ref{fig:BounceActionExitPoint}, we present the full approximated bounce action $S(T)\approx\min[S_4(T),S_3(T)/T]$ with respect to the temperatures $T$ at different cut-off scale $\Lambda$, which is chosen between $530\,\mathrm{GeV}$ and $750\,\mathrm{GeV}$ with interval $1\,\mathrm{GeV}$ from top-left to bottom-right, so does the right panel of Fig.~\ref{fig:BounceActionExitPoint}, which presents the exit point $\phi_\mathrm{exit}$ of bounce solution with respect to the temperature $T$ at different cut-off scale $\Lambda$. It is worth noting that, with decreasing cut-off scale in the left panel of Fig.~\ref{fig:BounceActionExitPoint}, the last monotonic curve has cut-off scale $\Lambda=594\,\mathrm{GeV}$, below which there is a near-flat platform for each choice of $\Lambda\leq593\,\mathrm{GeV}$ below some characteristic temperature $T<1/R_0$. These near-flat platforms of each bounce action curve with $\Lambda\leq593\,\mathrm{GeV}$ signify that their potential barriers could manifest themselves even at zero temperature, since the left and right endpoints of each bounce action curve represent the vanishing temperature $T_\mathrm{van}$ and transition temperature $T_\mathrm{tra}$. The characteristic temperature $T=1/R_0$ is defined as the dividing point, below/above which the bounce action is dominated by the $O(4)$/$O(3)$ bounce solution from vacuum/thermal decay. Therefore the final expression for nucleation rate is~\cite{Coleman:1977py,Callan:1977pt,Linde:1980tt,Linde:1981zj}
\begin{equation}\label{eq:GammaT}
\Gamma(T)=\left\{
    \begin{array}{ll}
     \frac{1}{R_0^4}\left(\frac{S_4(T)}{2\pi}\right)^2e^{-\min[S_4(T),S_3(T)/T]}, & \hbox{$T<1/R_0$;} \\
     T^4\left(\frac{S_3(T)}{2\pi T}\right)^\frac32e^{-\min[S_4(T),S_3(T)/T]}, & \hbox{$T>1/R_0$.}
    \end{array}
  \right.
\end{equation}
It is worth noting that, the approximate solution~\eqref{eq:GammaT} in the low-temperature regime $T<1/R_0$ is indeed in conflict with the effective potential~\eqref{eq:treehighT} with thermal correction under high-temperature approximation. However, for the sake of computational simplicity, the toy model we present here is only for the purpose of illustrating our unified description on the slow and fast first-order PTs. One can certainly choose any other effective potentials that encounter both slow and fast first-order PTs, for example, see~\cite{Kobakhidze:2017mru} for the cubic extension of SM.

Before closing the discussion of this subsection, some tricks on numerical calculations are commented here for reference. The boundary conditions at the center $r=0$ and spatial infinity $r=\infty$ of bubble should be replaced with some artificial cut-offs $r_\mathrm{min}$ and $r_\mathrm{max}$ during numerical evaluation. Is there an unified and sufficient choice of $r_\mathrm{min}$ and $r_\mathrm{max}$ at different temperatures $T$ and with different cut-off scales $\Lambda$ ? We propose to use the dimensionless normalized field $y(x)$ for $\phi(r)$ with $\phi=yM_\phi$ and $r=x/M_R$, where $M_\phi=vev$ is the vev of true vacuum and $M_R=\sqrt{V_\mathrm{max}}/\phi_\mathrm{max}$ is a rough estimation for the inverse scale of bubble radius at onset of nucleation. Our choices of $x_\mathrm{min}=10^{-2}$ for both $O(4)$- and $O(3)$-bubbles and $x_\mathrm{max}=20$ for $O(4)$-bubble and $x_\mathrm{max}=10$ for $O(3)$-bubble have been tested to be sufficient. Moreover, the initial value of exit point used during shooting algorithm becomes more and more sensitive when close to the vanishing temperature $T_\mathrm{van}$. A simple strategy is to use the last exit point at higher temperature to be the initial value of exit point at lower temperature, since the each curve of exit point in the right panel of Fig.~\ref{fig:BounceActionExitPoint} is continuous with respect to the temperature except for the jump at $T=1/R_0$.

\subsubsection{The characteristic temperatures}\label{subsubsec:temperature}

We here list all the characteristic temperatures involved in the first-order PT as follows.
\begin{description}
  \item[$T_\mathrm{inf}$]: inflection temperature at which the second minimum is about to appear;
  \item[$T_\mathrm{deg}$]: degeneration temperature at which the two minimums becomes degenerated;
  \item[$T_\mathrm{tra}$]: transition temperature at which the exit point sits exactly at the true vacuum;
  \item[$T_\mathrm{van}$]: vanishing temperature at which the potential barrier becomes vanishing;
  \item[$T_{\Gamma/H^4}$]: nucleation temperature at which the nucleation rate first catches up the Hubble rate;
  \item[$T_\mathrm{num}$]: nucleation temperature at which the first bubble is nucleated in the casual Hubble volume;
  \item[$T_\mathrm{per}$]: percolation temperature at which the true vacuum bubbles cover 30\% of space;
  \item[$T_\mathrm{min}$]: minimum temperature at which the non-monotonic bounce action minimizes itself;
  \item[$T_\mathrm{reh}$]: reheating temperature at which the plasma is reheated after PT is completed;
\end{description}
first four of which have been illustrated in Fig.~\ref{fig:shooting} of subsection~\ref{subsubsec:potential}, and the rest of which will be discussed in subsection~\ref{subsubsec:temperature}.

\begin{figure}
  \includegraphics[width=0.5\textwidth]{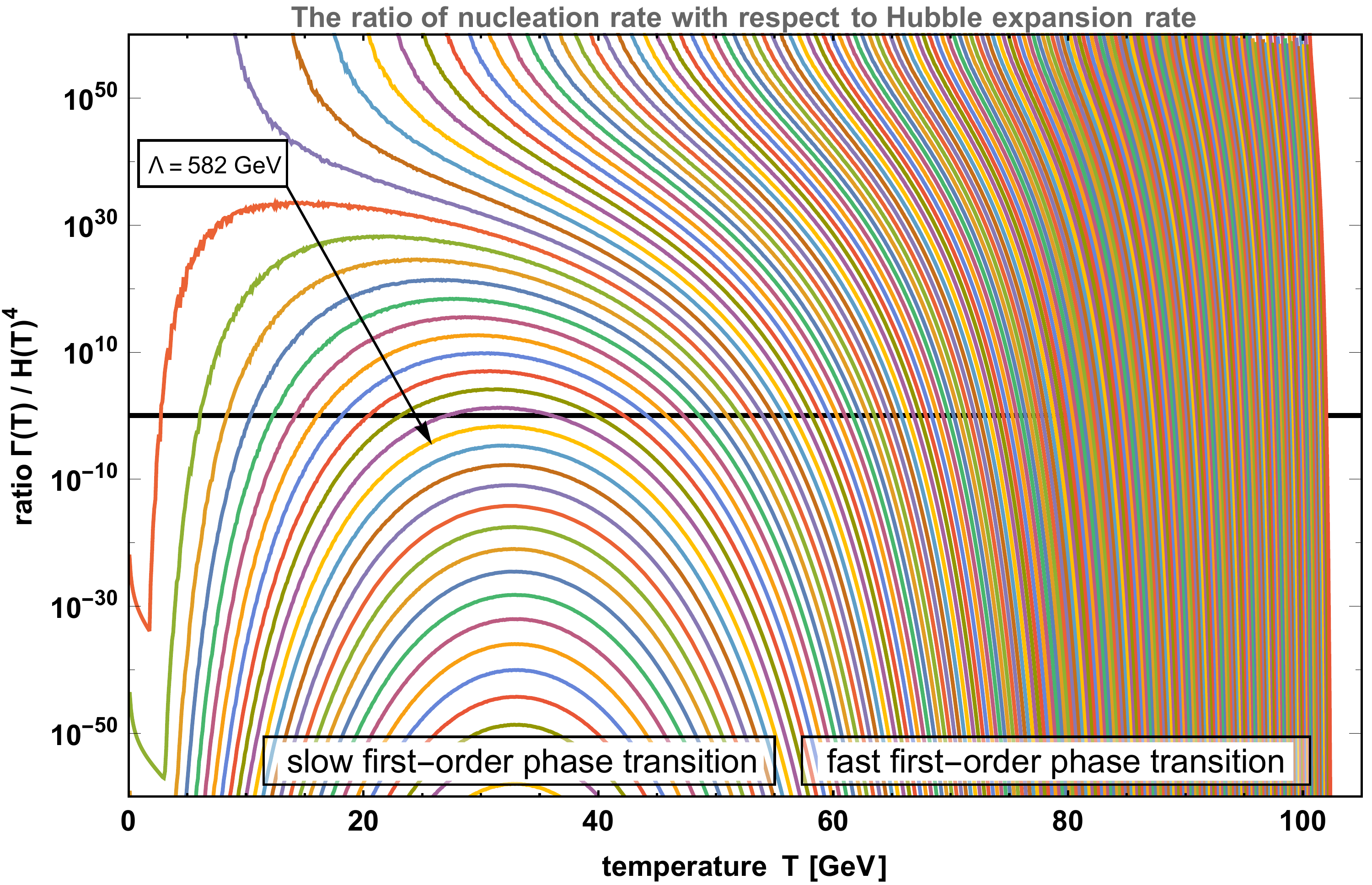}
  \includegraphics[width=0.5\textwidth]{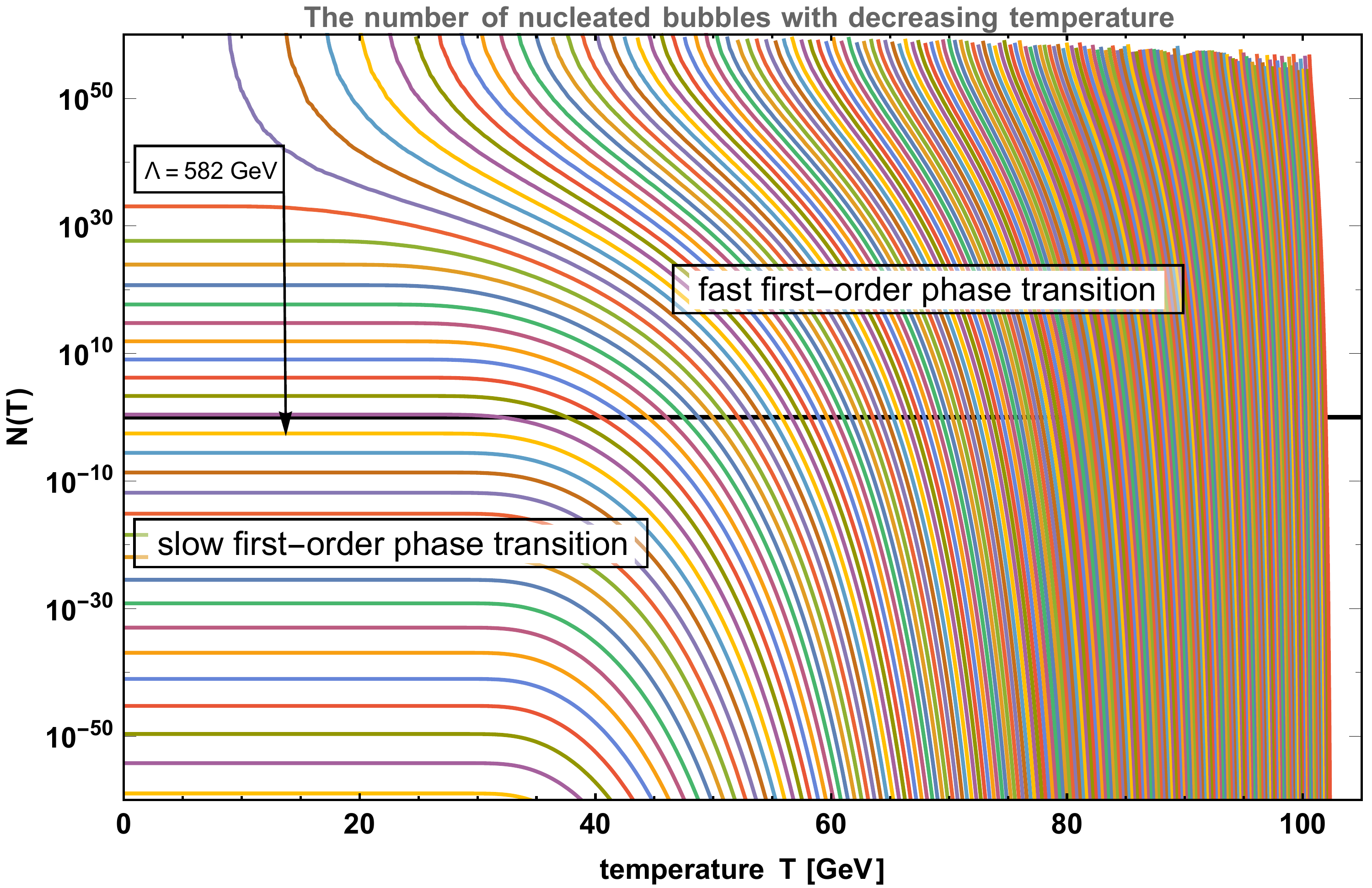}\\
  \caption{Left: The ratio of bubble nucleation rate with respect to Hubble expansion rate at given temperature. The nucleation temperature is defined by the moment when the bubble nucleation rate first catches up with the Hubble expansion rate. Right: The accumulated number of bubbles within our casual Hubble volume at given temperature since the beginning of PT at transition temperature. The nucleation temperature is defined by the moment when the number of bubbles within our causal Hubble volume first reaches the order of unity. In both panels, the nucleation temperature is found by the intersection point of each curve with the black horizontal line. Each curve is drawn from certain choice of cut-off scale $\Lambda$, ranging from $750\,\mathrm{GeV}$ at top-right to $530\,\mathrm{GeV}$ at bottom-left in both panels. However, both definitions of nucleation temperature break down for $\Lambda\leq582\,\mathrm{GeV}$.}\label{fig:NucleationTemperature}
\end{figure}

We start with the two commonly used definitions for the nucleation temperature. The nucleation temperature $T_\mathrm{nuc}$ is defined to approximate the reference temperature $T_*$ at which the GWs are most violently generated. Since the bubbles are nucleated with certain probability $\Gamma(T)$ per unit volume and per unit time, the moment when the accumulated number of bubbles within our casual Hubble volume first reaches the order of unity is given by
\begin{align}\label{eq:nucnum}
N(T)=\int_{t_\mathrm{tra}}^{t_\mathrm{nuc}}\mathrm{d}t\frac{\Gamma}{H^3}=\int_{T_\mathrm{nuc}}^{T_\mathrm{tra}}\frac{\mathrm{d}T}{T}\frac{\Gamma(T)}{H(T)^4}=1.
\end{align}
The nucleation temperature $T_\mathrm{nuc}$ solved from~\eqref{eq:nucnum} will be donated as $T_\mathrm{num}$. Since the nucleation rate $\Gamma(T)$ contains an exponential factor, the dominated contribution of the integration~\eqref{eq:nucnum} comes from the ratio of the bubble nucleation rate with respect to the Hubble expansion rate. Therefore the nucleation temperature can also be defined by the moment when the bubble nucleation rate first catches up with the Hubble expansion rate instead of being diluted by it, namely
\begin{align}\label{eq:GammaH4}
\frac{\Gamma(T_{\mathrm{nuc}})}{H(T_{\mathrm{nuc}})^4}=1.
\end{align}
The nucleation temperature $T_\mathrm{nuc}$ solved from~\eqref{eq:GammaH4} will be donated as $T_{\Gamma/H^4}$. For first-order PT at electroweak (EW) scale $T\sim\mathcal{O}(100)\,\mathrm{GeV}$, a thumb rule for the estimation of nucleation temperature can be simply deduced from~\eqref{eq:GammaH4},
\begin{align}
\frac{\Gamma(T)}{H(T)^4}=\frac{T^4}{H(T)^4}\left(\frac{S}{2\pi}\right)^\frac32e^{-S}=\left(\frac{90}{\pi^2g_\mathrm{dof}}\right)^2\frac{M_\mathrm{Pl}^4}{T^4}\left(\frac{S}{2\pi}\right)^\frac32e^{-S}=1
\end{align}
namely
\begin{align}\label{eq:thumbrule}
S=4\log\frac{M_\mathrm{Pl}}{T}+\frac32\log\frac{S}{2\pi}+2\log\frac{90}{\pi^2g_\mathrm{dof}}\approx4\log\frac{M_\mathrm{Pl}}{T}\approx130\sim140
\end{align}
Although extensively used in the literatures, the thumb rule~\eqref{eq:thumbrule} will not be used in this paper. The results from the two commonly used definitions~\eqref{eq:nucnum} and~\eqref{eq:GammaH4} of the nucleation temperature are presented in Fig.~\ref{fig:NucleationTemperature}.

In the left panel of Fig.~\ref{fig:NucleationTemperature}, each curve presents the ratio of bubble nucleation rate with respect to Hubble expansion rate at given temperature, of which the intersection point with the black horizontal line gives the nucleation temperature $T_{\Gamma/H^4}$. In the right panel of Fig.~\ref{fig:NucleationTemperature}, each curve presents the accumulated number of bubbles within our casual Hubble volume at given temperature since the beginning of PT, of which the intersection point with the black horizontal line gives the nucleation temperature $T_\mathrm{num}$. In both panels, each curve is drawn from certain choice of cut-off scale $\Lambda$, ranging from $750\,\mathrm{GeV}$ at top-right to $530\,\mathrm{GeV}$ at bottom-left in both panels. However, both definitions of nucleation temperature have no definition for $\Lambda\leq582\,\mathrm{GeV}$ since there is no intersection point with both black horizontal lines. Therefore the usually studied fast first-order PT is identified for $\Lambda\geq583\,\mathrm{GeV}$, since the PT will be quickly completed at a temperature just below the nucleation temperature $T_\mathrm{num}$ or $T_{\Gamma/H^4}$ due to the exponential growth of $N(T)$ or $\Gamma(T)/H(T)^4$ with respect to the decreasing temperature.
The slow first-order PT is identified for $\Lambda\leq582\,\mathrm{GeV}$, which will be our focus below. It is worth noting that the dividing value $\Lambda=582\,\mathrm{GeV}$ of slow/fast first-order PT should not be taken literally, because it is at least below $530\,\mathrm{GeV}$ when the full one-loop corrections are taken into account in section~\ref{sec:fixrun}.

\begin{figure}
  \includegraphics[width=0.5\textwidth]{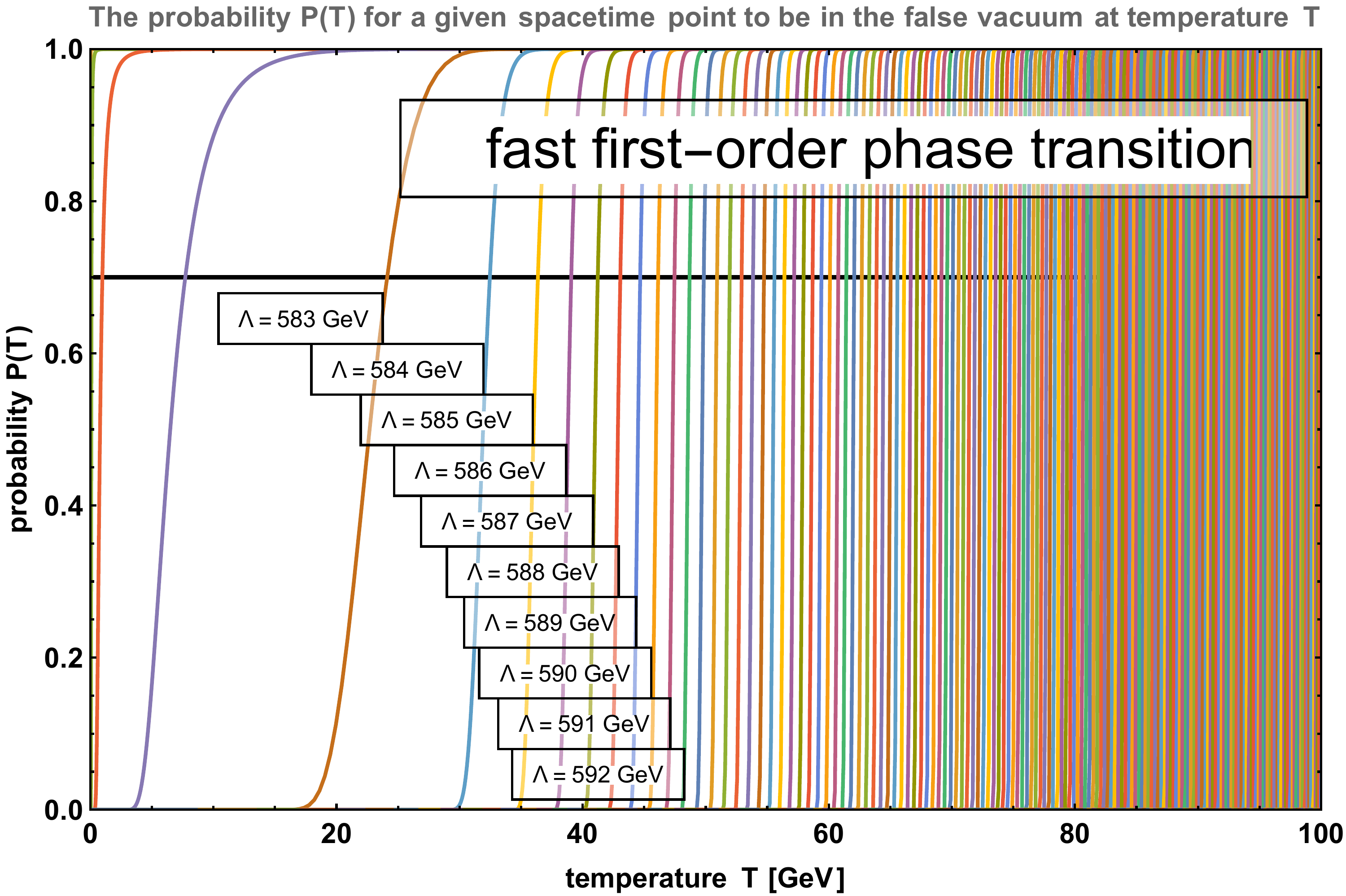}
  \includegraphics[width=0.5\textwidth]{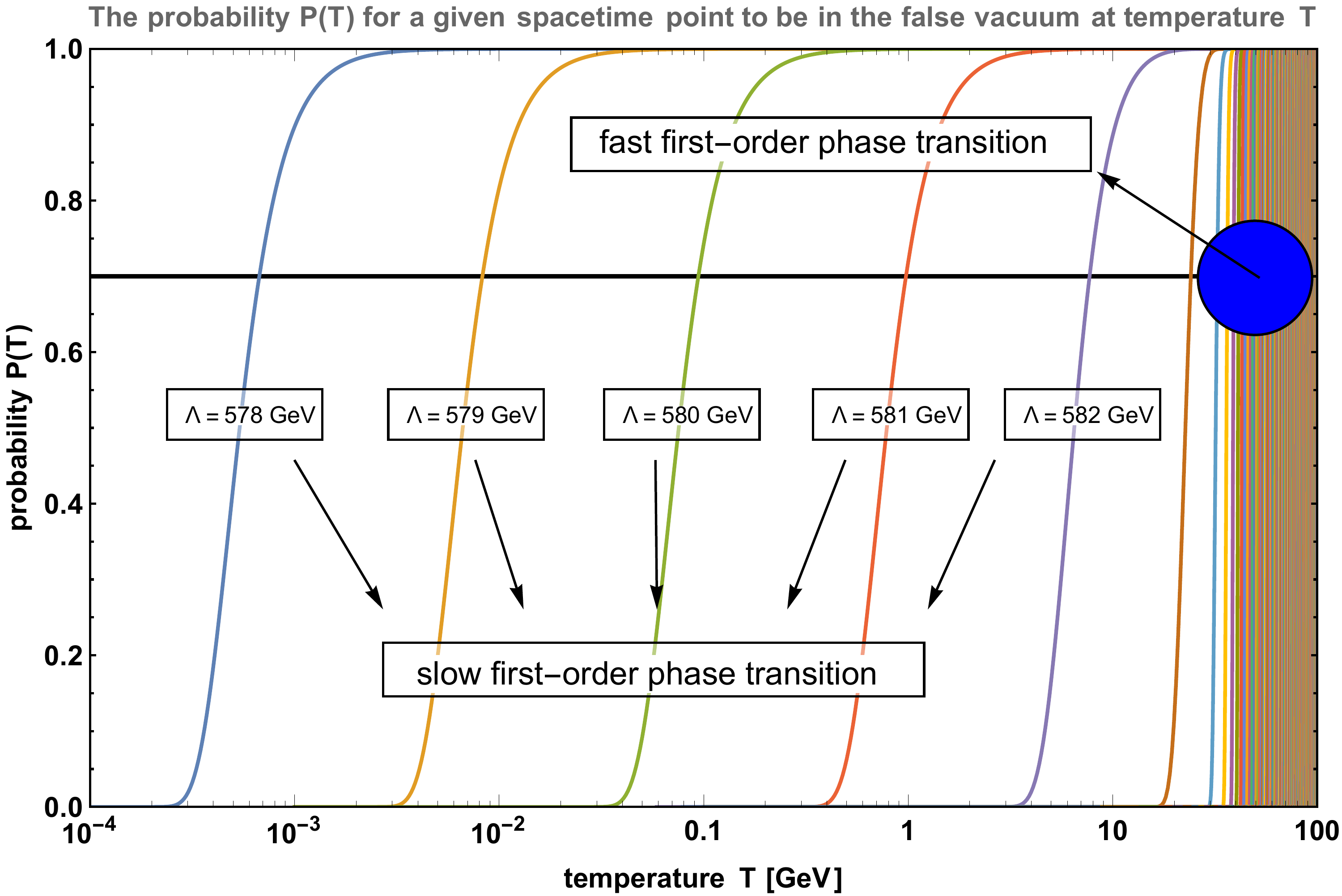}\\
  \caption{The probability $P(T)$ of staying at the false vacuum at given temperature $T$ with certain cut-off scale $\Lambda$ for the fast first-order PT (left) with $\Lambda\gtrsim583\,\mathrm{GeV}$ and slow first-order PT (right) with $\Lambda\lesssim582\,\mathrm{GeV}$. The percolation temperature $T_\mathrm{per}$ is given by the intersection point of $P(T)$ with the black horizontal line $P(T)\equiv0.7$. The slow first-order PT is only drawn for $\Lambda\gtrsim578\,\mathrm{GeV}$ to meet the BBN bound.}\label{fig:PercolationTemperature}
\end{figure}

\begin{figure}
\includegraphics[width=\textwidth]{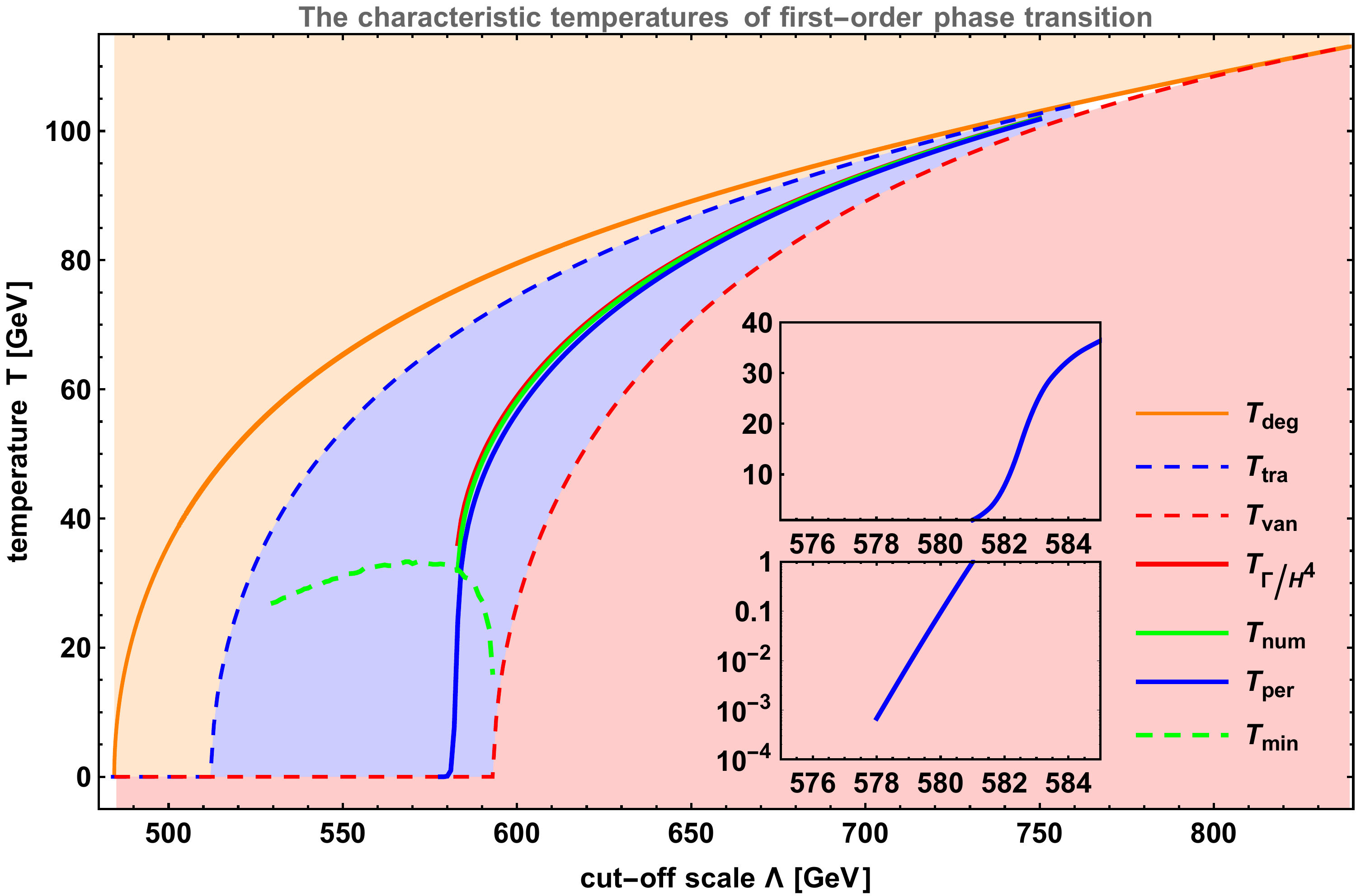}\\
\caption{The characteristic temperatures of first-order PT at a given cut-off scale. The orange solid line presents the degeneration temperature $T_\mathrm{deg}$, below which the newly emerged second minimum becomes the global minimum. However, the first-order PT can only possibly happen below the transition temperature $T_\mathrm{tra}$ presented as the blue dashed line, and the potential barrier disappears below the vanishing temperature $T_\mathrm{van}$ presented as the red dashed line, therefore only the blue-shaded region could possibly commit a first-order PT. The nucleation temperature is provided by $T_{\Gamma/H^4}/T_\mathrm{num}$ presented as the red/green solid line in the regime of fast first-order PT, and by $T_\mathrm{min}$ presented as the green dashed line in the regime of slow first-order PT. The percolation temperature $T_\mathrm{per}$ presented as the blue solid line is well-defined in both regimes of slow and fast first-order PTs, of which the regime of slow first-order PT is presented in the zooming windows. }\label{fig:CharacteristicTemperature}
\end{figure}

We next move to the definition of percolation temperature~\cite{Leitao:2012tx,Leitao:2015fmj}. As we show above, for fast first-order PT, the nucleation temperatures $T_\mathrm{num}$ and $T_{\Gamma/H^4}$ from~\eqref{eq:nucnum} and~\eqref{eq:GammaH4} are good approximation to the reference temperature $T_*$ at which the GWs are most violently generated. However, both~\eqref{eq:nucnum} and~\eqref{eq:GammaH4} have no definition for slow first-order PT. An essential question is how to define an alternative temperature to approximate $T_*$ in the regime of slow first-order PT, and most importantly, could this alternative temperature also approximate $T_\mathrm{num}$ and $T_{\Gamma/H^4}$ when extrapolated into the regime of fast first-order PT. In the literatures, the percolation temperature provides an appropriate candidate, which is defined as the moment when 30\% of space is covered by the true vacuum bubbles suggested by the numerical simulations~\cite{doi:10.1080/00018737100101261}. To calculate the percolation temperature, one starts with $P(t)$,
\begin{align}
P(t)=\exp\left[-\frac{4\pi}{3}\int_{t_\mathrm{tra}}^t\mathrm{d}t'\Gamma(t')a^3(t')r^3(t,t')\right]
\end{align}
which describes the probability of staying at the false vacuum~\cite{Turner:1992tz}, or simply the fraction of false vacuum. Here the scale factor accounts for the background expansion, and the exponential accounts for the overlap of bubbles. The comoving distance of bubble expansion from the onset of nucleation $t'$ to some later time $t>t'$ reads
\begin{align}
r(t,t')&=\frac{R_n(t')}{a(t')}+\int_{t'}^t\frac{\mathrm{d}t''}{a(t'')}.
\end{align}
Here the initial physical size $R_n(t')$ of nucleated bubble at $t'$ is usually neglected in the case of slow first-order PT with long-time expansion and in the case of runaway bubble with near speed-of-light expansion $v\approx1$. For a first-order PT in the radiation-dominated era with $t=\xi/T^2$, $\xi^2\equiv45M_\mathrm{Pl}^2/2\pi^2g_\mathrm{dof}$, $g_\mathrm{dof}=106.75$, $M_\mathrm{Pl}^2\equiv1/8\pi G_\mathrm{N}$, then the percolation temperature $T_\mathrm{per}$ is directly solved from
\begin{align}\label{eq:percolation}
P(T_\mathrm{per})=\exp\left[-\frac{64\pi}{3}\xi^4\int_T^{T_\mathrm{tra}}\mathrm{d}T'\frac{\Gamma(T')}{T'^6}\left(\frac{1}{T}-\frac{1}{T'}\right)^3\right]=0.7
\end{align}
as shown in Fig.~\ref{fig:PercolationTemperature}.
We will see in Fig.~\ref{fig:CharacteristicTemperature} that $T_\mathrm{per}$ indeed matches $T_\mathrm{num}$ and $T_{\Gamma/H^4}$ when extrapolated into the regime of fast first-order PT, which in turn solidifies the numerical choice of 70\%. It is worth noting that for the EWPT the percolation temperature should be at least bounded below from a successful Big-Bang-Nucleosynthesis (BBN), namely $T_\mathrm{per}\gtrsim\mathcal{O}(10^{-4})\,\mathrm{GeV}$.

We then move to the minimum temperature $T_\mathrm{min}$ that minimizes the non-monotonic bounce action in Fig.~\ref{fig:BounceActionExitPoint}, which is introduced to approximate the nucleation temperature in the regime of slow first-order PT. For the fast first-order PT, the nucleation temperature $T_\mathrm{nuc}$ can be defined as either $T_\mathrm{num}$ or $T_{\Gamma/H^4}$. The question is that what is the appropriate definition $T_\mathrm{nuc}$ for the slow first-order PT. In~\cite{Kobakhidze:2017mru}, it was suggested that, for the slow first-order PT, $T_\mathrm{nuc}$ can be defined as the nucleation temperature of those bubbles which become the bubbles of majority at percolation temperature.
To find the bubbles of majority at percolation temperature, one first notices that, the number density of bubbles at time $t$,
\begin{align}\label{eq:n(t)}
n(t)=\int_{t_\mathrm{tra}}^t\mathrm{d}t'\left(\frac{a(t')}{a(t)}\right)^3P(t')\Gamma(t'),
\end{align}
where the insertion of $P(t)$ accounts for the fact that the true vacuum bubbles can only be nucleated in the false vacuum. Rewriting the integration~\eqref{eq:n(t)} in terms of the physical radius of bubbles expanding from $t_R$ to $t$ in the  radiation-dominated era,
\begin{align}\label{eq:Radius}
R(t,t_R)=a(t)r(t,t_R)=2\sqrt{t}(\sqrt{t}-\sqrt{t_R})=\frac{2\xi}{T}\left(\frac{1}{T}-\frac{1}{T_R}\right),
\end{align}
namely,
\begin{align}
n(t)=\int_{R(t,t_\mathrm{tra})}^{R(t,t)\equiv0}\mathrm{d}R(t,t_R)\frac{\mathrm{d}t_R}{\mathrm{d}R(t,t_R)}\left(\frac{a(t_R)}{a(t)}\right)^3P(t_R)\Gamma(t_R),
\end{align}
one recognizes that the integral is nothing but the number density distribution~\cite{Turner:1992tz} of bubbles at time $t$ over the physical radius $R(t,t_R)=a(t)r(t,t_R)$ of bubbles that are nucleated at earlier time $t_R$, namely
\begin{align}\label{eq:dndR}
\frac{\mathrm{d}n}{\mathrm{d}R}(t,t_R)=\left(\frac{a(t_R)}{a(t)}\right)^4P(t_R)\Gamma(t_R).
\end{align}
Fixing the time $t$ at percolation time, one finds the bubbles of majority at percolation time have size $R(t_\mathrm{per},t_R)$ that maximizes the number distribution at percolation time from imposing the condition
\begin{align}
\left.\frac{\mathrm{d}}{\mathrm{d}t_R}\frac{\mathrm{d}n}{\mathrm{d}R}(t_\mathrm{per},t_R)\right|_{t_R=t_\mathrm{nuc}}=0,
\end{align}
where the corresponding $t_R$ or $T_R$ is defined as the nucleation temperature for slow first-order PT. To see that why the minimum temperature $T_\mathrm{min}$ provides a well approximation to the nucleation temperature $T_\mathrm{nuc}$ in the regime of slow first-order PT, one observes that any temperature that deviates from $T_\mathrm{min}$ would have larger bounce action and hence smaller nucleation rate of bubbles, therefore they can never dominate the number distribution at percolation except for those bubbles that are nucleated at $T_\mathrm{min}$. In what follows, we will use
\begin{align}\label{eq:nucleation}
T_\mathrm{nuc}=\left\{
                 \begin{array}{ll}
                   T_\mathrm{min}, & \hbox{slow first-order PT regime: $\Lambda\lesssim582\,\mathrm{GeV}$;} \\
                   T_\mathrm{num}\lesssim T_{\Gamma/H^4}, & \hbox{fast first-order PT regime: $\Lambda\gtrsim583\,\mathrm{GeV}$.}
                 \end{array}
               \right.
\end{align}

To conclude, we present all these characteristic temperatures involved in slow/fast first-order PT in Fig.~\ref{fig:CharacteristicTemperature} with respect to the cut-off scale $\Lambda$. The degeneration temperature $T_\mathrm{deg}$ is presented as the orange solid line, below which the newly emerged second minimum becomes the global minimum. However, the first-order PT can only possibly happen below the transition temperature $T_\mathrm{tra}$ presented as the blue dashed line due the Hubble friction term in~\eqref{eq:EOMO3}. The potential barrier disappears below the vanishing temperature $T_\mathrm{van}$ presented as red dashed line, therefore only the blue-shaded region could possibly commit a first-order PT in principle. The nucleation temperature is provided by $T_{\Gamma/H^4}/T_\mathrm{num}$ presented as the red/green solid line in the regime of fast first-order PT, and by $T_\mathrm{min}$ presented as the green dashed line in the regime of slow first-order PT. The percolation temperature $T_\mathrm{per}$ presented as blue solid line is well-defined in both regimes of slow and fast first-order PTs, of which the regime of slow first-order PT is presented in the zoomed windows. The first-order PT is usually completed soon after the percolation temperature $T_\mathrm{per}$ before reaching $T_\mathrm{van}$. After the PT is completed, there is a minor reheating due to the thermal energy released in the plasma. We will postpone the definition of reheating temperature $T_\mathrm{reh}$ until~\ref{subsubsec:GWcurves}, where the total redshift factor for the relic GW background is determined from the reheating temperature to the current temperature.

\subsubsection{The energy budget}\label{subsubsec:budget}

For vacuum decay at zero temperature in flat spacetime, all the released vacuum energy will be transited into accelerating the bubble wall, which is quickly approaching the speed-of-light after bubble nucleation. However, for thermal decay at finite temperature in plasma background, the background fluid would exert friction against the rapid expansion of bubble wall. Therefore it is vital to characterize the energy transfers of released vacuum energy among bubble wall and bulk fluid, which is summarized as an energy budget~\cite{Espinosa:2010hh} that would be presented at the end of this subsection.

\begin{figure}
\includegraphics[width=\textwidth]{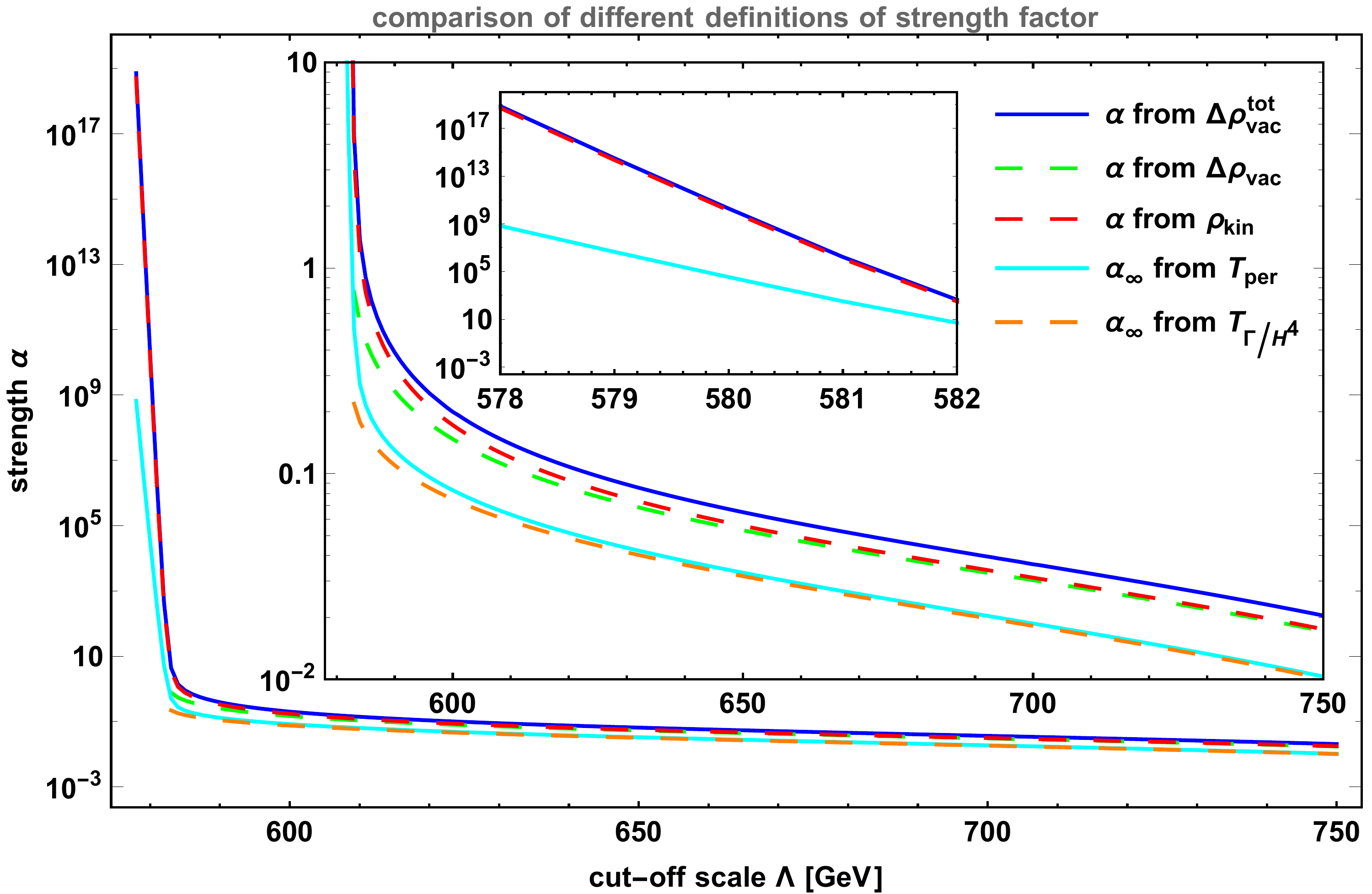}\\
\caption{The comparison of different definitions for the strength factor of first-order PT. The green dashed line is drawn from the definition~\eqref{eq:alphaI}, while the red dashed line is drawn from the definition~\eqref{eq:alphaII}. The red dashed line is approaching the green dashed line in the regime of the fast first-order PT for the reasons that: First, for the fast first-order PT the background spacetime can be treated as a flat spacetime without expansion, where the released vacuum energy can be transited into kinetic energy of bubble wall; Second, the bubbles of majority for the fast first-order PT are also those bubbles that are generated at the nucleation temperature. However, when compared with the blue solid line drawn from the definition~\eqref{eq:alphaIII} in the regime of fast first-order PT, both the green and red dashed lines underestimate about 20\% of the released vacuum energy density due to the ignorance of those bubbles generated before or around the nucleation temperature. In the regime of the slow first-order PT, the red dashed line also underestimates about 30\% of the blue solid line due to not only the ignorance of those bubbles generated around the nucleation temperature but also the non-ignorable effect from the background Hubble expansion. At last, the critical value of strength factor from~\eqref{eq:alphainfinity} suggests that the bubble is of runaway type for the considered regimes of both the slow and fast first-order PTs.}\label{fig:alpha}
\end{figure}

We start with the definition of the strength factor for the fast first-order PT, which measures the released vacuum energy density with respect to the background radiation energy density. The free energy density of scalar field is defined as the effective potential $\mathcal{F}(\phi,T)=V(\phi,T)$, which gives the energy density $\rho(\phi,T)=\mathcal{F}-T\mathcal{S}=V-T\frac{\mathrm{d}V}{\mathrm{d}T}$ and the pressure $p=-\mathcal{F}$. At the critical temperature, the energy can be transferred from first minimum to the degenerated second minimum without decreasing the temperature, which is described by the latent heat $\mathcal{L}(T_c)=\rho(0,T_c)-\rho(v(T_c),T_c)$. Below critical temperature, similar form of latent heat is used to define the released vacuum energy density,
\begin{align}
\Delta\rho_\mathrm{vac}(T)=\rho(0,T)-\rho(v(T),T).
\end{align}
Therefore the strength factor of PT is defined by
\begin{align}\label{eq:alphaI}
\alpha=\frac{\Delta\rho_\mathrm{vac}(T_*)}{\rho_\mathrm{rad}(T_*)},\quad \Delta\rho_\mathrm{vac}(T_*)=\Delta V_*-T_*\frac{\mathrm{d}\Delta V_*}{\mathrm{d}T}.
\end{align}
Here the difference of free energy density $\Delta V_*\equiv V(0,T_*)-V(v(T_*),T_*)\equiv -V(v(T_*),T_*)$ is computed at a reference temperature $T_*=T_{\Gamma/H^4}$, and we always normalize the false vacuum at the zero point of effective potential, namely $V(0,T)\equiv0$. To ensure that the vacuum energy density of bubbles never dominates the radiation energy density, we have checked that the condition $\rho(v(T_\mathrm{nuc}),T_\mathrm{nuc})-\rho(v(0),0)\ll\rho_\mathrm{rad}(T_\mathrm{nuc})$ is always fulfilled for the considered regimes of fast first-order PT. In Fig.~\ref{fig:alpha}, the strength factor from~\eqref{eq:alphaI} is presented as the green dashed line.

We next move to the strength factor for the slow first-order PT, which is defined at a reference temperature $T_*=T_\mathrm{per}$. In a recent paper~\cite{Kobakhidze:2017mru}, it is argued that, for runaway bubbles nucleated at $t_\mathrm{nuc}$, the kinetic energy of bubble walls at percolation time $t_\mathrm{per}$ is given by
\begin{align}
E_\mathrm{kin}(t_\mathrm{per})=4\pi\int_{t_\mathrm{nuc}}^{t_\mathrm{per}}
\mathrm{d}t\frac{\mathrm{d}R}{\mathrm{d}t}(t,t_\mathrm{nuc})R^2(t,t_\mathrm{nuc})\Delta\rho_\mathrm{vac}(t).
\end{align}
which can be expressed in terms of temperature as
\begin{align}
E_\mathrm{kin}(T_\mathrm{per})=32\pi\xi^3\int_{T_\mathrm{per}}^{T_\mathrm{nuc}}
\frac{\mathrm{d}T}{T}\left(2-\frac{T}{T_\mathrm{nuc}}\right)\left(\frac{1}{T^2}-\frac{1}{TT_\mathrm{nuc}}\right)^2\Delta\rho_\mathrm{vac}(T)
\end{align}
by using~\eqref{eq:Radius} in the radiation-dominated background. When divided by the physical volume of the bubble, the kinetic energy density
$\rho_\mathrm{kin}(T_\mathrm{per})=E_\mathrm{kin}(T_\mathrm{per})/\frac{4\pi}{3}R^3(t_\mathrm{per},t_\mathrm{nuc})$
could reproduce the released vacuum energy density $\rho_\mathrm{kin}(T_\mathrm{per})\approx\Delta\rho_\mathrm{vac}(T_\mathrm{nuc})$ for the fast first-order PT with $T_\mathrm{per}\approx T_\mathrm{nuc}$. Therefore, they define the strength factor for the slow first-order PT as
\begin{align}\label{eq:alphaII}
\alpha=\frac{\rho_\mathrm{kin}(T_\mathrm{per})}{\rho_\mathrm{rad}(T_\mathrm{per})},
\end{align}
which could also approach~\eqref{eq:alphaI} in the regime of fast first-order PT. The strength factor from~\eqref{eq:alphaII} is presented as the red dashed line in Fig.~\ref{fig:alpha}, which is approaching the green dashed line in the regime of fast first-order PT. This can be understood by noting two facts: First, for the fast first-order PT the background spacetime can be treated as a flat spacetime without Hubble expansion, where the released vacuum energy can be transited into kinetic energy of bubble wall; Second, the bubbles of majority for the fast first-order PT are also those bubbles that are generated at the nucleation temperature. At last, to ensure that the vacuum energy density of bubbles never dominates the radiation energy density, we have also checked that the condition $\rho(v(T_\mathrm{nuc}),T_\mathrm{nuc})-\rho(v(0),0)\ll\rho_\mathrm{rad}(T_\mathrm{nuc})$ is always fulfilled for the considered regimes of both the slow and fast first-order PTs.

Both definitions~\eqref{eq:alphaI} and~\eqref{eq:alphaII} for the strength factor have their limitations, since they all target on those bubbles generated at the nucleation temperature~\eqref{eq:nucleation}. For the fast first-order PT, most of bubbles are generated at the nucleation temperature $T_\mathrm{nuc}=T_{\Gamma/H^4}$, whose number density is large enough to complete the PT soon after $T_{\Gamma/H^4}$. Therefore it is reasonable to only consider the released vacuum energy density of those bubbles generated at $T_{\Gamma/H^4}$ as in the definition~\eqref{eq:alphaI}. For the slow first-order PT, the bubbles of majority at percolation temperature are generated at the nucleation temperature $T_\mathrm{nuc}=T_\mathrm{min}$, whose number density is not large enough so that one has to wait for a long time until those bubbles generated at $T_\mathrm{min}$ finally percolate at $T_\mathrm{per}$. Therefore it is reasonable to only consider the kinetic energy density of the bubbles of majority at percolation temperature as in the definition~\eqref{eq:alphaII}. However, for both the slow and fast first-order PTs, the number density of bubbles at time $t$ actually admits a distribution~\eqref{eq:dndR} over the sizes $R(t,t_R)$ of those bubbles that are nucleated at time $t_R$. We propose to take into account the full number density distribution~\eqref{eq:dndR} in order to estimate the total released vacuum energy density of all bubbles with different sizes and number densities at time $t$,
\begin{align}
\Delta\rho_\mathrm{vac}^\mathrm{tot}(t)=\frac{1}{1-P(t)}\int_{R(t,t)\equiv0}^{R(t,t_\mathrm{tra})}\mathrm{d}R(t,t_R)\frac{\mathrm{d}n}{\mathrm{d}R}(t,t_R)\frac{4\pi}{3}R(t,t_R)^3\Delta\rho_\mathrm{vac}(t_R),
\end{align}
where the prefactor accounts for the fact that only $1-P(t)$ fraction of physical volume has been covered by the bubbles at time $t$. Evaluated at percolation temperature, the total released vacuum energy density reads
\begin{align}
\Delta\rho_\mathrm{vac}^\mathrm{tot}(T_\mathrm{per})=\frac{1}{0.3}\int_{T_\mathrm{per}}^{T_\mathrm{tra}}\mathrm{d}T_R
\frac{\mathrm{d}R(T_\mathrm{per},T_R)}{\mathrm{d}T_R}\frac{\mathrm{d}n}{\mathrm{d}R}(T_\mathrm{per},T_R)\frac{4\pi}{3}R(T_\mathrm{per},T_R)^3\Delta\rho_\mathrm{vac}(T_R).
\end{align}
Here the physical radius of bubbles $R(T_\mathrm{per},T_R)$ at percolation temperature can be computed by~\eqref{eq:Radius} in the radiation-dominated background. We therefore define the strength factor as
\begin{align}\label{eq:alphaIII}
\alpha=\frac{\Delta\rho_\mathrm{vac}^\mathrm{tot}(T_\mathrm{per})}{\rho_\mathrm{rad}(T_\mathrm{per})},
\end{align}
for both the regimes of slow and fast first-order PTs, which is presented as the blue solid line in Fig.~\ref{fig:alpha}. In the regime of fast first-order PT, compared with current definition~\eqref{eq:alphaIII} from all kinds of bubbles with different sizes and number densities, the former definition~\eqref{eq:alphaI} underestimates about 20\% of the released vacuum energy density due to the ignorance for those bubbles nucleated before $T_{\Gamma/H^4}$. Therefore, the blue solid line in Fig.~\ref{fig:alpha} deviates from both the green and red dashed lines in the regime of fast first-order PT. Similarly, in the regime of the slow first-order PT, it seems that the blue solid line matches the red dashed line in Fig.~\ref{fig:alpha}, however, \eqref{eq:alphaII} actually underestimates about 30\% of~\eqref{eq:alphaIII} due to the ignorance for those bubbles nucleated around $T_\mathrm{min}$. The other reason for this underestimation of the released vacuum energy of bubble interior from the kinetic energy of bubble wall might be due to the non-ignorable effect from the background Hubble expansion.

The type of bubble expansion should be identified first before one could appreciate the energy budget of first-order PT. An expanding bubble can be of runaway or non-runaway types. A bubble wall will runaway if the Lorentz factor $\gamma=1/\sqrt{1-v^2}\rightarrow\infty$ when the friction is too small to prevent the bubble wall from approaching the speed-of-light, otherwise it is of non-runaway type, of which the bubble wall velocity is terminated at some finite value due to the eventual balance achieved between the plasma friction and net pressure on the bubble wall. A rigorous criterion~\cite{Bodeker:2009qy} credited from the mean field theory claimed that a bubble wall will runaway if the effective potential in the true vacuum remains deeper than the false vacuum even after replacing the thermal potential by its second-order Taylor expansion term in the false vacuum,
\begin{align}\label{eq:meanfield}
\widetilde{V}(\phi_\mathrm{true}(T_*),T_*)<\widetilde{V}(\phi_\mathrm{false}(T_*),T_*),
\end{align}
where $\widetilde{V}(\phi,T)=V_{T=0}(\phi)+\frac12V''_T(\phi_\mathrm{false}(T))(\phi-\phi_\mathrm{false}(T))^2$ is defined by replacing the thermal potential of $V(\phi,T)=V_{T=0}(\phi)+V_T(\phi)$ by its second-order Taylor expansion. Hence it is easy to verify that the toy model~\eqref{eq:treehighT} we studied in section~\ref{sec:slowfast} is of runaway type. Another commonly used criterion~\cite{Espinosa:2010hh} is to compare the strength factor of first-order PT with a critical value,
\begin{align}\label{eq:alphainfinity}
\alpha_\infty=4.9\times10^{-3}\left(\frac{\phi_*}{T_*}\right)^2,
\end{align}
which denotes the minimum value of $\alpha$ that bubble wall could runaway. Here the reference temperature $T_*$ can be chosen as either $T_{\Gamma/H^4}$ or $T_\mathrm{per}$. The results are shown in Fig.~\ref{fig:alpha} with different choices for the reference temperature $T_*$. It can be seen that the strength factor of first-order PT is always larger than the critical value~\eqref{eq:alphainfinity}. This agrees with our previous judgement from the criterion~\eqref{eq:meanfield} that the bubble walls of our toy model~\eqref{eq:treehighT} are all of runaway type for the cut-off scale that commits a first-order PT. It is worth noting that a recent paper~\cite{Bodeker:2017cim} casts the doubt on the runaway type of bubble expansion due to the additional friction from the subleading corrections arising from the transition radiation of wall-frame soft and massive vector bosons. In this paper, we will ignore this effect. We hope to come back to this issue in future.

\begin{figure}
\includegraphics[width=\textwidth]{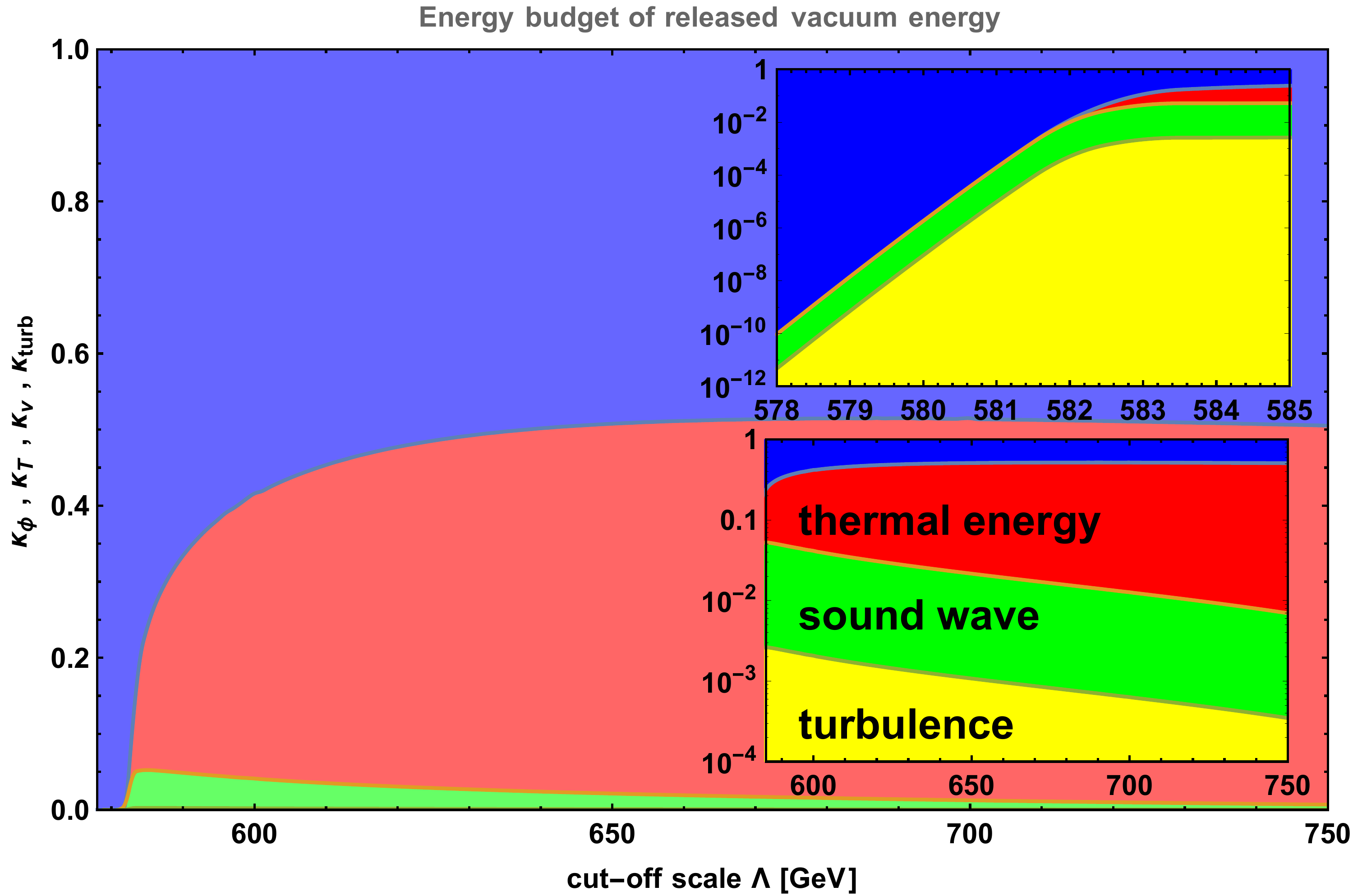}\\
\caption{The energy budget of released vacuum energy during the first-order PT at different cut-off scale. The blue shaded region accounts for the part that goes to accelerate the bubble wall, which is the dominated contribution for the regime of slow first-order PT. The red shaded region accounts for the part that goes to reheat the bulk plasma fluid, while the remaining parts from the sound waves and MHD turbulences of bulk fluid motions are denoted as the green and yellow shaded regions, respectively.}\label{fig:EnergyBudget}
\end{figure}

Now we are in the position to present the energy budget~\cite{Espinosa:2010hh} of first-order PT parameterized by various efficiency factors. For bubble expansion of runaway type, $\alpha>\alpha_\infty$, the surplus energy goes to accelerate bubble wall with efficiency factor
\begin{align}
\kappa_\phi=\frac{\alpha-\alpha_\infty}{\alpha},
\end{align}
and the saturated energy goes to bulk fluid with efficiency factor $\alpha/\alpha_\infty$, of which the fluid motion and fluid reheating weight with efficiency factors
\begin{align}
\kappa_v&=\frac{\alpha_\infty}{\alpha}\kappa_\infty,\\
\kappa_T&=\frac{\alpha_\infty}{\alpha}(1-\kappa_\infty), \quad \kappa_\infty=\frac{\alpha_\infty}{0.73+0.083\sqrt{\alpha_\infty}+\alpha_\infty}.
\end{align}
The fluid motion is usually dominated by the sound wave with efficiency factor $\kappa_\mathrm{sw}=(1-\epsilon)\kappa_v$ and subdominated by the MHD turbulence with efficiency factor $\kappa_\mathrm{turb}=\epsilon\kappa$, where the numerical factor $\epsilon$ is found to be around $5\%\sim10\%$. We take $\epsilon=0.05$ for concreteness in accordance with most of literatures. The results are summarized in Fig.~\ref{fig:EnergyBudget} with respect to the cut-off scale. The blue shaded region denotes the released vacuum energy that goes to accelerate bubble wall, and the red shaded region denotes the part that reheats the plasma, while the green and yellow shaded regions denote the remaining parts that transformed into the sound waves and MHD turbulences of bulk fluid motion, respectively. As shown in the regime of the slow first-order PT, almost all released vacuum energy has transited into the kinetic energy of bubble wall, which indicates that the dominated contribution of GW from slow first-order PT would be the bubble collision as we will see later in~\ref{subsubsec:treeresult}.

\subsubsection{The characteristic length scales}\label{subsubsec:length}

The characteristic length scale of first-order PT is usually determined by the bubble radius at collision, which gives rise to the peak frequency of GWs from first-order PT observed in the simulation of bubble collision.

\begin{figure}
\includegraphics[width=\textwidth]{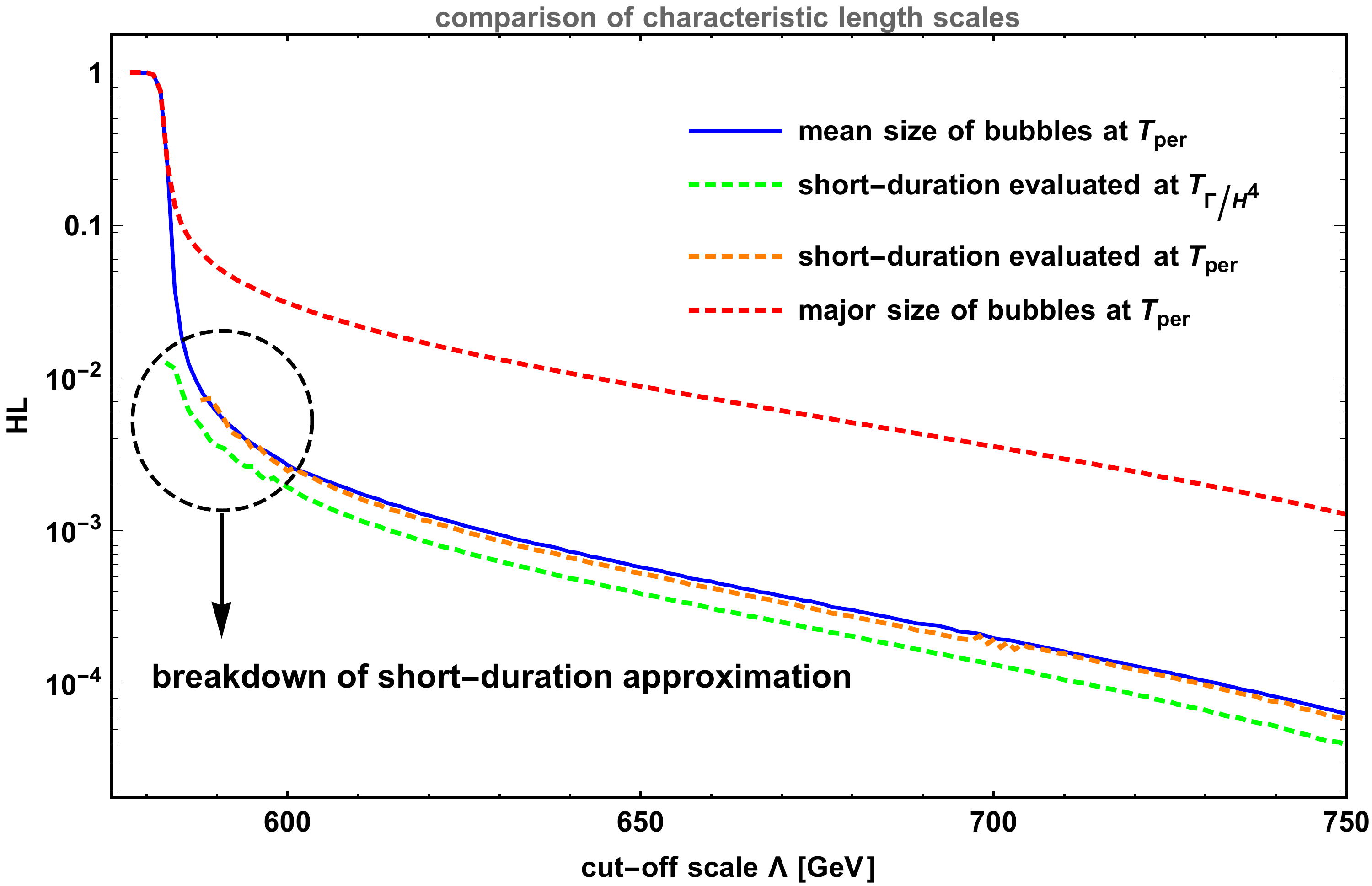}\\
\caption{The characteristic length scales of first-order PT normalized by the Hubble scale. In the regime of fast first-order PT, the characteristic length scale is estimated from the short time-duration by taking the time derivative of bounce action at a reference temperature $T_*=T_{\Gamma/H^4}$ (green dashed line) or $T_*=T_\mathrm{per}$ (orange dashed line). However, the short-duration approximation will break down in the regime of slow first-order PT when the bubble radius in unit of Hubble radius is near or larger than one percent. In the regime of slow first-order PT, it was suggested recently that the physical distance (red dashed line) expanded by the bubbles of majority at percolation since the time of their nucleation could play such role of the characteristic length scale. However, naively computing the red dashed line into the regime of fast first-order PT deviates significantly from the green/orange dashed line. An alternative unified approach for the characteristic length scale is proposed for both regimes of slow and fast first-order PTs from the mean radius of bubbles at percolation presented as the blue solid line. As you can see, the blue solid line perfectly interpolates between the short-duration (orange dashed line) in the regime of fast first-order PT and the major radius (red dashed line) of bubbles in the regime of slow first-order PT. }\label{fig:HLLambda}
\end{figure}

The commonly adopted estimation of the characteristic length scale comes from the short-duration approximation of fast first-order PT, so that the bounce action $S(t)=S(t_*)-\beta(t-t_*)+\mathcal{O}((t-t_*)^2)$ can be expanded at the leading order of duration $t-t_*\simeq\beta^{-1}$. Therefore, the characteristic length scale $L=v_w\beta^{-1}$ is determined by the bubble wall velocity and
\begin{align}\label{eq:betaH}
\beta=-\left.\frac{\mathrm{d}S}{\mathrm{d}t}\right|_{t=t_*}\simeq\frac{\dot{\Gamma}}{\Gamma},\quad\frac{\beta}{H_*}=T_*\left.\frac{\mathrm{d}S}{\mathrm{d}T}\right|_{T=T_*},
\end{align}
Here bubble wall velocity $v_w\approx1$ due to the runaway type of bubble expansion for the toy model we discussed in this section. The initial size of bubble at the onset of nucleation is neglected so that the total duration is sufficient for the estimation of mean radius of bubbles at collision, which is normalized by the Hubble radius at the reference temperature $T_*$. In Fig.~\ref{fig:HLLambda}, we present the normalized time-duration $H/\beta$ with green dashed line at reference temperature $T_*=T_{\Gamma/H^4}$ and with orange dashed line at reference temperature $T_*=T_\mathrm{per}$. The short-duration approximation breaks down when the time-duration is near or large than one percent of Hubble time $H/\beta\gtrsim10^{-2}$. See~\cite{Megevand:2016lpr} for their treatment on the breakdown of short-duration approximation by expanding the bounce action to the quadratic order with Gaussian ansatz.

However, for slow first-order PT, $\beta$ becomes negative at the reference temperature $T_*=T_\mathrm{per}$ if the definition~\eqref{eq:betaH} is insisted. To find a characteristic length scale for slow first-order PT, it was proposed in~\cite{Kobakhidze:2017mru} that the physical distance
\begin{align}\label{eq:majority}
R_\mathrm{major}\equiv R(T_\mathrm{per},T_\mathrm{nuc})
\end{align}
expanded by the bubbles of majority at percolation from the time of their nucleation gives rise to a good candidate. We present in Fig.~\ref{fig:HLLambda} with red dashed line for the normalized major radius of bubbles at percolation $H(T_\mathrm{per})R(T_\mathrm{per},T_\mathrm{nuc})$ for the slow first-order PT. When extrapolated into the regime of fast first-order PT, however, the red dashed line deviates from the green dashed line significantly. The failure of matching red dashed line with green dashed line in the regime of fast first-order PT motivates us to search for an unified characteristic length scale defined for both regimes of slow and fast first-order PTs.

We propose in this paper that the mean radius of bubbles~\cite{Megevand:2016lpr} at percolation could serve as an appropriate measure of the characteristic length scale for both regimes of slow and fast first-order PTs. To calculate the mean radius of bubbles at percolation, one simply averages bubbles radius over the full number density distribution~\eqref{eq:dndR} at percolation time $t_\mathrm{per}$,
\begin{align}\label{eq:mean}
R_\mathrm{mean}\equiv\overline{R}(t_\mathrm{per})\equiv \frac{1}{n(t_\mathrm{per})}\int_{t_\mathrm{tra}}^{t_\mathrm{per}}\mathrm{d}n(t_R)R(t_\mathrm{per},t_R).
\end{align}
We present in Fig.~\ref{fig:HLLambda} with the blue solid line for the normalized mean radius $H(t_\mathrm{per})\overline{R}(t_\mathrm{per})$ of bubbles at percolation. Note that the mismatch between the blue solid line and green dashed line in the regime of fast first-order PT is mainly due to the different choice of the reference temperature. After evaluating~\eqref{eq:betaH} at the reference temperature $T_*=T_\mathrm{per}$, the blue solid line perfectly matches the orange dashed line in the regime of fast first-order PT. Therefore our mean radius~\eqref{eq:mean} of bubbles is capable of interpolating between the short-duration~\eqref{eq:betaH} in the regime of fast first-order PT and the major radius~\eqref{eq:majority} of bubbles in the regime of slow first-order PT. The unified choice~\eqref{eq:mean} with a mean size of bubbles at percolation is more physical intuitive to meet the earlier insight~\cite{Witten:1984rs} that the peak frequency is roughly the inverse of the mean size of bubbles at collisions.

\subsection{Stochastic gravitational waves from numerical simulations}\label{subsec:simulation}

After completing the first step in subsection~\ref{subsec:model} of calculating those phenomenological parameters from the effective potential of a particular particle physics model, we will discuss in this subsection~\ref{subsec:simulation} the second step of obtaining the predictions of the relic GWs from first-order PT. We will first introduce in~\ref{subsubsec:GWcurves} the numerical fitting formulas of the energy density spectrum of the relic GWs from the first-order PT and the sensitivity curves from both space-based laser interferometer GWs detectors and pulsars timing arrays GWs detectors. The results will be presented in~\ref{subsubsec:treeresult} with different prescriptions for the slow and fast first-order PTs.

\subsubsection{The relic gravitational waves and sensitivity curves}\label{subsubsec:GWcurves}

After the PT is completed, there is a short reheating process of the plasma due to the thermal energy of amount $\kappa_T\alpha\rho_\mathrm{rad}$ released during the bubble expansion. Therefore the reheating temperature reads~\cite{Leitao:2015fmj}
\begin{align}\label{eq:Treh}
T_\mathrm{reh}^4=T_*^4(1+\kappa_T\alpha),
\end{align}
where the reference temperature $T_*=T_{\Gamma/H^4}$ for the fast first-order PT and $T_*=T_\mathrm{per}$ for both the slow/fast first-order PTs. It is worth noting that, besides the reheating near the completion of PT, there is another reheating process at the onset of bubble nucleation when relaxing the field value at bubble center from the exit point down to the true vacuum minimum. This extra reheating effect could change the energy budget of bubble expansion, which requires further modeling.

The total redshift factor from the reheating temperature $T_\mathrm{reh}$ to current measured CMB temperature $T_0=2.725\,\mathrm{K}$ is deduced by
\begin{align}
\begin{split}
\frac{a_\mathrm{reh}}{a_0}&=\frac{g_s^{1/3}(T_0)}{g_s^{1/3}(T_\mathrm{reh})}\frac{T_0}{T_\mathrm{reh}}\frac{H_\mathrm{reh}}{H_\mathrm{reh}},\\
&=\frac{g_s^{1/3}(T_0)}{g_s^{1/3}(T_\mathrm{reh})}\frac{T_0}{T_\mathrm{reh}}\left(\frac{\pi^2g_\mathrm{dof}(T_\mathrm{reh})}{90}\right)^{1/2}
\frac{T_\mathrm{reh}^2}{M_\mathrm{Pl}}\frac{1}{H_\mathrm{reh}},\\
&=g_s^{1/3}(T_0)\left(\frac{\pi^2}{90}\right)^{1/2}\frac{T_0}{M_\mathrm{Pl}}g_\mathrm{reh}^{1/6}\frac{T_\mathrm{reh}}{H_\mathrm{reh}},\\
&=1.65\times10^{-5}\,\mathrm{Hz}\left(\frac{g_\mathrm{reh}}{100}\right)^{1/6}\left(\frac{T_\mathrm{reh}}{100\mathrm{GeV}}\right)\frac{1}{H_\mathrm{reh}},\\
\end{split}
\end{align}
where the conservation $s(T_\mathrm{reh})a_\mathrm{reh}^3=s(T_0)a_0^3$ of entropy density $s(T)\sim g_s(T)T^3$ is used in the first line, the FRW equation $3M_\mathrm{Pl}^2H_\mathrm{reh}^2=\frac{\pi^2}{30}g_\mathrm{dof}(T_\mathrm{reh})T_\mathrm{reh}^4$ is used in the second line, an approximation on the effective d.o.f. of entropy density and of energy density $g_s(T_\mathrm{reh})\approx g_\mathrm{dof}(T_\mathrm{reh})\equiv g_\mathrm{reh}$ is used in the third line, and the current d.o.f. $g_s(T_0)=2+\frac{4}{11}\times\frac{7}{8}\times2 N_\mathrm{eff}$ with $N_\mathrm{eff}=3.046$ is used in the last line. The Natural Unit converters $1\,\mathrm{K}=8.617\times10^{-5}\,\mathrm{eV}$ and $1\,\mathrm{eV}=1.519\times10^{15}\,\mathrm{Hz}$ are used to get the final expression.
Therefore the redshifted peak frequency~\cite{Kosowsky:1991ua} should scale as $f(a_0)=\frac{a_\mathrm{reh}}{a_0}f(a_*)$, namely
\begin{align}
f_0=1.65\times10^{-5}\,\mathrm{Hz}\left(\frac{g_\mathrm{dof}}{100}\right)^{1/6}
\left(\frac{T_*/(1+\kappa_T\alpha)^{1/4}}{100\mathrm{GeV}}\right)\frac{f_*}{H_*}.
\end{align}
In what follows, we will abbreviate $T_\star\equiv T_*/(1+\kappa_T\alpha)^\frac14$ to take into account the reheating effect after the completion of first-order PT. In what follows, quantity with subscript ``$_*$'' will be evaluated at the reference temperature $T_*$, which should not be confused with the reheating temperature $T_\star$ encountered in the redshift factor.

The redshifted energy density~\cite{Kosowsky:1991ua} should scale as $\rho_\mathrm{GW}(a)\sim a^{-4}$, hence its normalized version $\Omega_\mathrm{GW}(a)\equiv\rho_\mathrm{GW}(a)/\rho_\mathrm{crit}(a)$ with respect to the critical energy density scales as
\begin{align*}
\Omega_\mathrm{GW}(a_0)=1.67\times10^{-5}h^{-2}\left(\frac{100}{g_\mathrm{dof}}\right)^{1/3}\Omega_\mathrm{GW}(a_*).
\end{align*}
Here the scaling behavior of the energy density $\Omega_\mathrm{GW}(a_*)$ of GWs from the first-order PT can be estimated by simple dimensional analysis~\cite{Kosowsky:1991ua}: the dimensionless energy density $\Omega_\mathrm{GW}\sim\rho_\mathrm{GW}/\rho_\mathrm{crit}$ is defined by the critical energy density $\rho_\mathrm{crit}\sim(1+\alpha)\rho_\mathrm{rad}$ and the GW energy density $\rho_\mathrm{GW}\sim E_\mathrm{GW}/(v\beta^{-1})^3$, where $(v_w\beta^{-1})^3$ measures the volume of bubble with bubble wall velocity $v_w$ after an effective duration $\beta^{-1}$. The GW energy $E_\mathrm{GW}\sim P_\mathrm{GW}\beta^{-1}$ can be estimated from the power $P_\mathrm{GW}\sim G_\mathrm{N}\dot{E}_\mathrm{kin}^2$, where the Newtonian constant $G_\mathrm{N}\sim H^2/(1+\alpha)\rho_\mathrm{rad}$ and the time derivative of kinetic energy $\dot{E}_\mathrm{kin}\sim\beta E_\mathrm{kin}$. After inserting the kinetic energy $E_\mathrm{kin}\sim\kappa\alpha\rho_\mathrm{rad}(v_w\beta^{-1})^3$, the scaling behavior of the dimensionless energy density is estimated as
\begin{align}
\Omega_\mathrm{GW}(a_*)\sim\left(\frac{H_*}{\beta}\right)^2\left(\frac{\kappa\alpha}{1+\alpha}\right)^2v_w^3.
\end{align}
Since the GW relic from the first-order PT is a stochastic background, there is no definite wave form for the observed strain. Therefore one usually adopts the energy density spectrum
\begin{align}
h^2\Omega_\mathrm{GW}(f)\equiv \frac{h^2}{\rho_\mathrm{crit}}\frac{\mathrm{d}\rho_\mathrm{GW}}{\mathrm{d}\log f}
\end{align}
to characterize the energy density per logarithmic frequency interval normalized to the critical density of the Universe. The outcomes from the simulations of bubble collisions usually use an analytic power-law fitting for the shape of the energy density spectrum around the peak frequency. It is worth noting that, the validity of all the fitting results below have not been verified in the case of slow first-order PT with Hubble scale bubbles of runaway type, therefore a larger simulation lasting longer than several Hubble times is required for future works. Here we only naively adopt the fitting results from the state-of-art simulations.

The first source of GWs from first-order PT comes from the uncollided envelop of thin bubble walls during the bubble collision, while the collided thin bubble walls are assumed to disappear instantly after two bubbles overlap. This is the widely used envelop approximation that contributes to both numerical simulations~\cite{Kosowsky:1991ua,Kosowsky:1992rz,Kosowsky:1992vn,Kamionkowski:1993fg,Huber:2008hg} (see also~\cite{Child:2012}) and analytic estimations~\cite{Jinno:2016vai}. The dimensionless energy density spectrum is fitted as~\cite{Huber:2008hg}
\begin{align}\label{eq:Omegaenv}
h^2\Omega_\phi=\underbrace{1.67\times10^{-5}\left(\frac{100}{g_\mathrm{dof}}\right)^\frac13}
\underbrace{(HL)_*^2\left(\frac{\kappa_\phi\alpha}{1+\alpha}\right)^2\frac{0.11v_w^3}{0.42+v_w^2}}
\underbrace{\frac{3.8\left(f/f_\phi\right)^{2.8}}{1+2.8\left(f/f_\phi\right)^{3.8}}},
\end{align}
where the first underbrace accounts for the redshift effect, the second one reflects its scaling behavior, and the third one parameterizes the spectral shape of the GW radiation. The peak frequency involved in the spectral shape is fitted as~\cite{Huber:2008hg}
\begin{align}\label{eq:fenv}
f_\phi=\underbrace{1.65\times10^{-5}\,\mathrm{Hz}\left(\frac{g_\mathrm{dof}}{100}\right)^\frac16\frac{T_\star}{100\mathrm{GeV}}}
\underbrace{\frac{0.62}{1.8-0.1v_w+v_w^2}\frac{1}{(HL)_*}},
\end{align}
where the first underbrace accounts for the redshift effect, while the second one approximates the peak frequency with respect to the Hubble scale at the reference temperature. Here the characteristic normalized length scale $(HL)_*$ is estimated by the normalized time-duration $H_*/\beta$ for the fast first-order PT. For the slow first-order PT in~\cite{Kobakhidze:2017mru}, $(HL)_*$ is estimated by the size of bubbles of majority at percolation temperature. For our unified description of both the slow and fast first-order PT, $(HL)_*$ is estimated by the mean radius of bubbles at percolation temperature. The numerical fitting formulas~\eqref{eq:Omegaenv} and~\eqref{eq:fenv} from envelop approximation~\cite{Weir:2016tov} provide an adequate modeling for the scalar field contribution to the GWs from first-order PT, especially for the slow first-order PT where the most of the released vacuum energy has deposited into accelerating the bubble walls.

The second source of GWs from first-order PT comes from the sound waves of bulk fluid motion~\cite{Hogan:1986qda}, of which the numerical simulations~\cite{Hindmarsh:2013xza,Hindmarsh:2015qta} are carried out for the scalar-fluid system without explicitly using the envelop approximation. The sound waves of bulk fluid motion come from the collision of the long-lasting shell of fluid kinetic energy in the plasma that steamed from the bubble expansion in the fluid. The dimensionless energy density spectrum is fitted as~\cite{Hindmarsh:2015qta}
\begin{align}\label{eq:Omegasw}
h^2\Omega_\mathrm{sw}=\underbrace{2.65\times10^{-6}\left(\frac{100}{g_\mathrm{dof}}\right)^\frac13}
\underbrace{(HL)_*\left(\frac{\kappa_\mathrm{sw}\alpha}{1+\alpha}\right)^2}
\underbrace{\frac{7^\frac72\left(f/f_\mathrm{sw}\right)^3}{\left(4+3\left(f/f_\mathrm{sw}\right)^2\right)^\frac72}},
\end{align}
where the first underbrace accounts for the redshift effect, the second one reflects its scaling behavior, and the third one parameterizes the spectral shape of the GW radiation. The peak frequency involved in the spectral shape is fitted as~\cite{Hindmarsh:2015qta}
\begin{align}\label{eq:fsw}
f_\mathrm{sw}=\underbrace{1.65\times10^{-5}\,\mathrm{Hz}\left(\frac{g_\mathrm{dof}}{100}\right)^\frac16\frac{T_\star}{100\mathrm{GeV}}}
\underbrace{\frac{2}{\sqrt{3}(HL)_*}},
\end{align}
where the first underbrace accounts for the redshift effect, while the second one approximates the peak frequency with respect to the Hubble scale at the reference temperature. It is worth noting that, the numerical fitting formulas~\eqref{eq:Omegasw} and~\eqref{eq:fsw} are limited to the case of fast first-order PT with weak release of vacuum energy, whereas the sound waves can last for several Hubble times to enhance its contribution to the total GWs radiations. However, its extrapolation into the regime of slow first-order PT with strong release of vacuum energy requires further investigations.

The third source of GWs from first-order PT comes from the MHD turbulence~\cite{Kamionkowski:1993fg,Kosowsky:2001xp,Dolgov:2002ra,Nicolis:2003tg,Caprini:2006jb,Gogoberidze:2007an,Caprini:2009yp} of bulk plasma motion, of which the both the velocity field of bulk fluid and the magnetic field of fully ionized plasma are stirred up due to the bubble expansion in the fluid. Analytically modeling the MHD turbulence as the Kolmogorov-type turbulence exactly derives the spectral shape of the dimensionless energy density spectrum~\cite{Caprini:2009yp},
\begin{align}\label{eq:Omegaturb}
h^2\Omega_\mathrm{turb}=\underbrace{3.35\times10^{-4}\left(\frac{100}{g_\mathrm{dof}}\right)^\frac13}
\underbrace{(HL)_*\left(\frac{\kappa_\mathrm{turb}\alpha}{1+\alpha}\right)^\frac32}
\underbrace{\frac{\left(f/f_\mathrm{turb}\right)^3}{\left(1+f/f_\mathrm{turb}\right)^\frac{11}{3}\left(1+8\pi f/H_0\right)}},
\end{align}
where the first underbrace accounts for the redshift effect, the second one reflects its scaling behavior, and the third one parameterizes the spectral shape of the GW radiation, and the redshift Hubble constant today is
\begin{align}
H_0=1.65\times10^{-5}\,\mathrm{Hz}\left(\frac{g_\mathrm{dof}}{100}\right)^\frac16\frac{T_\star}{100\mathrm{GeV}}.
\end{align}
The peak frequency involved in the spectral shape is fitted as~\cite{Caprini:2009yp}
\begin{align}\label{eq:fturb}
f_\mathrm{turb}=\underbrace{1.65\times10^{-5}\,\mathrm{Hz}\left(\frac{g_\mathrm{dof}}{100}\right)^\frac16\frac{T_\star}{100\mathrm{GeV}}}
\underbrace{\frac{3.5}{2(HL)_*}},
\end{align}
where the first underbrace accounts for the redshift effect, while the second one approximates the peak frequency with respect to the Hubble scale at the reference temperature. The characteristic length scale of turbulence is the stirring scale, which is two times of the mean radius of bubbles $2(HL)_*$ at collisions. The numerical fitting formulas~\eqref{eq:Omegaturb} and~\eqref{eq:fturb} with analytically estimated spectral shape is also limited to the case of fast first-order PT that lasts for less than one eddy turn-over time, whereas the turbulence can last for several Hubble times before dissipation. Moreover, the extra conversion of sound waves to turbulence flows from the development of shocks at $t_\mathrm{shock}\sim L_\mathrm{f}/\overline{U_\mathrm{f}}$ would necessarily enhance the turbulence contribution for fast first-order PT but suppress the turbulence contribution for slow first-order PT, because the physical bubble radius $L_\mathrm{f}$ is so large with respect to the r.m.s. fluid velocity $\overline{U_\mathrm{f}}$ for the slow first-order PT.

To characterize the detectability of the relic GWs from a certain model that commits a first-order PT, it is instructive to compare its energy density spectrums with the sensitivity curves of various GWs detectors. See~\cite{Moore:2014lga} for a pedagogical introduction on the GWs sensitivity curves. It is worth noting that, the sensitivity curves we will present below for several GWs detectors serve as a conservative estimations due to the increase in sensitivity for the power-law spectrum of stochastic GW backgrounds~\cite{Thrane:2013oya}. The energy density spectrum $\Omega_\mathrm{GW}(f)$, the power spectral density $S_h(f)$, and the characteristic strain $h_c(f)$ are related via
\begin{align}
H_0^2\Omega_\mathrm{GW}(f)=\frac{2\pi^2}{3}f^3S_h(f)=\frac{2\pi^2}{3}f^2h_c^2(f),
\end{align}
where the characteristic strain of signal $h_c^2(f)=4f^2|\widetilde{h}(f)|^2$ is defined in terms of the Fourier transformation $\widetilde{h}(f)$ of the observed signal $h(t)$ in $s(t)=n(t)+h(t)$. To obtain the expression for the noise, one simply substitutes $S_n(f)$ for $S_h(f)$ and $h_n(f)$ for $h_c(f)$, where the characteristic strain of noise $h_n^2(f)=fS_n(f)$ is defined by its power spectral density $S_n(f)$ that is computed from the $\langle\widetilde{n}(f)\widetilde{n}^*(f')\rangle=\frac12\delta(f-f')S_n(f)$. For reference, we list below the sensitivity curves from three space-based laser interferometer GWs detectors and three pulsars timing array GWs detectors.

The analytic fitting to the six sky-averaged sensitivity curves for evolving Laser Interferometer Space Antenna (eLISA) is of the form~\cite{Klein:2015hvg}
\begin{align}
S_n(f)=\frac{20}{3}\frac{4S_{n,acc}(f)+S_{n,sn}(f)+S_{n,omn}(f)}{L^2}\left[1+\left(\frac{f}{0.41\frac{c}{2L}}\right)^2\right],
\end{align}
where the noise from low-frequency acceleration,
\begin{align}
\begin{split}
S_{n,acc}(f)=\left\{
               \begin{array}{ll}
                 9\times10^{-28}\frac{1}{(2\pi f)^4}\left(1+\frac{10^{-4}\mathrm{Hz}}{f}\right)\mathrm{m}^2\mathrm{Hz}^{-1}, & \hbox{N1;} \\
                 9\times10^{-30}\frac{1}{(2\pi f)^4}\left(1+\frac{10^{-4}\mathrm{Hz}}{f}\right)\mathrm{m}^2\mathrm{Hz}^{-1}, & \hbox{N2.}
               \end{array}
             \right.
\end{split}
\end{align}
the noise from shot noise,
\begin{align}
\begin{split}
S_{n,sn}(f)=\left\{
              \begin{array}{ll}
                1.98\times10^{-23}\mathrm{m}^2\mathrm{Hz}^{-1}, & \hbox{A1;} \\
                2.22\times10^{-23}\mathrm{m}^2\mathrm{Hz}^{-1}, & \hbox{A2;} \\
                2.96\times10^{-23}\mathrm{m}^2\mathrm{Hz}^{-1}, & \hbox{A5.}
              \end{array}
            \right.
\end{split}
\end{align}
and the noise from other measurement noise,
\begin{align}
S_{n,omn}=2.65\times10^{-23}\mathrm{m}^2\mathrm{Hz}^{-1}
\end{align}
The preliminary configurations of eLISA~\cite{Caprini:2015zlo} consists of the noise level (N1: LISA Pathfinder required, N2: LISA Pathfinder expected), arm length (A1: 1 million km, A2: 2 million km, A5: 5 million km), mission duration (M2: 2 years, M5: 5 years) and laser links (L4: 4 links, L6: 6 links), which will be labeled by N$i$A$j$M$k$L$l$. We adopt the configurations C1 (N2A5M5L6), C2 (N2A1M5L6), C3 (N2A2M5L4), C4 (N1A1M2L4) in our study.

The characteristic strain of the Deci-Hertz Interferometer Gravitational wave Observatory (DECIGO) is given in by~\cite{Yagi:2011wg}
\begin{align}
S_h(f)=S_{n,sn}(f)+S_{n,rp}(f)+S_{n,acc}(f)\quad\mathrm{Hz}^{-1},
\end{align}
where the shot noise, the radiation pressure, and the acceleration noise are
\begin{align}
S_{n,sn}(f)&=7.05\times10^{-48}\left[1+\left(\frac{f}{f_p}\right)^2\right],\\
S_{n,rp}(f)&=4.8\times10^{-51}\left(\frac{f}{1\mathrm{Hz}}\right)^{-4}\frac{1}{1+\left(\frac{f}{f_p}\right)^2},\\
S_{n,acc}(f)&=5.33\times10^{-52}\left(\frac{f}{1\mathrm{Hz}}\right)^{-4},\quad f_p=7.36\,\mathrm{Hz}
\end{align}
respectively. The characteristic strain of the Big Bang Observer (BBO) is given in by~\cite{Yagi:2011wg}
\begin{align}
S_h(f)=2.00\times10^{-49}\left(\frac{f}{1\mathrm{Hz}}\right)^2+4.58\times10^{-49}+1.26\times10^{-51}\left(\frac{f}{1\mathrm{Hz}}\right)^{-4}\,\mathrm{Hz}^{-1}
\end{align}

The characteristic strain of a pulsar timing array (PTA) is given in by~\cite{Moore:2014lga}
\begin{align}
h_c(f)=\{f=1/T,\sqrt{24\pi^2\Delta t}\sigma f^\frac32\},
\end{align}
where $f=1/T$ is a vertical line and $\sqrt{24\pi^2\Delta t}\sigma f^\frac32$ is a simple power-law fitting line started at $f=1/T$ and ended at $f=1/\Delta t$. We will use the configurations of the European Pulsar Timing Array (EPTA) that assumes 5 pulsars timed every $\Delta t=2$ weeks for $T=10$ years with an r.m.s. error in each timing residual of $\sigma=100$ ns, and the International Pulsar Timing Array (IPTA) that assumes 20 pulsars timed every $\Delta t=2$ weeks for $T=15$ years with an r.m.s. error in each timing residual of $\sigma=50$ ns, and the Square Kilometre Array (SKA) that assumes 50 pulsars timed every $\Delta t=2$ weeks for $T=20$ years with an r.m.s. error in each timing residual of $\sigma=30$ ns.

\subsubsection{The results from slow/fast first-order phase transition}\label{subsubsec:treeresult}

\begin{figure}
  \centering
  \includegraphics[width=0.48\textwidth]{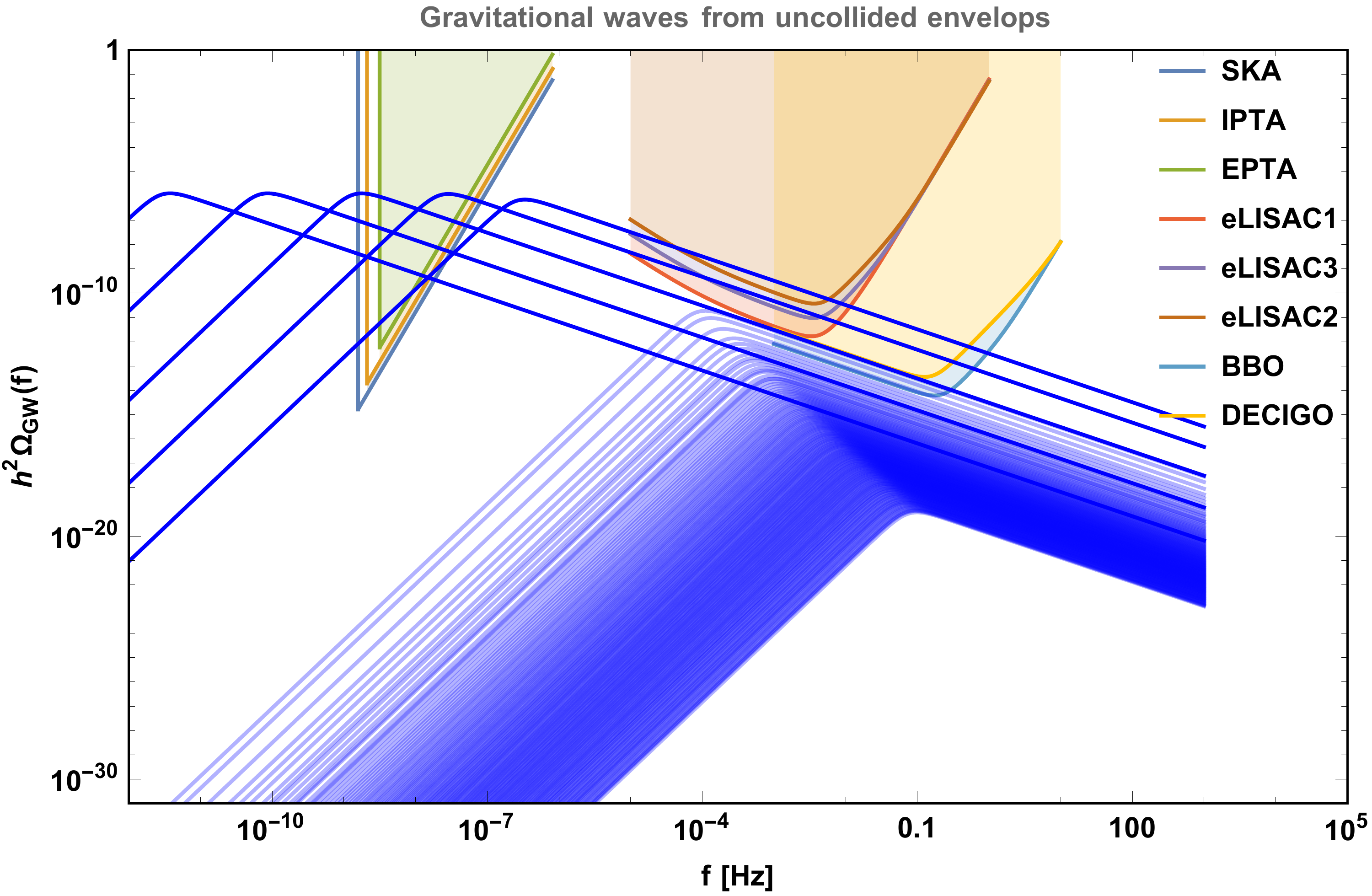}
  \includegraphics[width=0.48\textwidth]{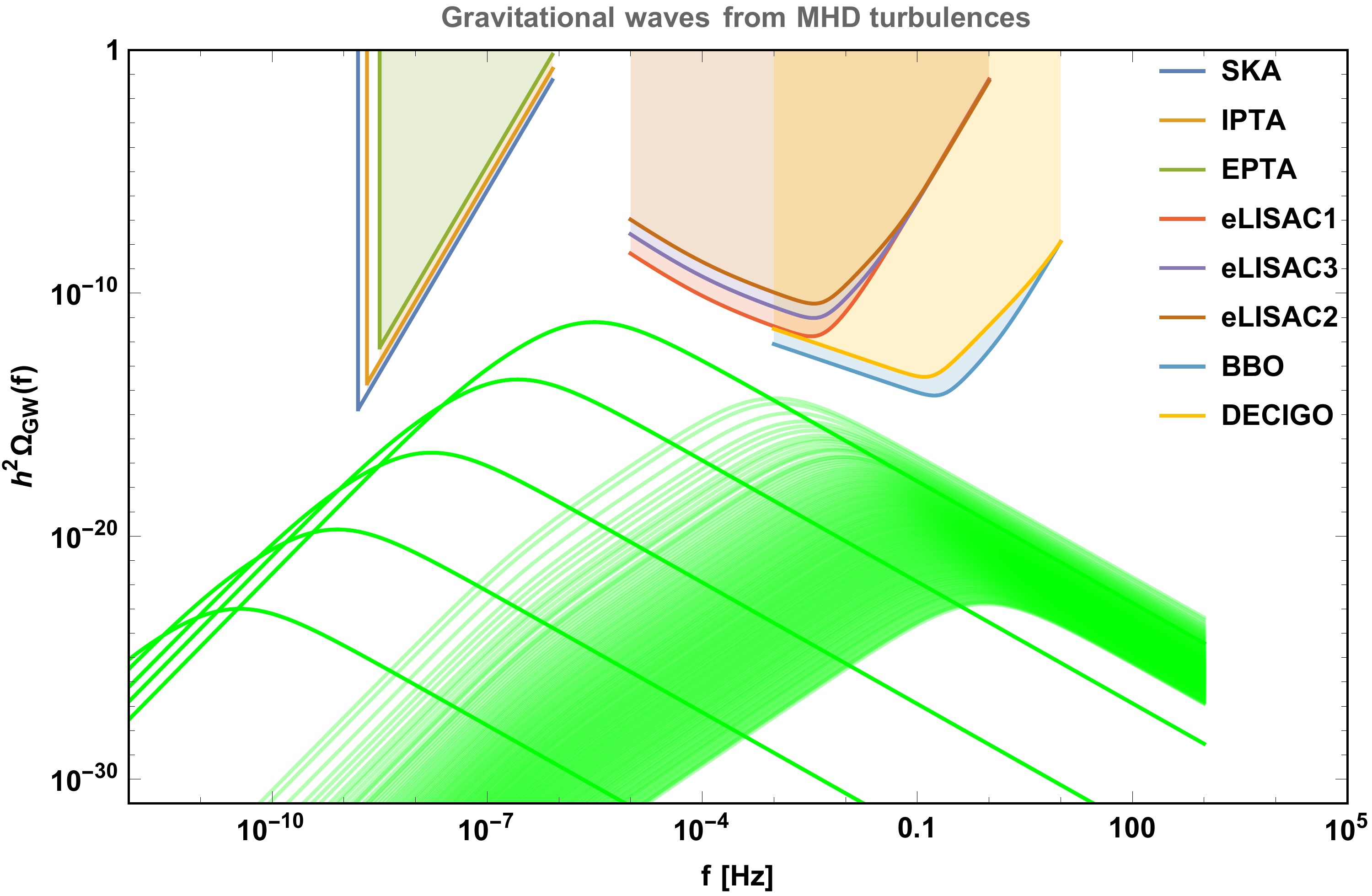}\\
  \includegraphics[width=0.48\textwidth]{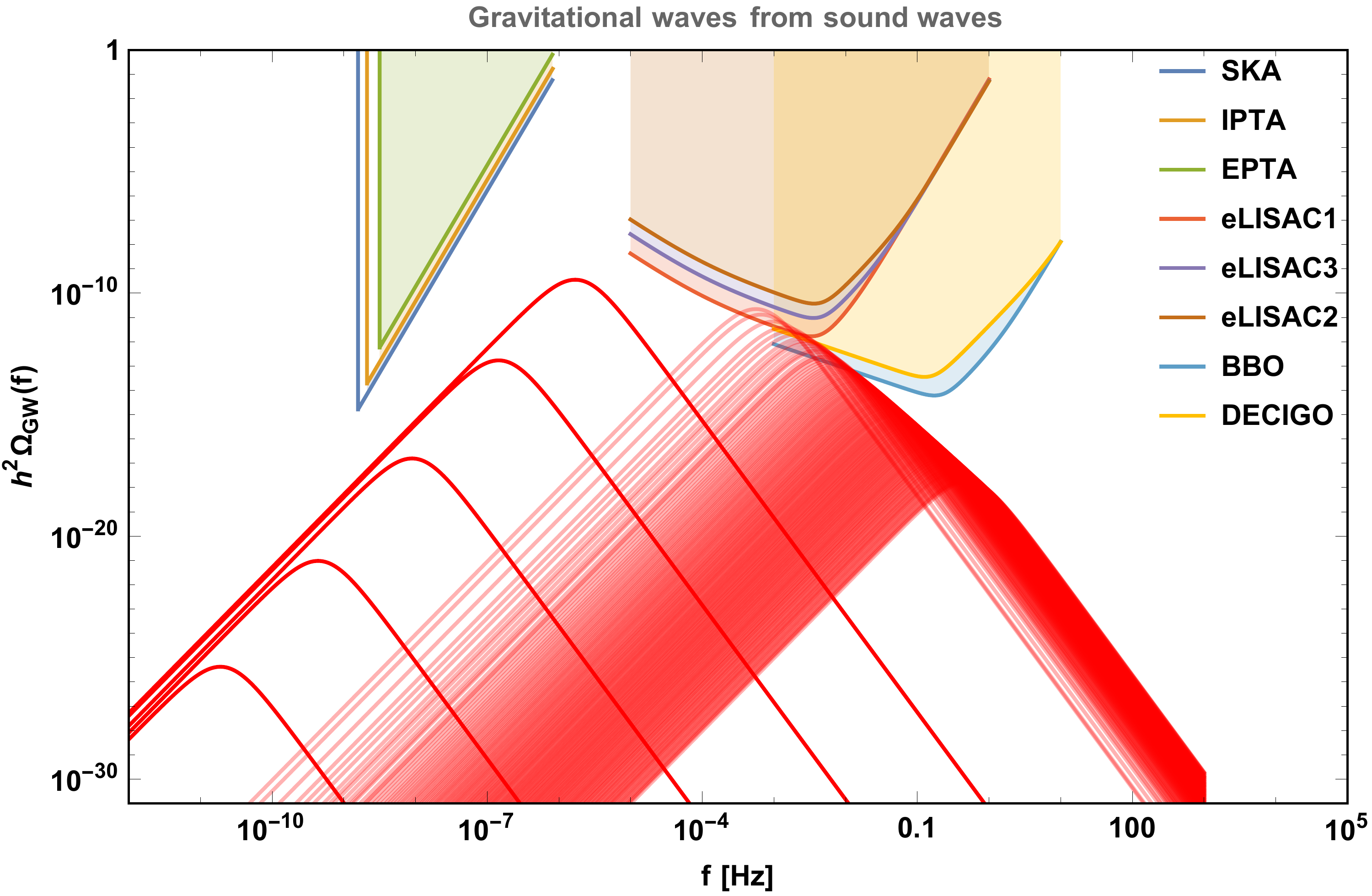}
  \includegraphics[width=0.48\textwidth]{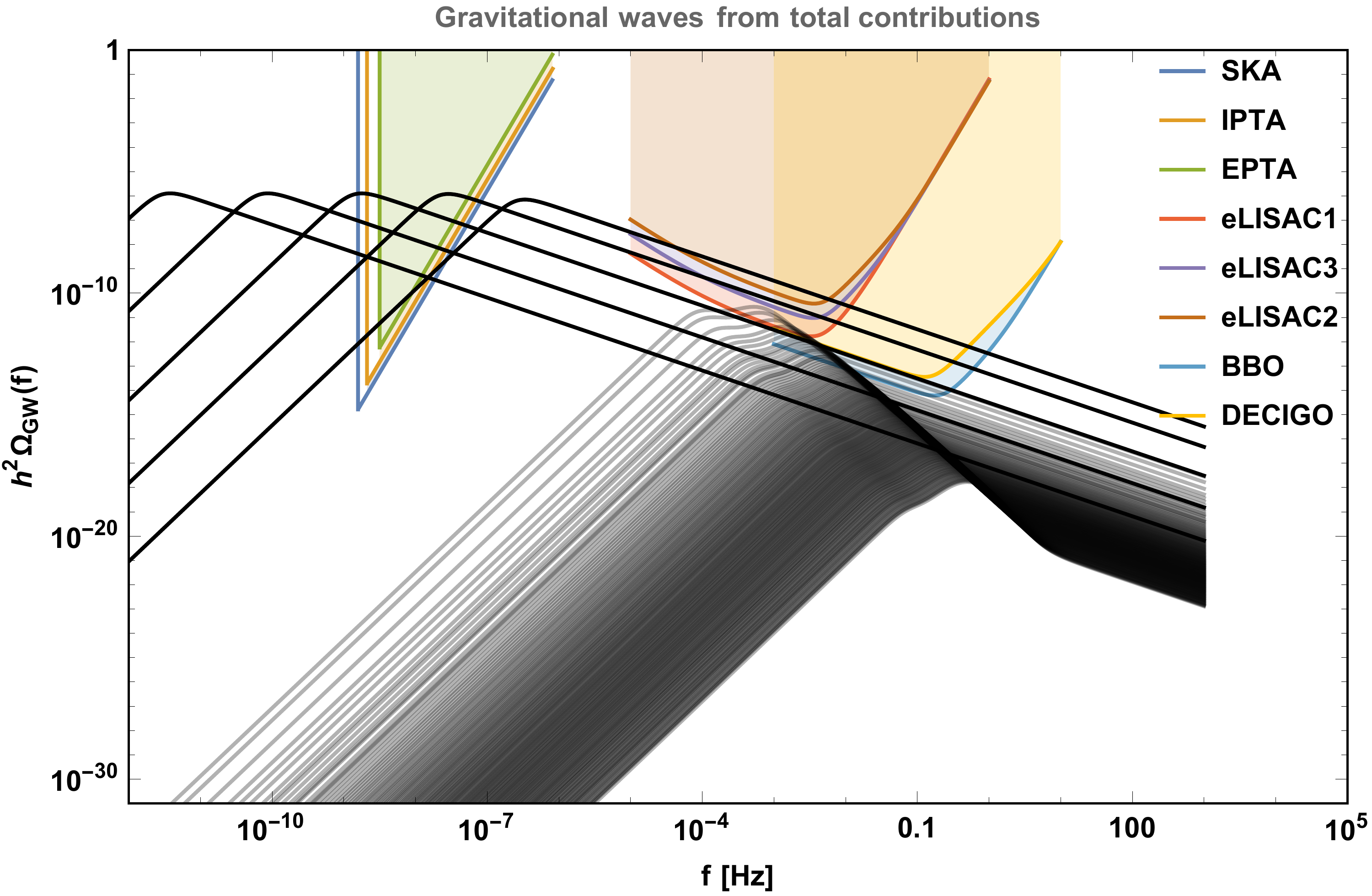}\\
  \includegraphics[width=0.48\textwidth]{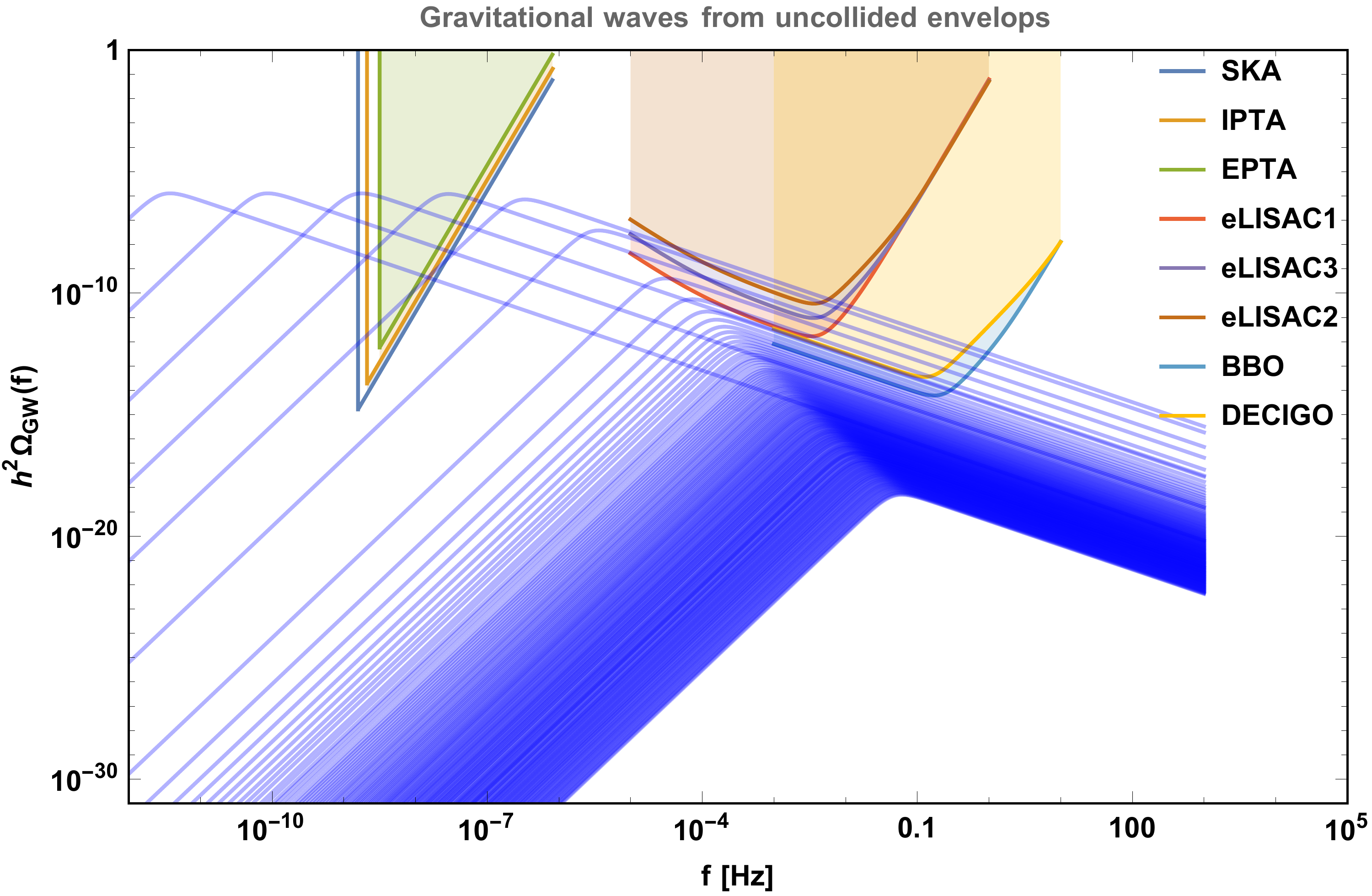}
  \includegraphics[width=0.48\textwidth]{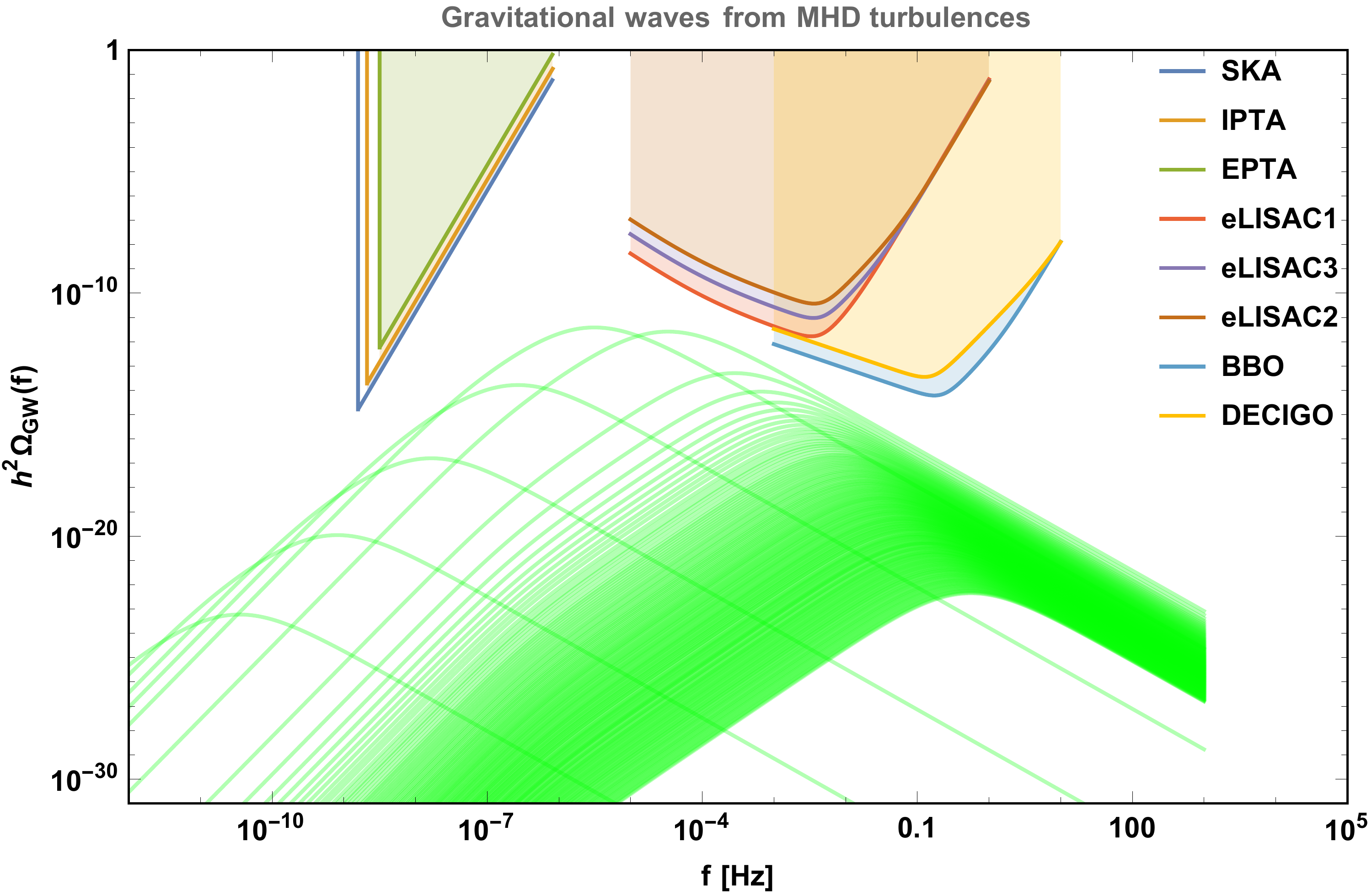}\\
  \includegraphics[width=0.48\textwidth]{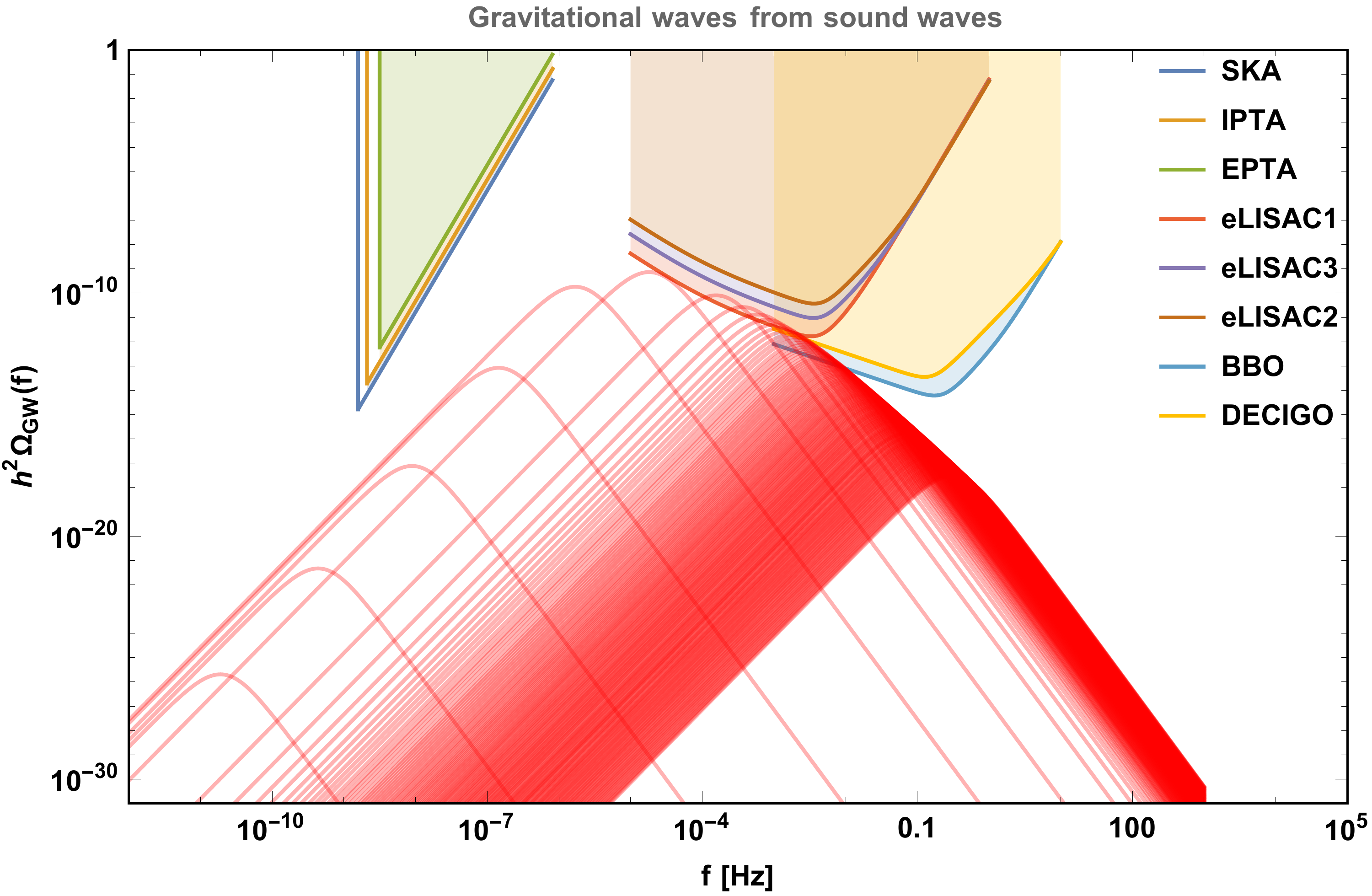}
  \includegraphics[width=0.48\textwidth]{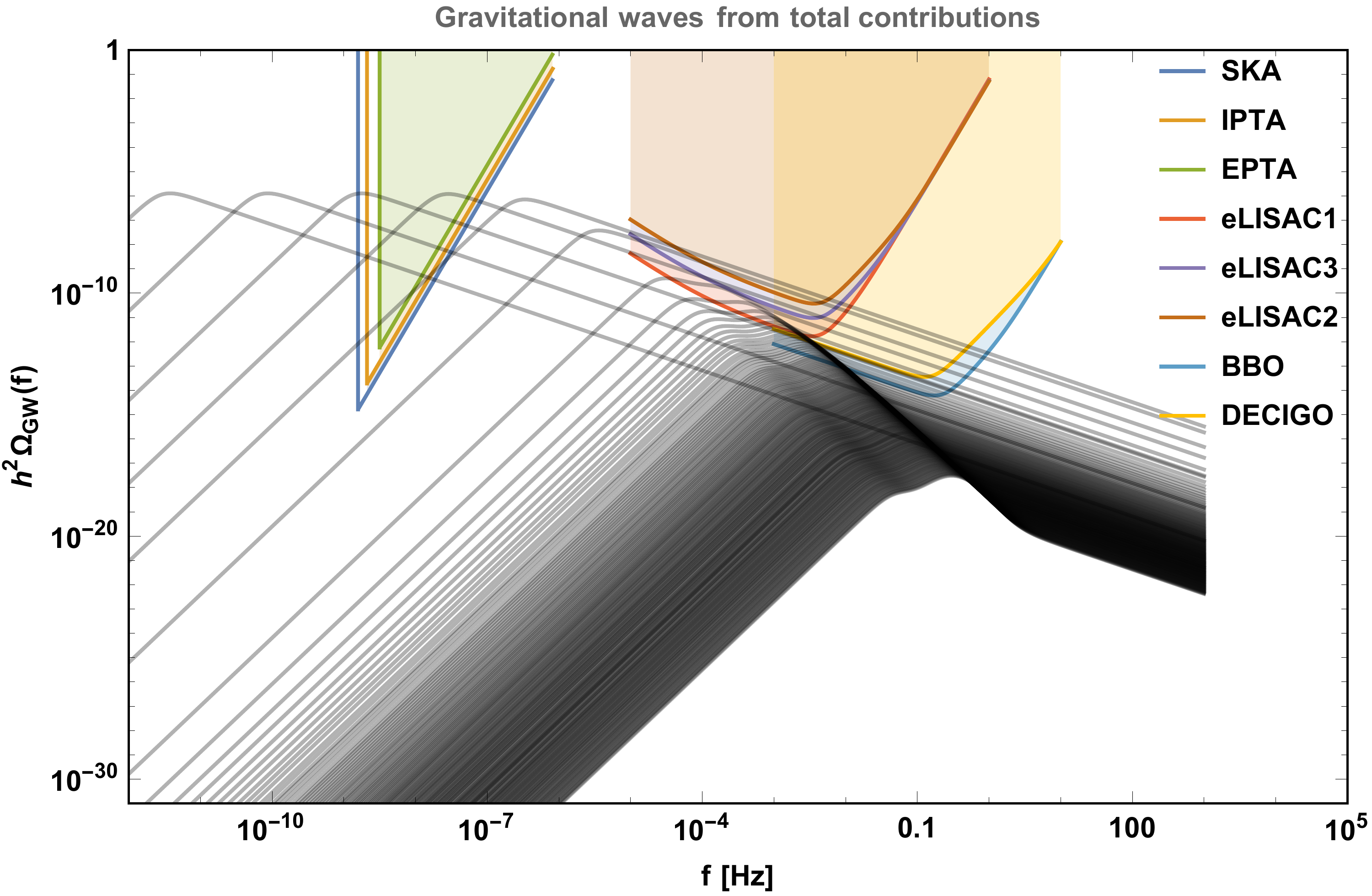}\\
  \caption{The energy density spectrum of the relic GWs from the first-order PT with respect to the cut-off scale $\Lambda$ ranging from $578\,\mathrm{GeV}$ (low peak frequency) to $750\,\mathrm{GeV}$ (high peak frequency) with interval $1\,\mathrm{GeV}$, where the first four panels present a piecewise combination of prescriptions~\eqref{eq:FFPT} and~\eqref{eq:SFPT}, while the last four panels present our unified prescription~\eqref{eq:UFPT}. }\label{fig:toyGW}
\end{figure}

Now we will carry out the predictions of the relic GWs from the first-order PT. The reference temperature $T_*$, the strength factor $\alpha$, and the characteristic length scale $HL$ involved in~\eqref{eq:Omegaenv},~\eqref{eq:Omegasw} and~\eqref{eq:Omegaturb} are summarized as following three prescriptions:
\begin{enumerate}
  \item For fast first-order PT,
      \begin{align}\label{eq:FFPT}
      \begin{split}
       T_*=T_{\Gamma/H^4}\quad&\hbox{from}~\eqref{eq:GammaH4},\\
       \alpha=\frac{\Delta\rho_\mathrm{vac}(T_*)}{\rho_\mathrm{rad}(T_*)}\quad&\hbox{from}~\eqref{eq:alphaI},\\
       (HL)_*=H_*/\beta\quad&\hbox{from}~\eqref{eq:betaH}.
      \end{split}
      \end{align}
  \item For slow first-order PT,
      \begin{align}\label{eq:SFPT}
      \begin{split}
       T_*=T_\mathrm{per}\quad&\hbox{from}~\eqref{eq:percolation},\\
       \alpha=\frac{\rho_\mathrm{kin}(T_*)}{\rho_\mathrm{rad}(T_*)}\quad&\hbox{from}~\eqref{eq:alphaII},\\
       (HL)_*=H_*R_\mathrm{major}\quad&\hbox{from}~\eqref{eq:majority}.
      \end{split}
      \end{align}
  \item For our unified description for both slow and fast first-order PTs,
      \begin{align}\label{eq:UFPT}
      \begin{split}
       T_*=T_\mathrm{per}\quad&\hbox{from}~\eqref{eq:percolation},\\
       \alpha=\frac{\Delta\rho_\mathrm{vac}^\mathrm{tot}(T_*)}{\rho_\mathrm{rad}(T_*)}\quad&\hbox{from}~\eqref{eq:alphaIII},\\
       (HL)_*=H_*R_\mathrm{mean}\quad&\hbox{from}~\eqref{eq:mean}.
      \end{split}
      \end{align}
\end{enumerate}
The results are summarized in Fig.~\ref{fig:toyGW}, where the black solid lines present the total contributions from the uncollided envelops (blue solid lines), the MHD turbulences (green solid lines) and the sound waves (red solid lines) with respect to the cut-off scale $\Lambda$ ranging from $578\,\mathrm{GeV}$ (low peak frequency) to $750\,\mathrm{GeV}$ (high peak frequency) with interval $1\,\mathrm{GeV}$. The first four panels present a piecewise combination of prescription~\eqref{eq:FFPT} and prescription~\eqref{eq:SFPT}, while the last four panels present our unified prescription~\eqref{eq:UFPT} for both slow and fast first-order PTs. Although not significantly announced, for a given cut-off scale, the peak frequency from our unified prescription is lower than that from the combined prescriptions in the regime of fast first-order PT, while the energy density spectrum at peak frequency from our unified prescription is higher than that from the combined prescriptions in the regime of fast first-order PT. Most importantly, the transition between the slow and fast first-order PTs with our unified prescription is smoother than that from the combined prescription.

\section{The effect from a fixed/running renormalization-group scale}\label{sec:fixrun}

In this section, after taking into account the full one-loop corrections instead of the toy model~\eqref{eq:treehighT}, we will explore the effects on the predictions of the relic GWs from the first-order PT with a fixed/running RG scale. Although the potential barrier from the sextic term~\eqref{eq:tree} is already manifested at the tree-level, the predictions of the relic GWs from the first-order PT deviate significantly when the full one-loop corrections are taken into account. In most of the literatures concerning the GWs from first-order PTs, the RG scale involved in the effective potential with quantum corrections is usually chosen at a fixed scale. However, those coupling constants that conspire to give rise to a potential barrier could flip signs when considering a running RG scale. Therefore it is necessarily important to investigate how a running RG scale could affect the predictions compared to a fixed RG scale. In this section, we will closely follow the treatment of~\cite{Delaunay:2007wb} for a fixed RG scale and the one in \cite{Rose:2015lna} for a running RG scale.

\subsection{The full one-loop analysis}\label{subsec:looppotential}

The full one-loop corrections will be given in this subsection, where the vacuum corrections under $\overline{MS}$ scheme are given in~\ref{subsubsec:loopvacuum}, the renormalization conditions with pole matching initial conditions are given in~\ref{subsubsec:RGcondition}, the thermal corrections with resumed ring diagrams are given in~\ref{subsubsec:loopthermal}, and the results of the RG improvement from one-loop RG equations with sextic term are given in~\ref{subsubsec:RGimprovement}. The infrared (IR) divergences from massless Goldstone bosons encountered when posing the renormalization condition have been taken into account by appropriate pole matching conditions~\cite{Buttazzo:2013uya}, and the imaginary parts from the negative effective masses of Higgs boson and Goldstone bosons when penetrating through the potential barrier have been maximally canceled by adding thermal corrections with ring corrections, since we have retained the presence of both Higgs boson and Goldstone bosons.

\subsubsection{The one-loop vacuum corrections}\label{subsubsec:loopvacuum}

According to the background field method, expanding the classical action on the classical value of Higgs field gives rise to an effective potential $V_\mathrm{eff}(\phi,T)=V_\mathrm{tree}(\phi)+V_\mathrm{1-loop}(\phi,T)$, of which the tree-level part is still given by the classical potential~\eqref{eq:tree}, while the full one-loop part, after resumming in the IR limit only the zero-mode of the propagator shifted by the Debye mass, is split into two parts~\cite{Delaunay:2007wb},
\begin{align}\label{eq:oneloop}
\begin{split}
V_\mathrm{1-loop}(\phi,T)&=\sum\limits_{i=h,\chi,W,Z,t}\frac{n_iT}{2}\sum\limits_{0\neq n=-\infty}^{+\infty}
\int\frac{\mathrm{d}^3\vec{k}}{(2\pi)^3}\log\left[\vec{k}^2+\omega_n^2+m_i^2(\phi)\right]\\
&+\sum\limits_{i=h,\chi,W,Z,t}\frac{n_iT}{2}\int\frac{\mathrm{d}^3\vec{k}}{(2\pi)^3}\log\left[\vec{k}^2+m_i^2(\phi)+\Pi_i(\phi,T)\right].
\end{split}
\end{align}
The Euclidean loop four-momentum $k_E=(\omega_n,\vec{k})$ consists of the Matsubara frequencies $\omega_n=2n\pi T$ for boson and $\omega_n=(2n+1)\pi T$ for fermion. The Debye mass is added to boson sectors in order to resum the IR divergent ring diagrams reorganized as the second line of~\eqref{eq:oneloop}, of which the concrete forms will be given in~\ref{subsubsec:loopthermal}. The first line of~\eqref{eq:oneloop} can be further split into the zero-temperature vacuum potential and the finite-temperature thermal potential, therefore the full one-loop potential~\eqref{eq:oneloop} consists of three parts~\cite{Delaunay:2007wb}, $V_\mathrm{1-loop}(\phi,T)=V_\mathrm{1-loop}^{T=0}(\phi)+V_\mathrm{1-loop}^{T\neq0}(\phi,T)+V_\mathrm{1-loop}^\mathrm{ring}(\phi,T)$, namely
\begin{align}
V_\mathrm{1-loop}^{T=0}(\phi)&=\sum\limits_{i=h,\chi,W,Z,t}\frac{n_i}{2}\int\frac{\mathrm{d}^4k_\mathrm{E}}{(2\pi)^4}\log\left[k_\mathrm{E}^2+m_i^2(\phi)\right],\label{eq:oneloopvacuum}\\
V_\mathrm{1-loop}^{T\neq0}(\phi,T)&=\sum\limits_{i=h,\chi,W,Z,t}\frac{n_iT^4}{2\pi^2}\int_0^\infty\mathrm{d}kk^2\log\left[1\mp\exp\left(-\sqrt{k^2+m_i^2(\phi)/T^2}\right)\right],\label{eq:oneloopthermal}\\
V_\mathrm{1-loop}^\mathrm{ring}(\phi,T)&=\sum\limits_{i=h,\chi,W_L,Z_L,\gamma_L}\frac{\bar{n}_iT}{4\pi^2}\int_0^\infty\mathrm{d}kk^2\log\left[1+\frac{\Pi_i(\phi,T)}{k^2+m_i^2(\phi)}\right]\label{eq:oneloopring}
\end{align}
where the one-loop vacuum potential~\eqref{eq:oneloopvacuum} will be our focus in~\ref{subsubsec:loopvacuum} and~\ref{subsubsec:RGcondition}, while the one-loop thermal potential~\eqref{eq:oneloopthermal} and ring part~\eqref{eq:oneloopring} will be treated afterwards in subsection~\ref{subsubsec:loopthermal}. Here $n_i$ denotes the degrees of freedom (d.o.f.) of particle species $i$. The subscript ``$L$'' denotes the longitudinal modes of that particle, of which the longitudinal d.o.f. are counted by $\bar{n}_i$.

Under minimal substraction ($\overline{MS}$) renormalization scheme used to remove the ultraviolet divergent constants from the dimensional regularization, the zero-temperature vacuum potential~\eqref{eq:oneloopvacuum} takes the well-known form of the Coleman-Weinberg (CW) potential,
\begin{align}\label{eq:CWpotential}
V_\mathrm{1-loop}^{T=0}(\phi)\equiv V_\mathrm{CW}(\phi,g_i,\mu)=\sum\limits_{i=h,\chi,W,Z,t}\frac{n_i}{64\pi^2}m_i^4(\phi)\left[\log\frac{m_i^2(\phi)}{\mu^2}-c_i\right],
\end{align}
where the d.o.f. $n_i$, the constant $c_i$, and the effective mass $m_i(\phi)$ of the Higgs boson $h$, the Goldstone bosons $\chi_{1,2,3}$, the $W^\pm$ bosons, the $Z$ boson, and the top quark $t$ are
\begin{align}
\begin{split}
&n_h=1, \; n_\chi=3, \; n_W=6, \; n_Z=3, \; n_t=-12,\\
&c_h=\frac32, \; c_\chi=\frac32, \; c_W=\frac56, \; c_Z=\frac56, \; c_t=\frac32,\\
&m_h^2=3\lambda\phi^2+m^2+\frac{15}{4}\frac{\kappa}{\Lambda^2}\phi^4, \; m_\chi^2=\lambda\phi^2+m^2+\frac34\frac{\kappa}{\Lambda^2}\phi^4,\\
&m_W^2=\frac14g^2\phi^2, \quad m_Z^2=\frac14(g'^2+g^2)\phi^2, \quad m_t^2=\frac12y_t^2\phi^2.
\end{split}
\end{align}

\subsubsection{The renormalization conditions}\label{subsubsec:RGcondition}

Although the $\overline{MS}$ renormalization scheme enjoys the merit of gauge-invariant RG equations, the $\overline{MS}$ parameters should be determined in terms of physical observables, here, at one-loop level. We adopt in this section the strategy that is outlined in~\cite{Buttazzo:2013uya} by expressing the $\overline{MS}$ parameters in terms of the on-shell (OS) parameters. Therefore the so-called pole matching processes at, for example, the top pole mass scale for the $U(1)_Y$ gauge coupling $g'$, the $SU(2)_L$ gauge coupling $g$, the $SU(3)_C$ gauge coupling $g_s$, the top Yukawa coupling $y_t$, the SM mass parameter $m_\mathrm{SM}$, and the SM quartic coupling $\lambda_\mathrm{SM}$, give in the following the numerical values~\cite{Buttazzo:2013uya},
\begin{align}\label{eq:polematch}
\begin{split}
g'_0&=0.35830+0.00011\left(\frac{M_t}{\mathrm{GeV}}-173.34\right)-0.0020\left(\frac{M_W-80.384\,\mathrm{GeV}}{0.014\,\mathrm{GeV}}\right),\\
g_0&=0.64779+0.00004\left(\frac{M_t}{\mathrm{GeV}}-173.34\right)+0.00011\left(\frac{M_W-80.384\,\mathrm{GeV}}{0.014\,\mathrm{GeV}}\right),\\
g_{s0}&=1.1666+0.00314\left(\frac{\alpha_s(M_Z)-0.1184}{0.0007}\right)-0.00046\left(\frac{M_t}{\mathrm{GeV}}-173.34\right),\\
y_{t0}&=0.93690+0.00556\left(\frac{M_t}{\mathrm{GeV}}-173.34\right)-0.00042\left(\frac{\alpha_s(M_Z)-0.1184}{0.0007}\right),\\
\lambda_\mathrm{SM}&=0.12604+0.00206\left(\frac{M_h}{\mathrm{GeV}}-125.15\right)-0.00004\left(\frac{M_t}{\mathrm{GeV}}-173.34\right),\\
m_\mathrm{SM}&=\frac{1}{\sqrt{2}}\left(131.55+0.94\left(\frac{M_h}{\mathrm{GeV}}-125.15\right)+0.17\left(\frac{M_t}{\mathrm{GeV}}-173.34\right)\right),
\end{split}
\end{align}
where measured values of the input parameters~\cite{Buttazzo:2013uya}
\begin{align}\label{eq:input}
\begin{split}
v=246.21971\pm0.00006\,\mathrm{GeV},   \quad & \hbox{From Fermi constant for $\mu$ decay},\\
M_h=125.15\pm0.24\,\mathrm{GeV},       \quad & \hbox{Pole mass of the Higgs boson},\\
M_W=80.384\pm0.014\,\mathrm{GeV},      \quad & \hbox{Pole mass of the W boson},\\
M_Z=91.1876\pm0.0021\,\mathrm{GeV},    \quad & \hbox{Pole mass of the Z boson},\\
M_t=173.34\pm0.76\pm0.3\,\mathrm{GeV}, \quad & \hbox{Pole mass of the top quark},\\
\alpha_s(M_Z)=0.1184\pm0.0007,         \quad & \hbox{$\overline{MS}$ gauge $SU(3)_C$ coupling}.
\end{split}
\end{align}

For the case with the fixed RG scale at the top pole mass $\mu_*=M_t$, the renormalization conditions for the effective potential with vacuum corrections $V_\mathrm{eff}^{T=0}(\phi,g_i,\mu)\equiv V_\mathrm{tree}(\phi,g_i)+V_\mathrm{CW}(\phi,g_i,\mu)$ are given by fixing the position of true EW vacuum and the Higgs boson mass at the measured value of vev, namely~\cite{Delaunay:2007wb}
\begin{align}\label{eq:fixRGcon}
\begin{split}
\left.\frac{\mathrm{d}}{\mathrm{d}\phi}V_\mathrm{eff}^{T=0}(\phi=v,g_{i,*},\mu_*)\right|_{n_\chi=0}&=0,\\
\left.\frac{\mathrm{d^2}}{\mathrm{d}\phi^2}V_\mathrm{eff}^{T=0}(\phi=v,g_{i,*},\mu_*)\right|_{n_\chi=0}&=M_h^2-(\Sigma(p^2=M_h^2)-\Sigma(p^2=0))\equiv2m_\mathrm{SM}^2,
\end{split}
\end{align}
where the mismatch between the zero-momentum Higgs mass $V_\mathrm{eff}^{''T=0}(\phi=v,g_i(\mu_*),\mu_*)$ and the physical Higgs pole mass $M_h^2$ has been compensated by the difference of the one-loop Higgs self-energy $\Sigma(p^2=M_h^2)-\Sigma(p^2=0)$. Since the pole matching process~\eqref{eq:polematch} for $m_\mathrm{SM}$ has eliminated the divergent part, therefore one has to move away by hand the contribution from the massless Goldstone bosons at the left hand side of~\eqref{eq:fixRGcon} that could cause the IR divergence at vev. Solving the renormalization conditions~\eqref{eq:fixRGcon} with input cut-off scale $\Lambda$ and fixed $\kappa\equiv1$ allows us to express the Lagrangian parameters $m_*(v,M_h,M_t,\cdots)$ and $\lambda_*(v,M_h,M_t,\cdots)$ in terms of the measured observables. Hence a renormalized effective potential $V_\mathrm{eff}(\phi,g_*,\mu_*)$ is obtained, which will be used later in the next subsection~\ref{subsubsec:RGbounce} for solving the renormalized bounce equation. Although not used in this paper, there is a simplified version of the renormalization conditions by requiring that the physical observables are not shifted by vacuum corrections, namely, $V_\mathrm{CW}(\phi=v)=V'_\mathrm{CW}(\phi=v)=V''_\mathrm{CW}(\phi=v)=0$, which gives rise to the widely used form of the renormalized vacuum potential~\cite{Leitao:2015fmj},
\begin{align}\label{eq:fixRGconSim}
V_\mathrm{CW}(\phi)=\sum\limits_i\frac{n_i}{64\pi^2}\left[m_i^4(\phi)\left(\log\frac{m_i^2(\phi)}{m_i^2(v)}-\frac32\right)+2m_i^2(\phi)m_i^2(v)-\frac{m_i^4(\phi)}{2}\right].
\end{align}
However, the above renormalized effective potential cannot include the presence of the Goldstone bosons due to the occurrence of IR divergence in~\eqref{eq:fixRGconSim} with $m_\chi^2(v)=0$. In~\cite{Leitao:2015fmj}, the Higgs sector is also ignored. For the sake of generality and completeness, the renormalized effective potential~\eqref{eq:fixRGconSim} will not be used in the following. Instead, we will always use the renormalization conditions~\eqref{eq:fixRGcon} for a fixed RG scale.

For the case with a running RG scale $\mu(t)$, the renormalized effective potential with vacuum corrections  $V_\mathrm{eff}^{T=0}(\phi(t),g_i(t),\mu(t))$ should obey the following renormalization conditions,
\begin{align}\label{eq:runRGcon}
\begin{split}
\left.\frac{\partial}{\partial\phi(t)}V_\mathrm{eff}^{T=0}(\phi(t),g_i(t),\mu(t))\right|_{n_\chi=0,\;\phi(t)=\langle\phi(t)\rangle,\;t=0}&=0,\\
\left.\frac{\partial^2}{\partial\phi(t)^2}V_\mathrm{eff}^{T=0}(\phi(t),g_i(t),\mu(t))\right|_{n_\chi=0,\;\phi(t)=\langle\phi(t)\rangle,\;t=0}&=2m_\mathrm{SM}^2
\end{split}
\end{align}
at the initial point $t=0$ of RG running where the initial vev $\langle\phi(0)\rangle=v$ is fixed by the measured vev. It is worth noting that, unlike the case of~\eqref{eq:fixRGconSim}, the effective potentials renormalized from both~\eqref{eq:fixRGcon} and~\eqref{eq:runRGcon} have reserved the contributions from Goldstone bosons. The contributions from Goldstone bosons have only been eliminated from the left hand side of~\eqref{eq:runRGcon} at the initial point of RG running for the same reasoning we given for~\eqref{eq:fixRGcon}. Solving the renormalization conditions~\eqref{eq:runRGcon} with input cut-off scale $\Lambda$ and fixed $\kappa\equiv1$ will give us the initial values $m(0)$ and $\lambda(0)$ for the RG running of the Lagrangian parameters $m(t)$ and $\lambda(t)$.

\subsubsection{The one-loop thermal corrections}\label{subsubsec:loopthermal}

The one-loop thermal potential~\eqref{eq:oneloopthermal},
\begin{align}\label{eq:thermalpotential}
V_\mathrm{1-loop}^{T\neq0}\equiv V_T(\phi,T)=\sum\limits_{i=\mathrm{Bosons}}\frac{n_iT^4}{2\pi^2}J_B\left(\frac{m_i^2(\phi)}{T^2}\right)+\sum\limits_{i=\mathrm{Fermions}}\frac{n_iT^4}{2\pi^2}J_F\left(\frac{m_i^2(\phi)}{T^2}\right),
\end{align}
can be abbreviated in terms of the $J$-functions,
\begin{align}
J_B(x)=\int_0^\infty\mathrm{d}y y^2\log\left[1-\exp\left(-\sqrt{y^2+x}\right)\right]=\int_{\sqrt{x}}^\infty\mathrm{d}t t\sqrt{t^2-x}\log\left(1-\mathrm{e}^{-t}\right),\\
J_F(x)=\int_0^\infty\mathrm{d}y y^2\log\left[1+\exp\left(-\sqrt{y^2+x}\right)\right]=\int_{\sqrt{x}}^\infty\mathrm{d}t t\sqrt{t^2-x}\log\left(1+\mathrm{e}^{-t}\right),
\end{align}
of which the expansions around $x=0$,
\begin{align}
J_B(x)&=\frac{\pi^2}{12}x-\frac{\pi}{6}x^\frac32-\frac{x^2}{32}\log\frac{x}{\mathrm{cst.}}+\mathcal{O}\left(x^3\log\frac{x^\frac32}{\mathrm{cst.}}\right),\\
J_F(x)&=-\frac{\pi^2}{24}x-\frac{x^2}{32}\log\frac{x}{\mathrm{cst.}}+\mathcal{O}\left(x^3\log\frac{x^\frac32}{\mathrm{cst.}}\right),
\end{align}
give the thermal part of~\eqref{eq:treehighT} at the high-temperature limit $T\gg m_i(\phi)$. In general, we transform the integrals of $J$-functions with variable $t$ to obtain a better numerical behavior when Higgs boson and Goldstone bosons become tachyons below some scales.

The ring part of one-loop finite-temperature potential can be explicitly integrated out~\cite{Delaunay:2007wb} as
\begin{align}\label{eq:ringpotential}
V_\mathrm{1-loop}^\mathrm{ring}\equiv V_\mathrm{ring}(\phi,T)=\sum\limits_{i=h,\chi,W_L,Z_L,\gamma_L}\frac{\bar{n}_iT}{12\pi}\left[m_i^3(\phi)-\left(m_i^2(\phi)+\Pi_i(\phi,T)\right)^\frac32\right],
\end{align}
where only scalar field and longitudinal mode of vector field contribute so that the reduced d.o.f. $\bar{n}_{h,\chi,W_L,Z_L,\gamma_L}=\{1,3,2,1,1\}$. The thermal masses $M_i^2(\phi)\equiv m_i^2(\phi)+\Pi_i(\phi,T)$ with Debye masses $\Pi_i(\phi,T)$ have the following forms~\cite{Delaunay:2007wb,Rose:2015lna}
\begin{align}
\begin{split}
M_W^2&=\left(m_W^2+\frac{11}{6}g^2T^2\right)\\
M_h^2&=\left(m_h^2+\left(\frac{g'^2+3g^2}{16}+\frac{\lambda}{2}+\frac{y_t^2}{4}\right)T^2\right)-\frac34\frac{\kappa v^2}{\Lambda^2}T^2\\
M_\chi^2&=\left(m_\chi^2+\left(\frac{g'^2+3g^2}{16}+\frac{\lambda}{2}+\frac{y_t^2}{4}\right)T^2\right)-\frac34\frac{\kappa v^2}{\Lambda^2}T^2\\
M_Z^2&=\frac12\left(m_Z^2+\frac{11}{6}(g'^2+g^2)T^2+\Delta(\phi,T)\right)\\
M_\gamma^2&=\frac12\left(m_Z^2+\frac{11}{6}(g'^2+g^2)T^2-\Delta(\phi,T)\right)\\
\end{split}
\end{align}
with
\begin{align*}
\Delta^2(\phi,T)&=m_Z^4+\frac{11}{3}\frac{(g^2-g'^2)^2}{g^2+g'^2}\left(m_Z^2+\frac{11}{12}(g^2+g'^2)T^2\right)T^2
\end{align*}

\subsubsection{The renormalization-group improvement}\label{subsubsec:RGimprovement}

The RG improvement is implemented for the full one-loop effective potential $V_\mathrm{eff}(\phi,g_i,\mu,T)=V_\mathrm{tree}+V_\mathrm{CW}+V_\mathrm{T}+V_\mathrm{ring}$ given in~\eqref{eq:CWpotential},~\eqref{eq:thermalpotential}, and~\eqref{eq:ringpotential} by replacing the RG scale $\mu$ with the sliding scale $\mu(t)\equiv\mu_0\exp(t)$, where the initial running point is usually chosen at the top pole mass scale $\mu_0\equiv M_t$. The classical Higgs field is also redefined by the running Higgs field $\phi(t)=Z(t)\phi_\mathrm{cal}$ in terms of the wavefunction renormalization factor $Z(t)=\exp(\Gamma(t))$ that is defined by $\Gamma(t)=\int_0^t\mathrm{d}t'\gamma(t')$ in terms of the Higgs field anomalous dimension $\gamma(t)=\phi'(t)/\phi(t)$, at one-loop order, namely
\begin{align}
\gamma=\frac{1}{(4\pi)^2}\left(\frac34g'^2+\frac94g^2-3y_t^2\right).
\end{align}
Similarly, all the coupling constants under RG improvement start running according to the beta functions $\beta_{g_i}\equiv g'_i(t)$ that obey the following one-loop RG equations~\cite{Delaunay:2007wb,Rose:2015lna},
\begin{align}\label{eq:RGEqs}
\begin{split}
\beta_{g'}&=\frac{g'^3}{(4\pi)^2}\frac{41}{6};\\
\beta_g&=\frac{g^3}{(4\pi)^2}\left(-\frac{19}{6}\right);\\
\beta_{g_s}&=\frac{g_s^3}{(4\pi)^2}(-7);\\
\beta_{y_t}&=\frac{y_t}{(4\pi)^2}\left(-\frac{17}{12}g'^2-\frac94g^2-8g_s^2+\frac92y_t^2\right);\\
\beta_{m^2}&=\frac{2m^2}{(4\pi)^2}\left(-\frac34g'^2-\frac94g^2+3y_t^2+6\lambda\right);\\
\beta_\lambda&=\frac{1}{(4\pi)^2}\left(24\lambda^2-6y_t^4+\frac38(2g^4+(g'^2+g^2)^2)+(-3g'^2-9g^2+12y_t^2)\lambda-24\frac{\kappa}{\Lambda^2}m^2\right);\\
\beta_\kappa&=\frac{6\kappa}{(4\pi)^2}\left(-\frac34g'^2-\frac94g^2+3y_t^2+18\lambda\right),
\end{split}
\end{align}
where the initial conditions of above one-loop RG equations are given by the pole matching conditions~\eqref{eq:polematch} and the initial values of $m(0)$ and $\lambda(0)$ are solved from the renormalization conditions~\eqref{eq:runRGcon}. The initial value $\kappa(0)\equiv1$ and we treat the cut-off scale $\Lambda$ as the model-dependent input parameter.

\begin{figure}
  \includegraphics[width=0.5\textwidth]{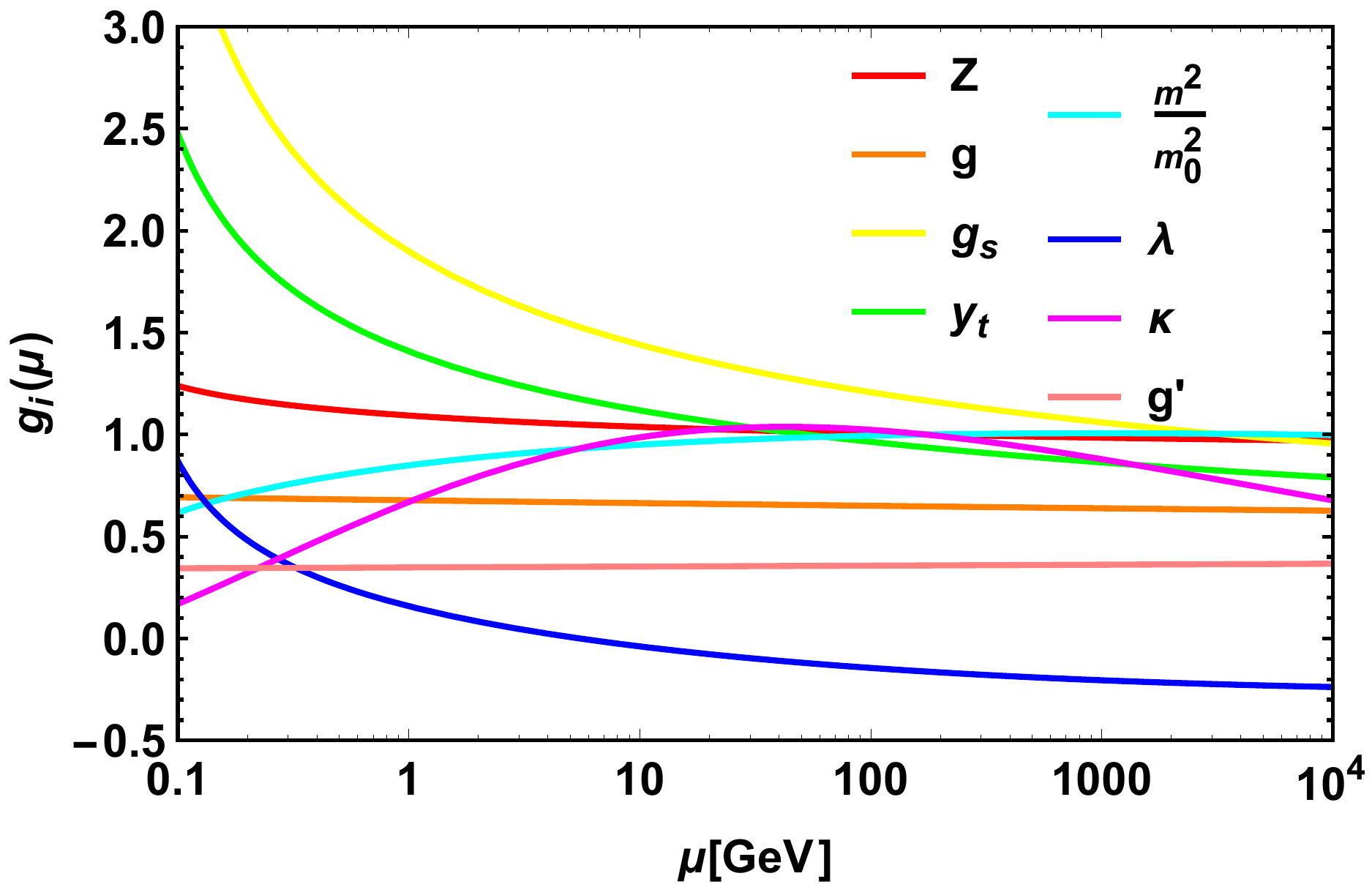}
  \includegraphics[width=0.5\textwidth]{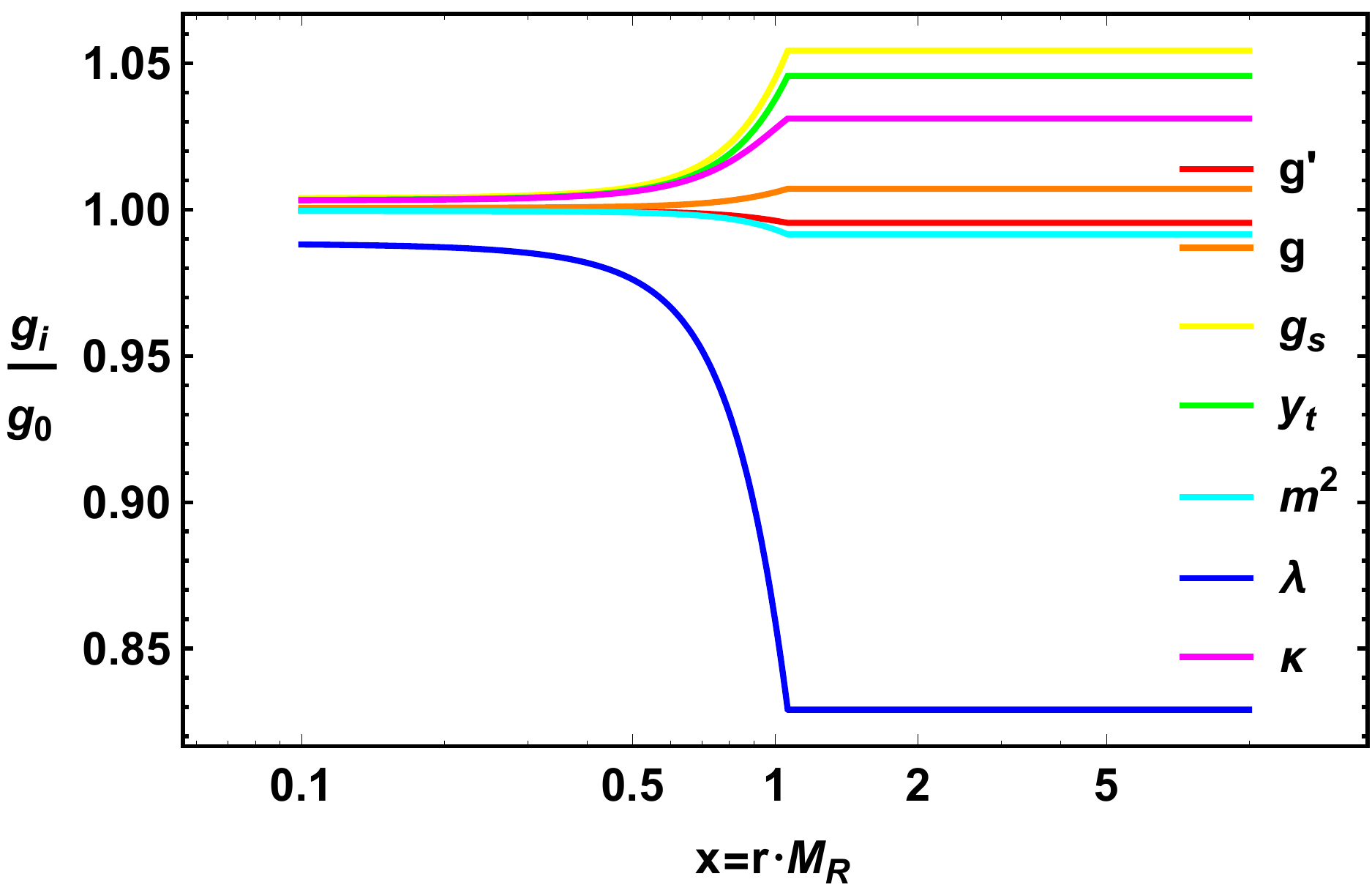}\\
  \includegraphics[width=0.5\textwidth]{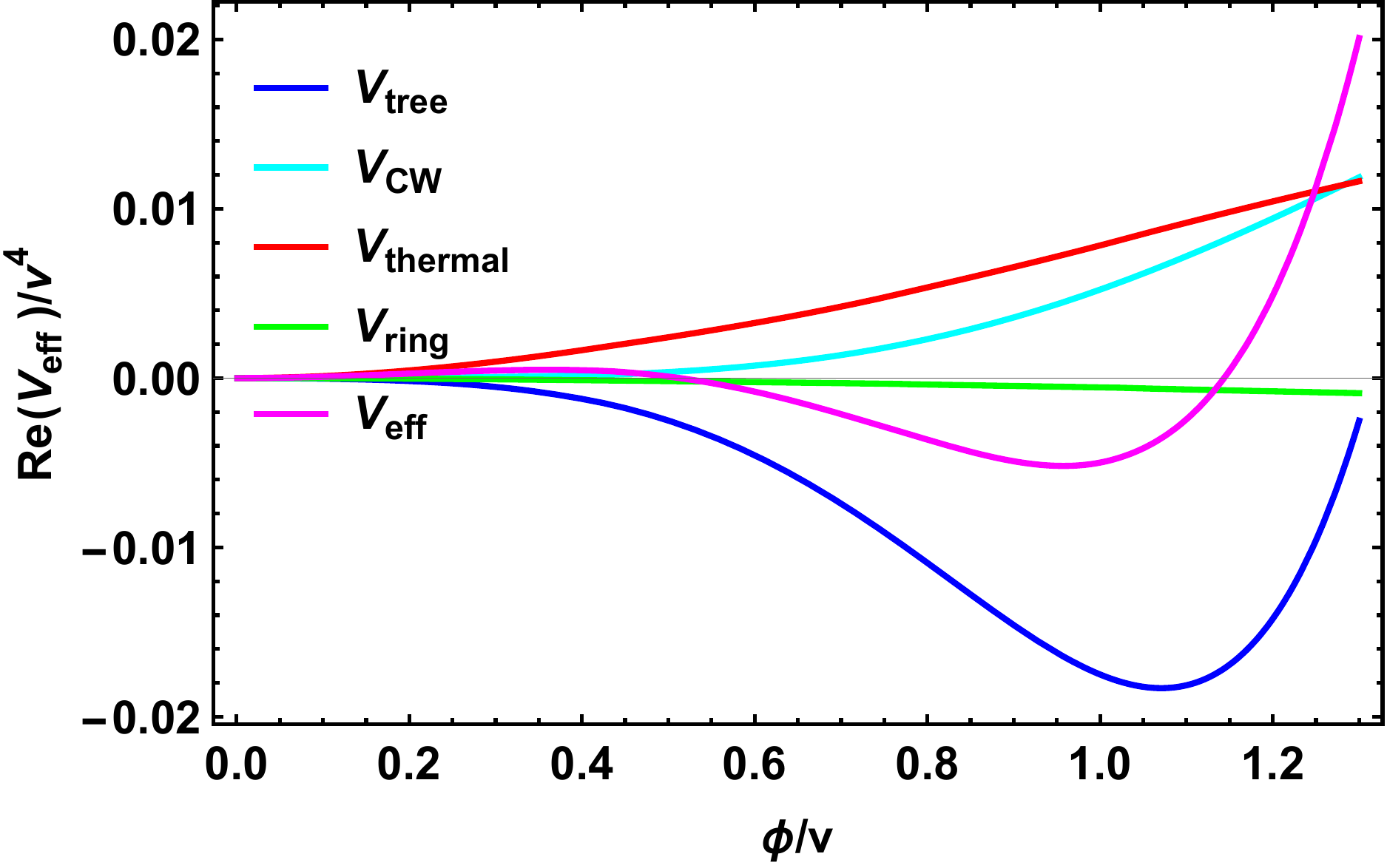}
  \includegraphics[width=0.5\textwidth]{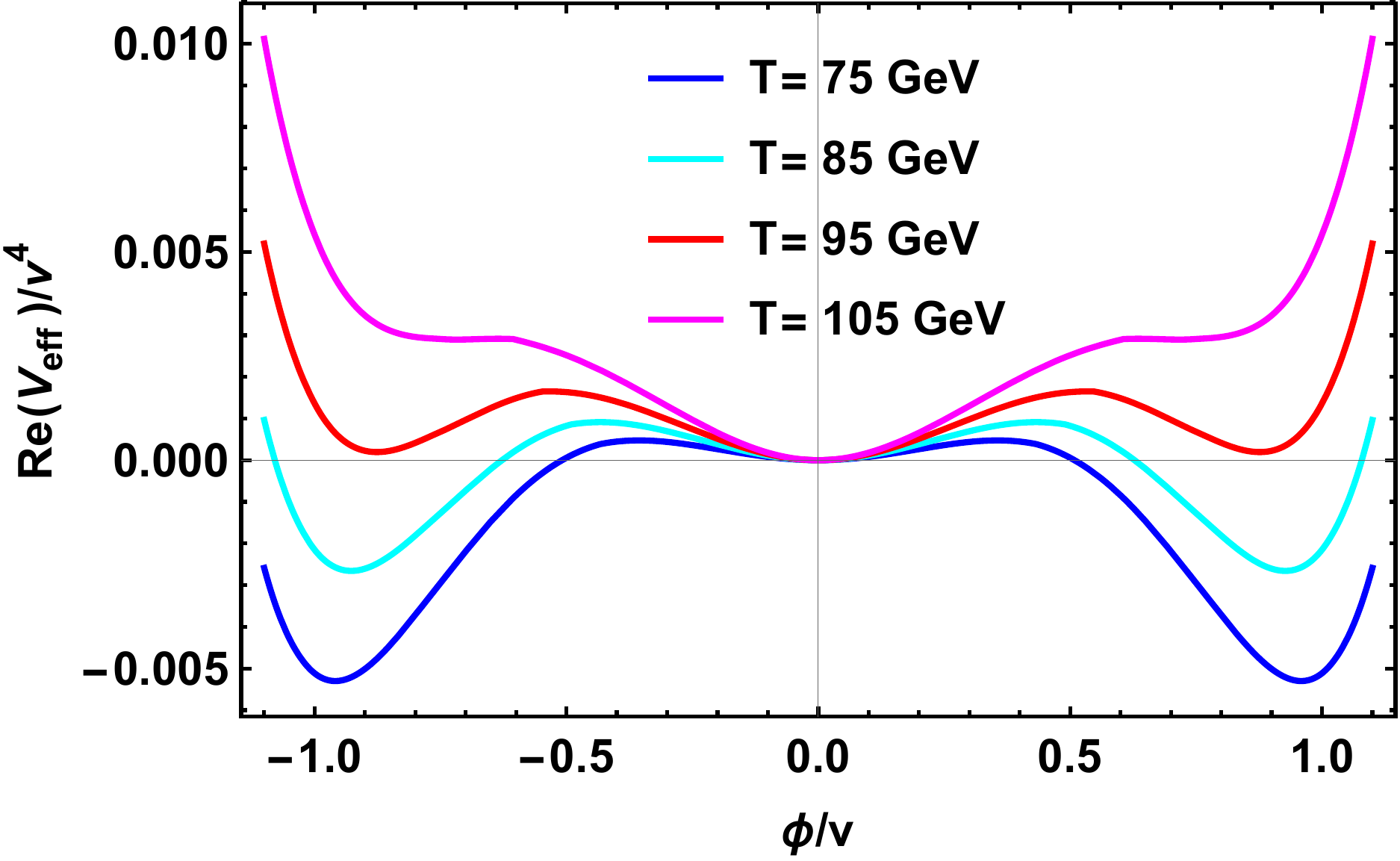}\\
  \caption{The involvement of RG improvement for the full one-loop effective potential at an illustrative cut-off scale $\Lambda=540\,\mathrm{GeV}$. The RG equations are solved for the running couplings in the first panel. In the second panel, the scale-dependence of running couplings are evaluated at the different spatial distances of the nucleated bubble. In the third panel, the RG-improved effective potential is presented for its real part, of which the tree-level part, the vacuum corrections, the thermal corrections, and the ring part are compared. In the last panel, the temperature-dependence of the RG-improved effective potential is presented at some illustrative temperatures.}\label{fig:RGVeff}
\end{figure}

The last step left to carry out the RG improvement is to choose an appropriate form for the running RG scale $\mu(t)$~\cite{Rose:2015lna}. To have a canonically normalized bounce equation we will discuss below in~\ref{subsubsec:RGbounce}, one needs to introduce the canonical field $\phi_\mathrm{can}$ that is implicitly defined by $\mathrm{d}\phi_\mathrm{can}/\mathrm{d}\phi_\mathrm{cal}=Z(\phi_\mathrm{cal})$. An approximate solution $\phi_\mathrm{can}=Z(\phi_\mathrm{cal})\phi_\mathrm{cal}$ satisfying $\mathrm{d}\phi_\mathrm{can}/\mathrm{d}\phi_\mathrm{cal}=Z(\phi_\mathrm{cal})(1+\gamma(\phi_\mathrm{cal}))$ is valid as along as the anomalous dimension $\gamma(\phi_\mathrm{cal})$ remains small enough during the RG evolution. As we will see in Fig.~\ref{fig:RGVeff}, this is indeed the case. Therefore, we will choose the running RG scale to be the running Higgs field $\mu(t)=\phi(t)$ so that the logarithmic terms in the vacuum potential~\eqref{eq:CWpotential} can be minimized. Furthermore, there should be a multiplied factor in $\mu(t)=(M_t/v)\phi(t)$ in order to compensate the mismatch between the initial vev of running Higgs field $\langle\phi(0)\rangle\equiv v$ and the initial value of running RG scale $\mu(0)\equiv M_t$. However, the background temperature provides another cut-off scale for the running RG scale, therefore the final form of the running RG scale reads
\begin{align}\label{eq:RGscale}
\mu(t)=\max\left(\frac{M_t}{v}\phi(t),T,k=10^3\right),
\end{align}
where the function
\begin{align}
\max(x,y,k)=\frac{x}{1+\exp(-k(x-y))}+\frac{y}{1+\exp(-k(y-x))}
\end{align}
is aimed to smooth the usual maximum function, so that the derivative of the full one-loop effective potential could be well-behaved when posing the shooting algorithm for the renormalized bounce equation detailed in the next subsection~\ref{subsubsec:RGbounce}.

We take $\Lambda=540\,\mathrm{GeV}$ as an example in Fig.~\ref{fig:RGVeff} to illustrate the outcome of invoking the RG improvement. In the first panel, the RG equations~\eqref{eq:RGEqs} are solved for the running couplings. It is worth noting that, the strong coupling will eventually blow up in the IR limit if one naively solves the RG equations at arbitrarily low energy scale. The full treatment of strong coupling in the IR limit would require lattice computations due to the non-perturbative nature of quark confinement. However, for the fast first-order PT we are considering in this section, the nucleation temperatures are never running into the non-perturbative regime as we will see in Fig.~\ref{fig:RGPara}. Therefore, the RG equations~\eqref{eq:RGEqs} are still perturbatively valid for our analysis. It is also worth noting that, the threshold effects from decoupling of heavy particles can be safely ignored for our choice of RG scale~\eqref{eq:RGscale}, where we have tested that the decoupling conditions $\mu(t)\ll m_i(t)$ are never fulfilled for the considered particle species within the full range $0\leq\phi(t)\leq v$ of transited field. In the second panel, the bounce equation is solved as we will see in the next subsection~\ref{subsubsec:RGbounce}, where the scale-dependence of running couplings are visualized as the spatial-dependence of the nucleated bubble, that is, the running couplings will exhibit different values at different distances to the bubble center. In the third panel, the RG-improved effective potential is presented for its real part, of which the tree-level part, the vacuum corrections, the thermal corrections, and the ring part are compared. In the last panel, the temperature-dependence of the RG-improved effective potential is presented at some illustrative temperatures.

\subsection{The relic gravitational waves}\label{subsec:loopGW}

After choosing the running Higgs field as the physical field for describing the nucleated bubbles with the presence of the running RG scale in~\ref{subsubsec:RGbounce}, the predictions of relic GWs are carried out in~\ref{subsubsec:loopresult} for both the fixed and running RG scales. It turns out as a surprise that, the presence of running RG scale could gently amplify the amplitude by amount of one order of magnitude while at the same time shift the peak frequency to the lower frequency regime. Therefore, the effect from the RG improvement cannot be simply neglected in future for the precise prediction of energy density spectrum of GWs from first-order PT. However, the perspective for the detection of the relic GWs from the first-order PT with the full one-loop effective potential armed with the sextic term is not as promising as shown in the literature~\cite{Leitao:2015fmj}.

\subsubsection{The renormalized bounce equation and bounce solutions}\label{subsubsec:RGbounce}

For a fixed RG scale $\mu_*=M_t$, the bounce equation deduced from the full one-loop effective potential armed with the sextic term is not changed as along as that the mass coupling $m_*$ and quartic coupling $\lambda_*$ take the values solved from the renormalization condition~\eqref{eq:fixRGcon}, while all other coupling constants $g_{i,*}$ take the values from the pole matching conditions~\eqref{eq:polematch}. The derivative of the effective potential with respect to the Higgs field can be computed without difficulty. For a running RG scale~\eqref{eq:RGscale}, the nucleated bubble should be described by the running field $\phi(t)\equiv Z(t)\phi_\mathrm{cal}$ instead of the classical field $\phi_\mathrm{cal}$ since we have chosen the approximated solution for the canonically normalized field $\phi_\mathrm{can}$ to be the running field $\phi(t)\equiv\phi_t$. For every input value of the running field $\phi_t$, the sliding scale $t(\mu(\phi_t,T))$ is itself a function of $\phi_t$. Therefore all of the running couplings $g_i(t)$ are also a function of $\phi_t$. The resulted effective potential $V_\mathrm{eff}(\phi_t,g_i(t(\mu(\phi_t,T))),\mu(t(\mu(\phi_t,T))),T)\equiv V_\mathrm{eff}(\phi_t,T)$ is well-defined when it is taken its derivative in the following renormalized bounce equation,
\begin{align}\label{eq:RGbounce}
\frac{\mathrm{d}^2\phi_t}{\mathrm{d}r^2}+\frac{2}{r}\frac{\mathrm{d}\phi_t}{\mathrm{d}r}=\frac{\partial}{\partial\phi_t}V_\mathrm{eff}(\phi_t,T),
\end{align}
where the nucleated bubbles described by the renormalized bounce solution $\phi_t(r)$ will be scale-dependent, because at different radius $r$ of the nucleated bubble $\phi_t(r)$, the sliding scale $t(\mu(\phi_t(r),T))\equiv t(r,T)$ is different, so are the running couplings $g_i(t(r,T))\equiv g_i(r,T)$. The scale-dependence of the running couplings $g_i(\mu)$ is therefore visualized as the spatial-dependence of the nucleated bubbles, of which the running couplings evaluated beyond the bubble wall will eventually be screened by the temperature due to the choice of running RG scale~\eqref{eq:RGscale}. This effect is new to our knowledge that has not been discovered in the literatures, which should merit further investigation in future.

\begin{figure}
  \includegraphics[width=0.5\textwidth]{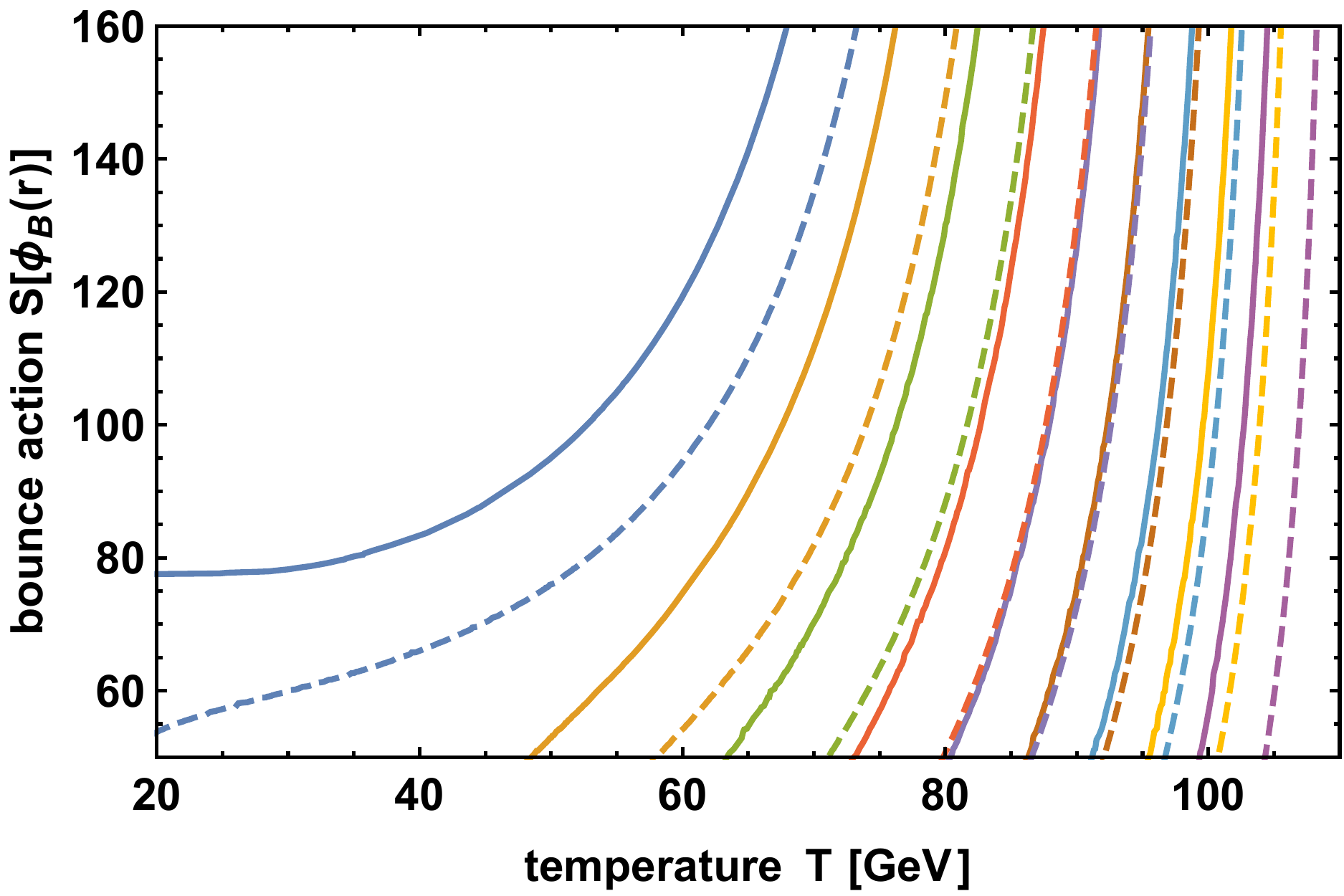}
  \includegraphics[width=0.5\textwidth]{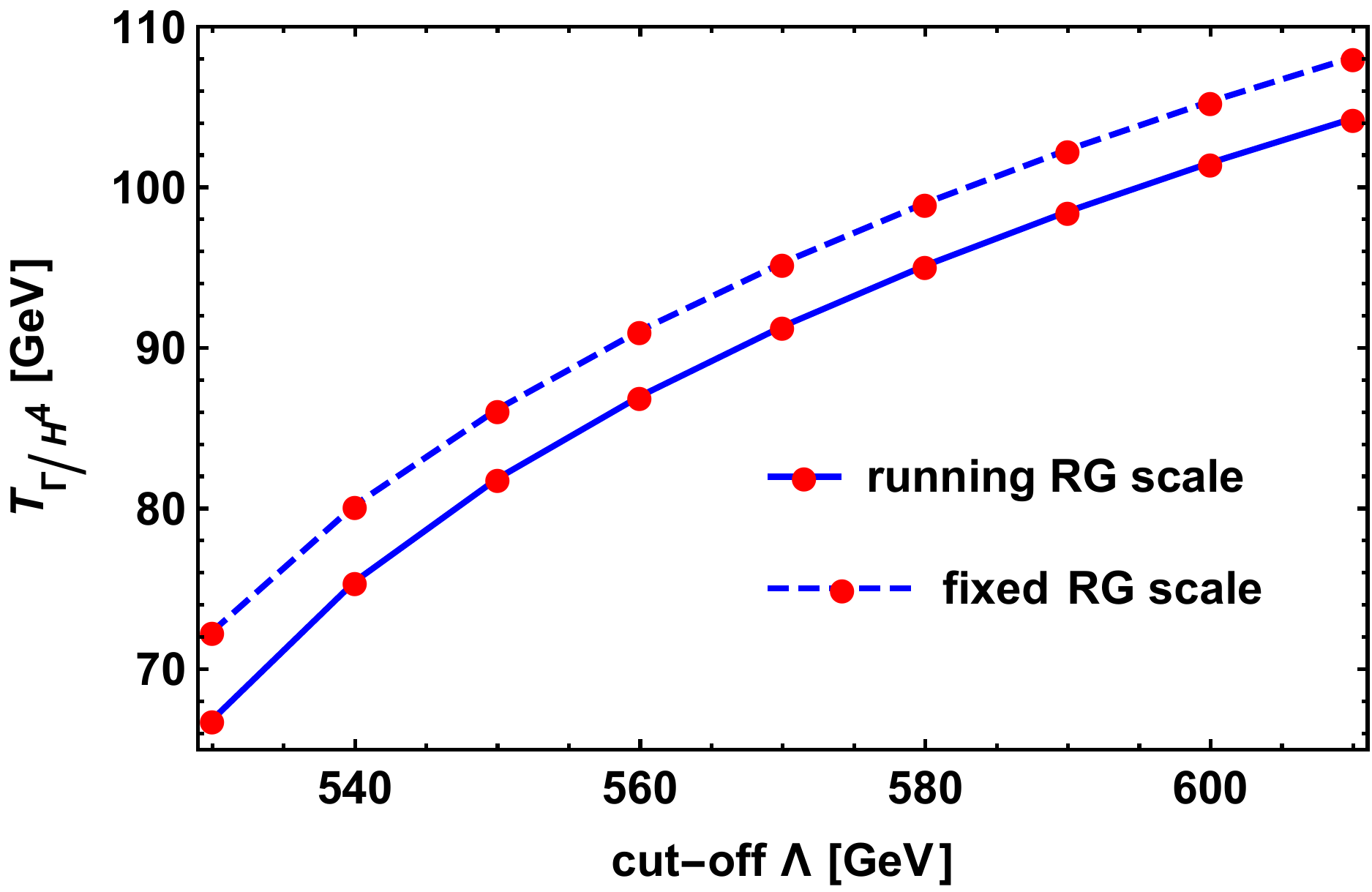}\\
  \includegraphics[width=0.5\textwidth]{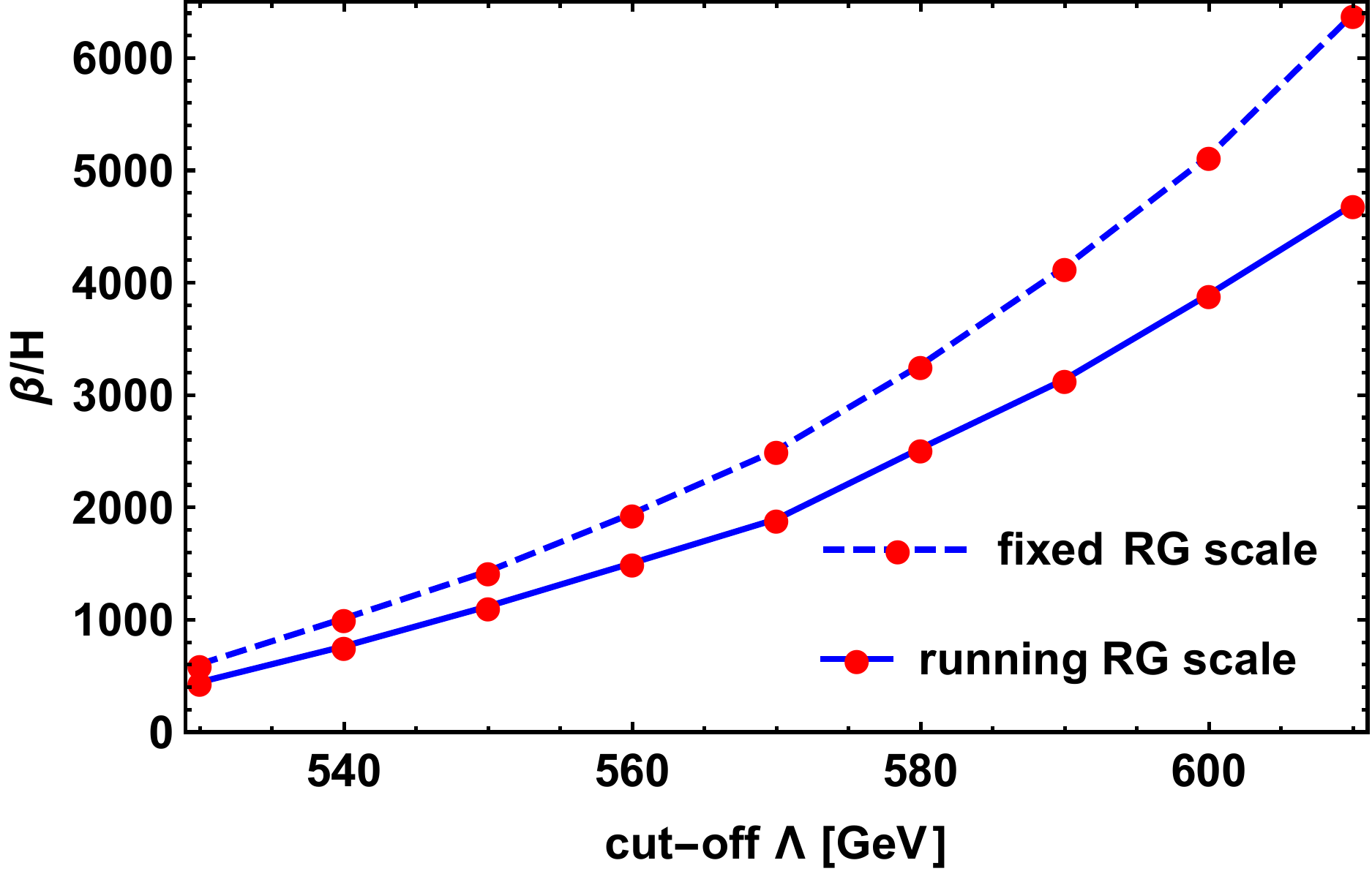}
  \includegraphics[width=0.5\textwidth]{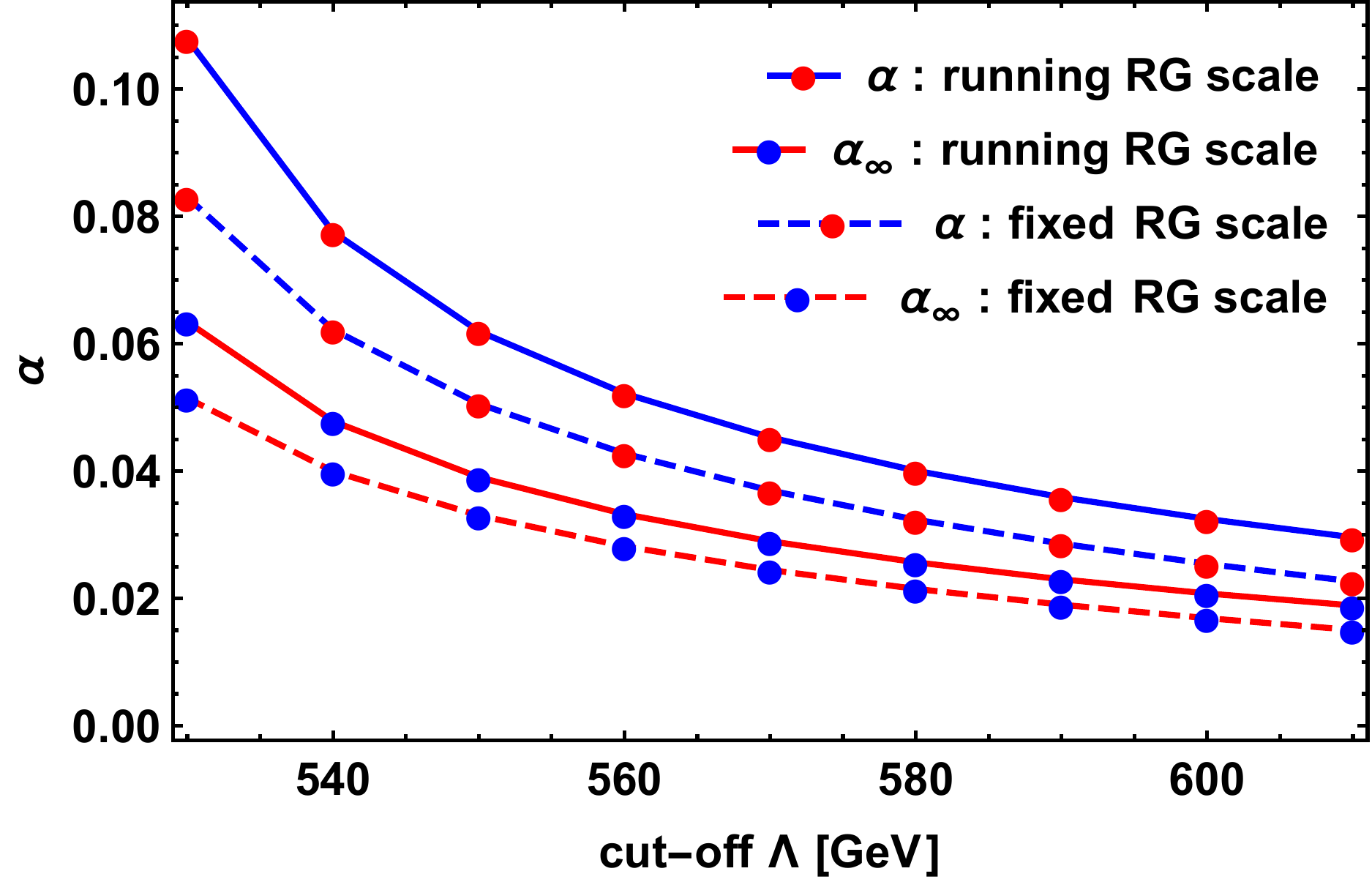}\\
  \caption{The phenomenological parameters of first-order PT from the full one-loop effective potential in the cases with fixed and running RG scales. In the first panel, the bounce actions are presented from left to right with respect to the cut-off scale $\Lambda$ ranging from $530\,\mathrm{GeV}$ to $610\,\mathrm{GeV}$ with interval $10\,\mathrm{GeV}$ for fixed (dashed) and running (solid) RG scales. In the second panel, the nucleation temperature $T_{\Gamma/H^4}$ from a running RG scale (solid) decreases by amount of $5\,\mathrm{GeV}$ compared with a fixed RG scale (dashed). In the third panel, the inverse of the short-duration of the first-order PT $\beta/H$ also decreases by amount of hundreds with (solid) or without (dashed) RG improvement. However, in the last panel, both the strength factor $\alpha$ and its critical value $\alpha_\infty$ with a running RG scale (solid) are lifted with respect to those from a fixed RG scale (dashed). The bubble walls are always running away for the considered ranges of the cut-off scale with or without RG improvement.}\label{fig:RGPara}
\end{figure}

We present in Fig.~\ref{fig:RGPara} the phenomenological parameters of first-order PT from the full one-loop effective potential in the cases with fixed and running RG scales. In the first panel, the bounce actions are shown from left to right with respect to the cut-off scale $\Lambda$ ranging from $530\,\mathrm{GeV}$ to $610\,\mathrm{GeV}$ with interval $10\,\mathrm{GeV}$ for fixed (dashed) and running (solid) RG scales. In the second panel, the nucleation temperature $T_{\Gamma/H^4}$ is drawn from the cut-off scale $\Lambda$, where the running RG scale (solid) could lower the nucleation temperature by amount of $5\,\mathrm{GeV}$ compared with the fixed RG scale (dashed). Similarly, the inverse of the short-duration of our first-order PT $\beta/H$ also decreases by amount of hundreds as shown in the third panel. However, in the last panel, the strength factor $\alpha$ along with its critical value $\alpha_\infty$ with a running RG scale are higher than those from a fixed RG scale. The bubble walls are always running away for the considered ranges of the cut-off scale with fixed/running RG scale. This is different from~\cite{Leitao:2015fmj} where the runaway behavior only manifests itself when $\Lambda\lesssim580\,\mathrm{GeV}$.

\subsubsection{The GW from fixed/running renormalization-group scale}\label{subsubsec:loopresult}

\begin{figure}
  \centering
  \includegraphics[width=0.48\textwidth]{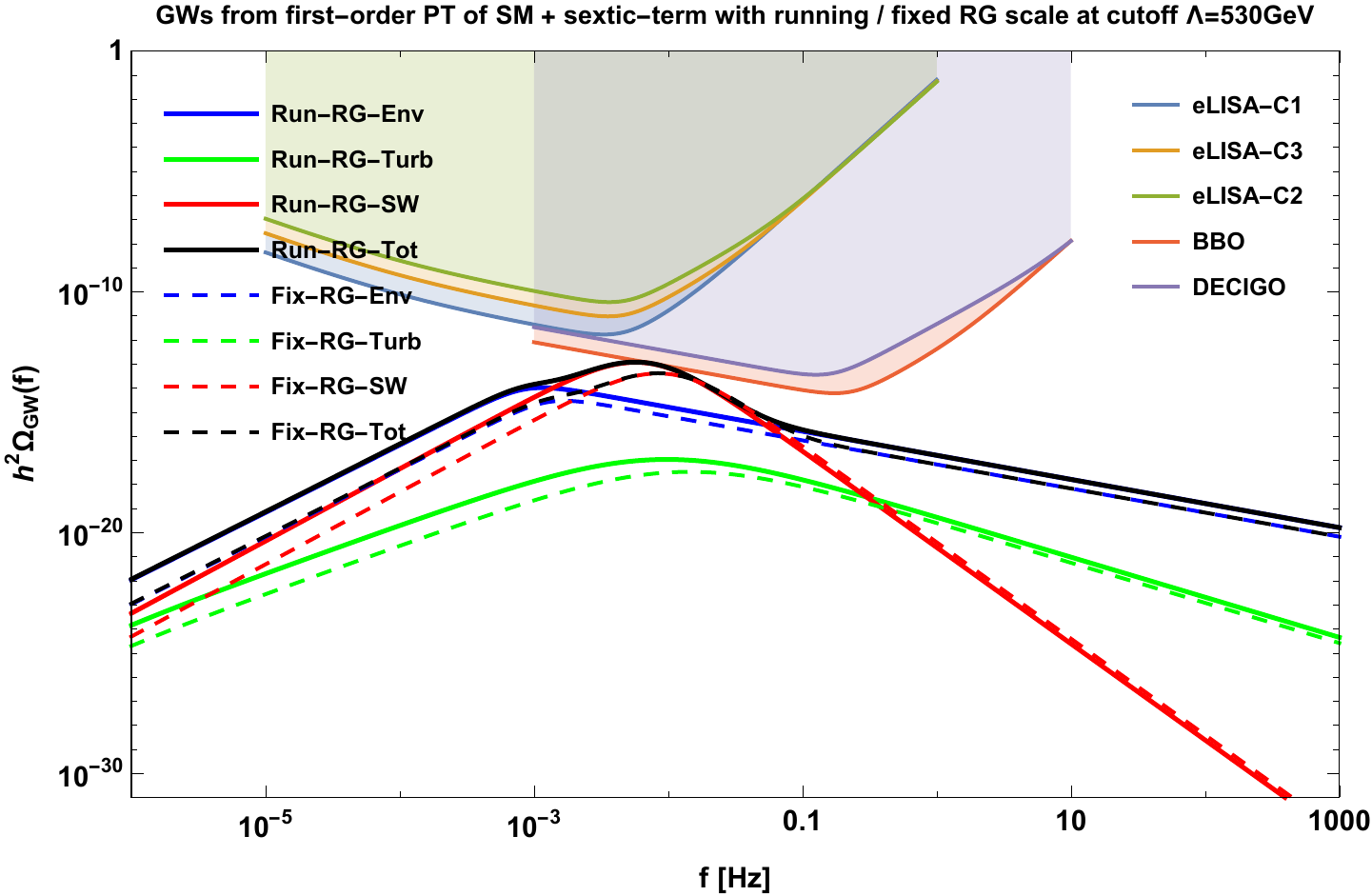}
  \includegraphics[width=0.48\textwidth]{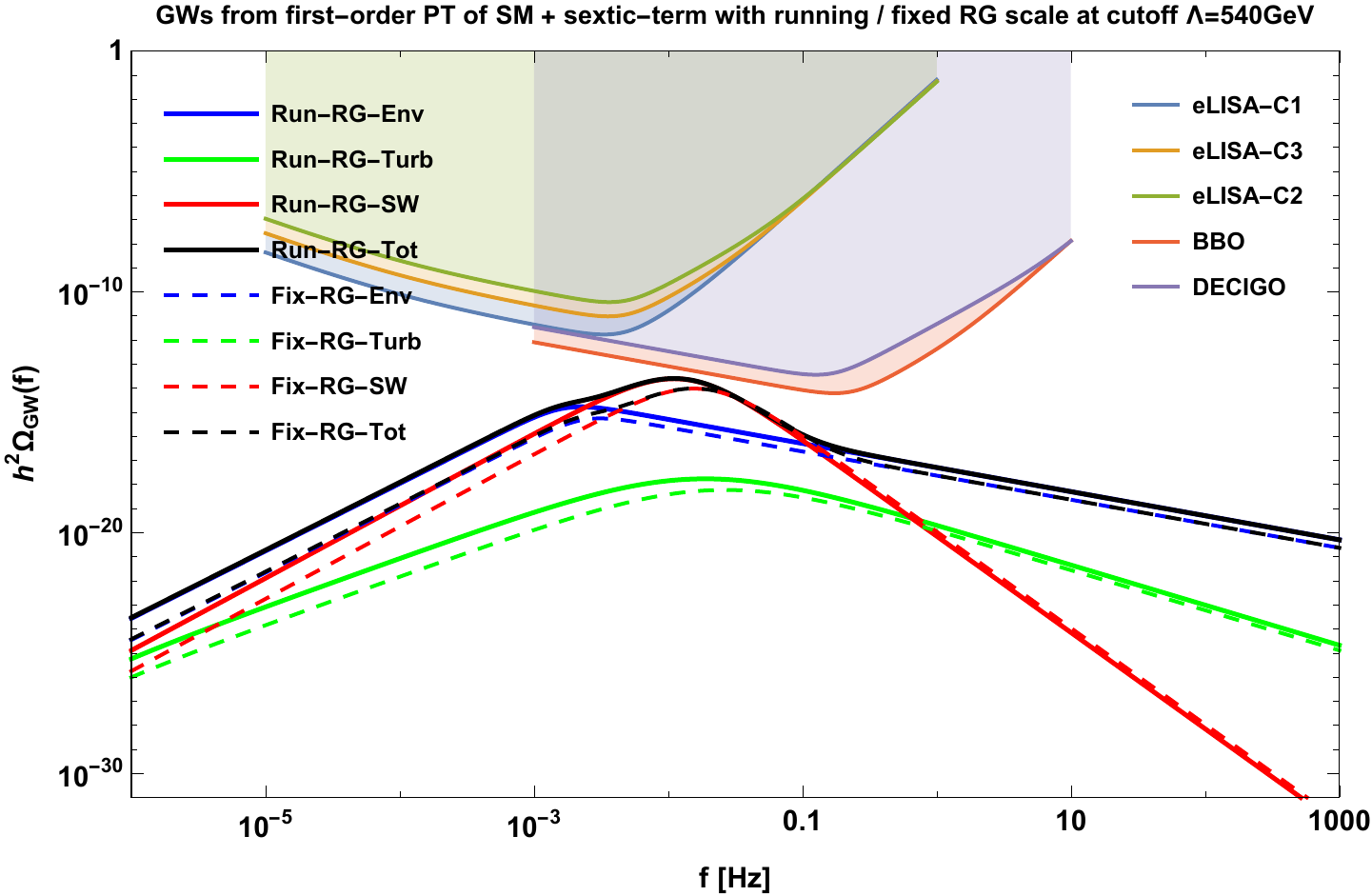}\\
  \includegraphics[width=0.48\textwidth]{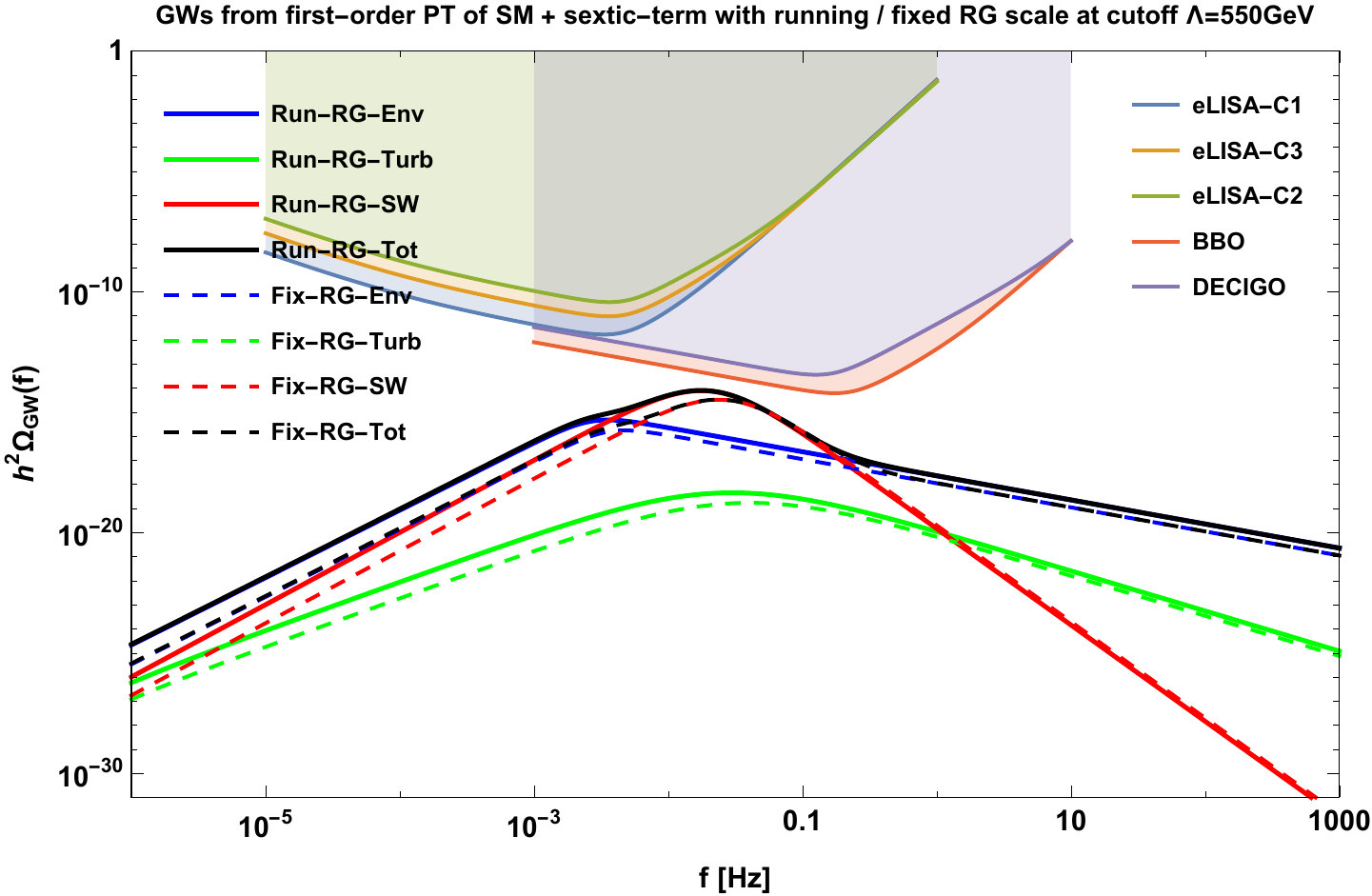}
  \includegraphics[width=0.48\textwidth]{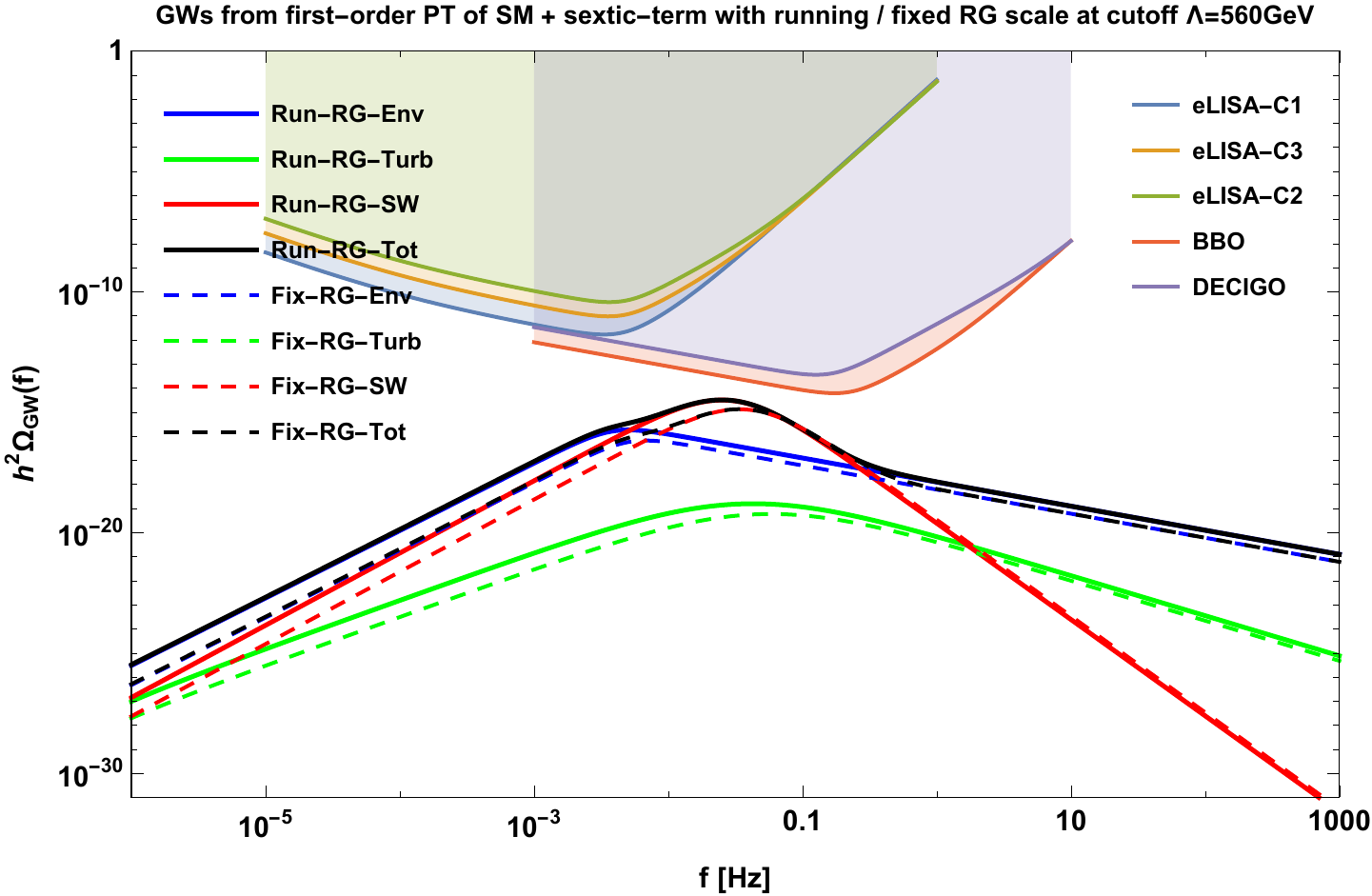}\\
  \includegraphics[width=0.48\textwidth]{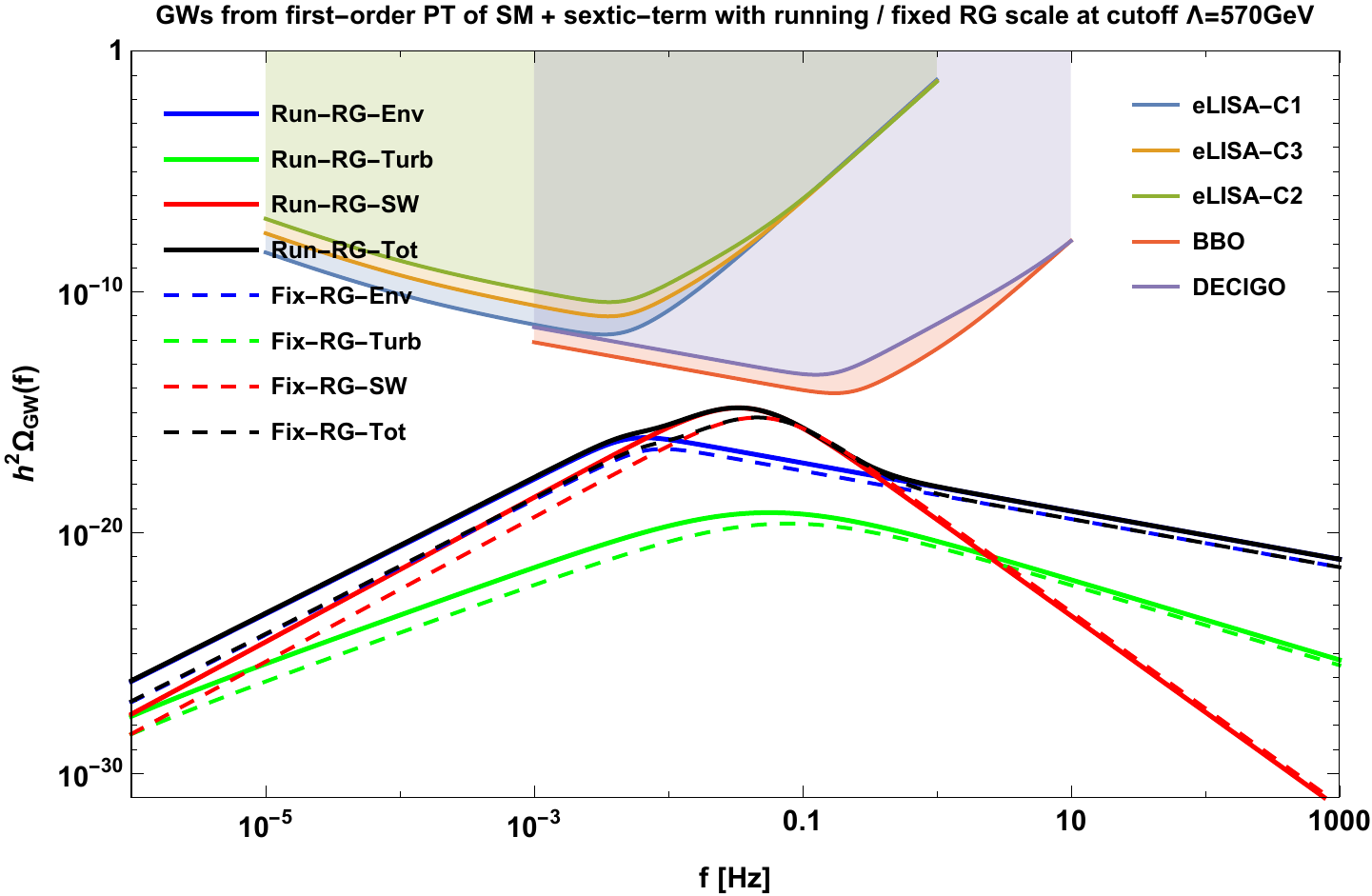}
  \includegraphics[width=0.48\textwidth]{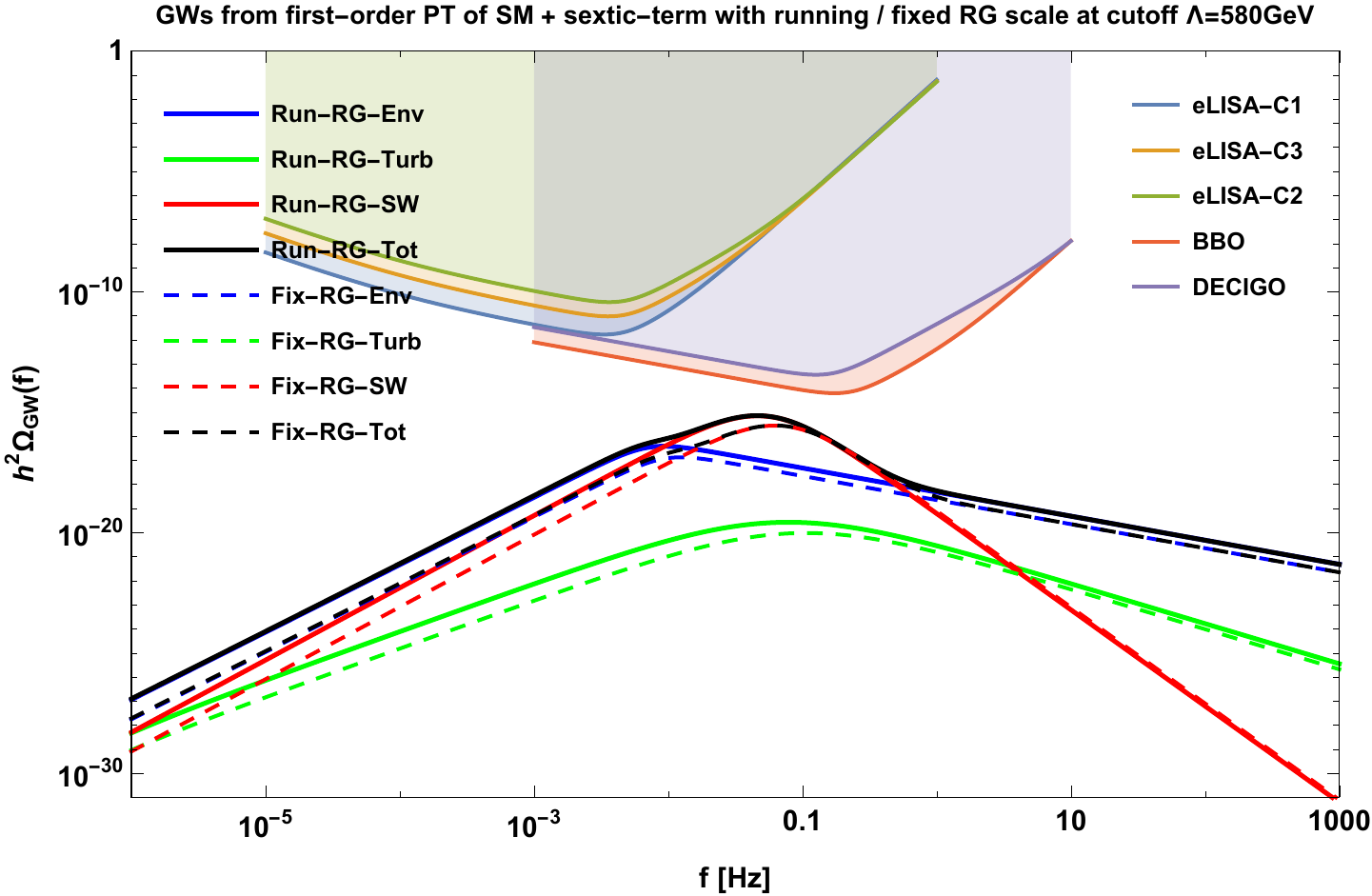}\\
  \includegraphics[width=0.48\textwidth]{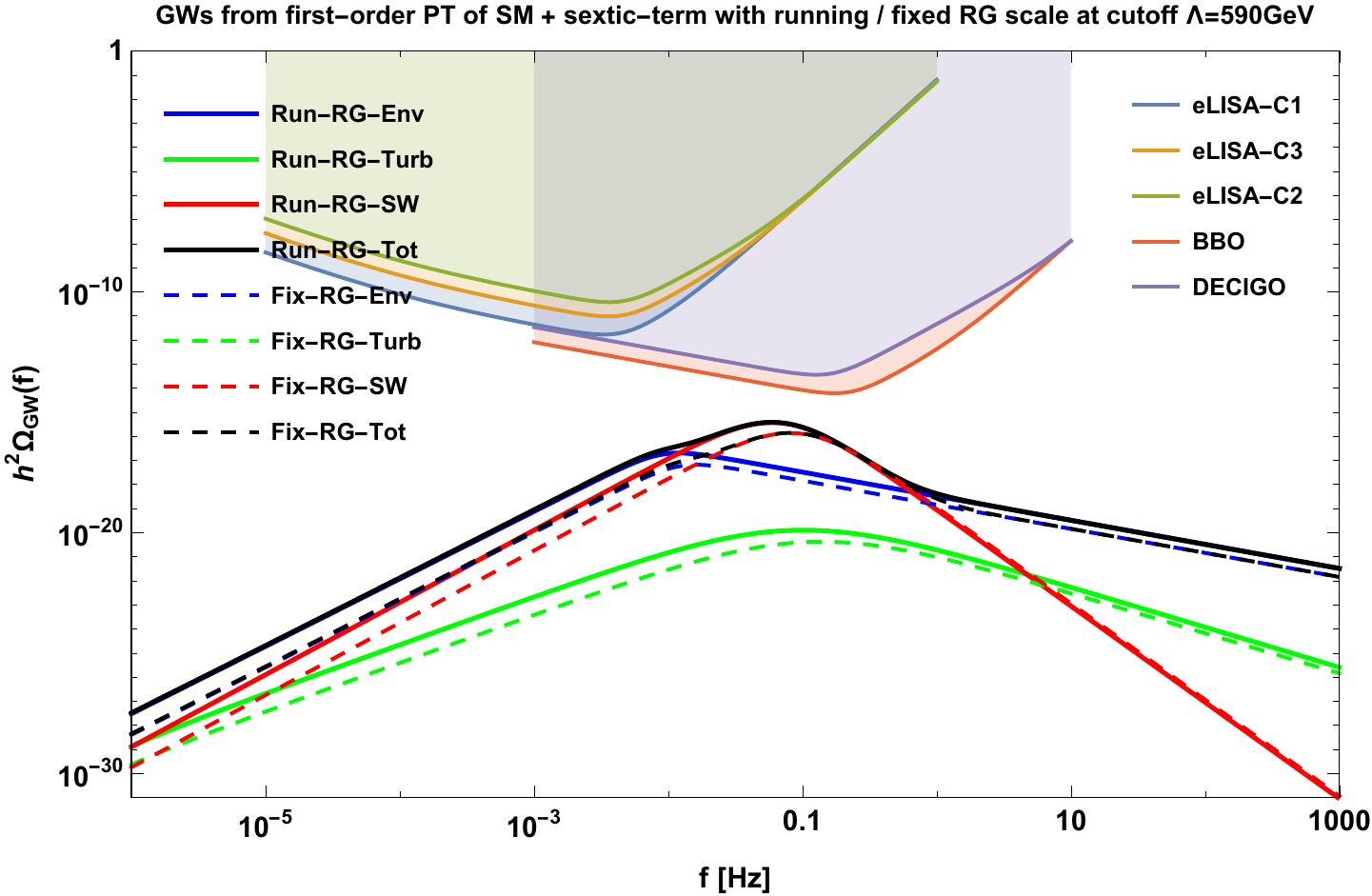}
  \includegraphics[width=0.48\textwidth]{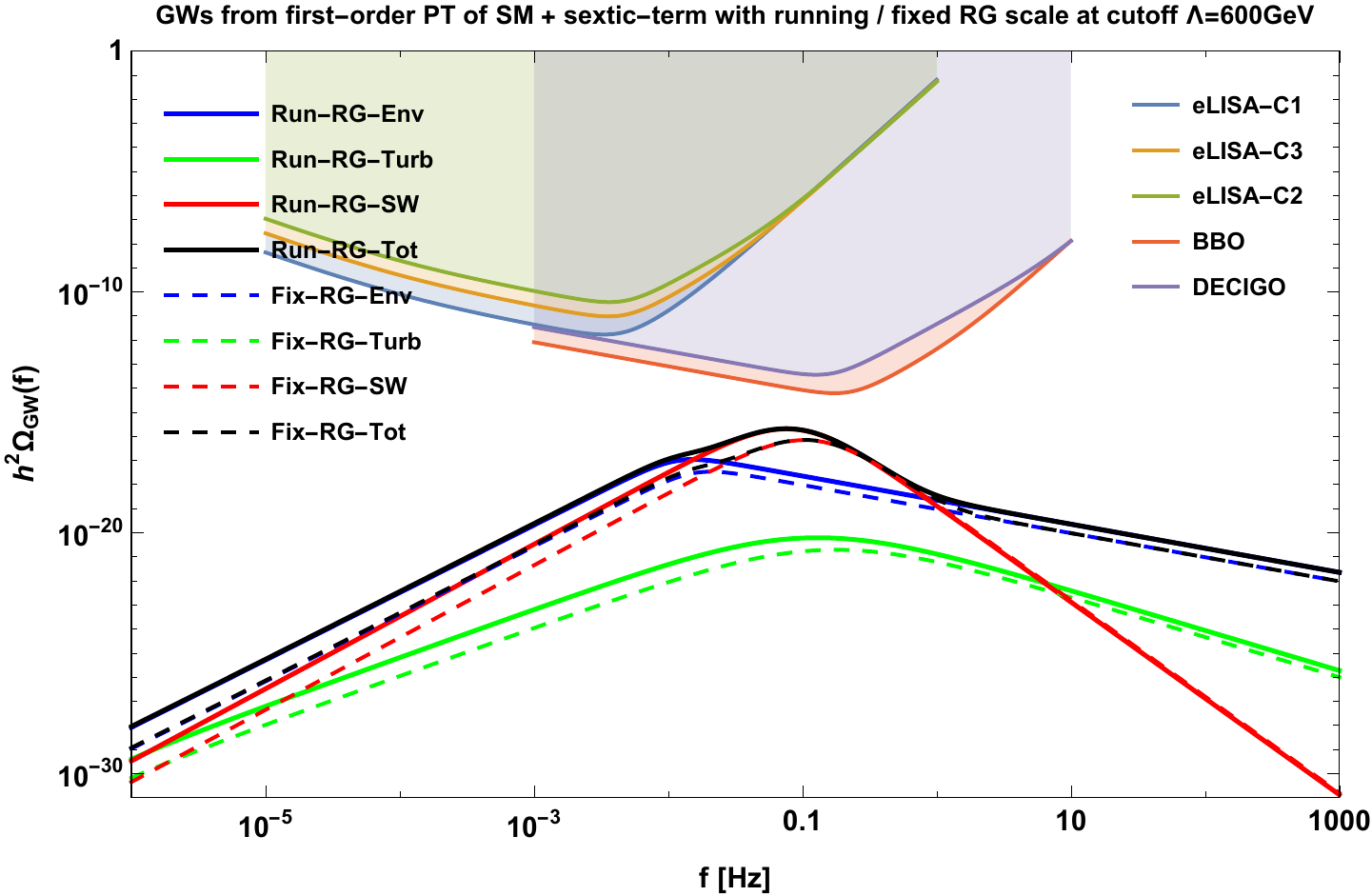}\\
  \caption{The energy density spectrum of the relic GWs from the uncollided envelops (blue), sound waves (red), MHD turbulences (green) and their total contribution (black) for the full one-loop effective potential with (solid) and without (dashed) RG improvement, where the panels are presented in order with the cut-off scale ranging from $530\,\mathrm{Gev}$ to $600\,\mathrm{GeV}$ with interval $10\,\mathrm{GeV}$.}\label{fig:RGGW}
\end{figure}

\begin{figure}
  \includegraphics[width=1.0\textwidth]{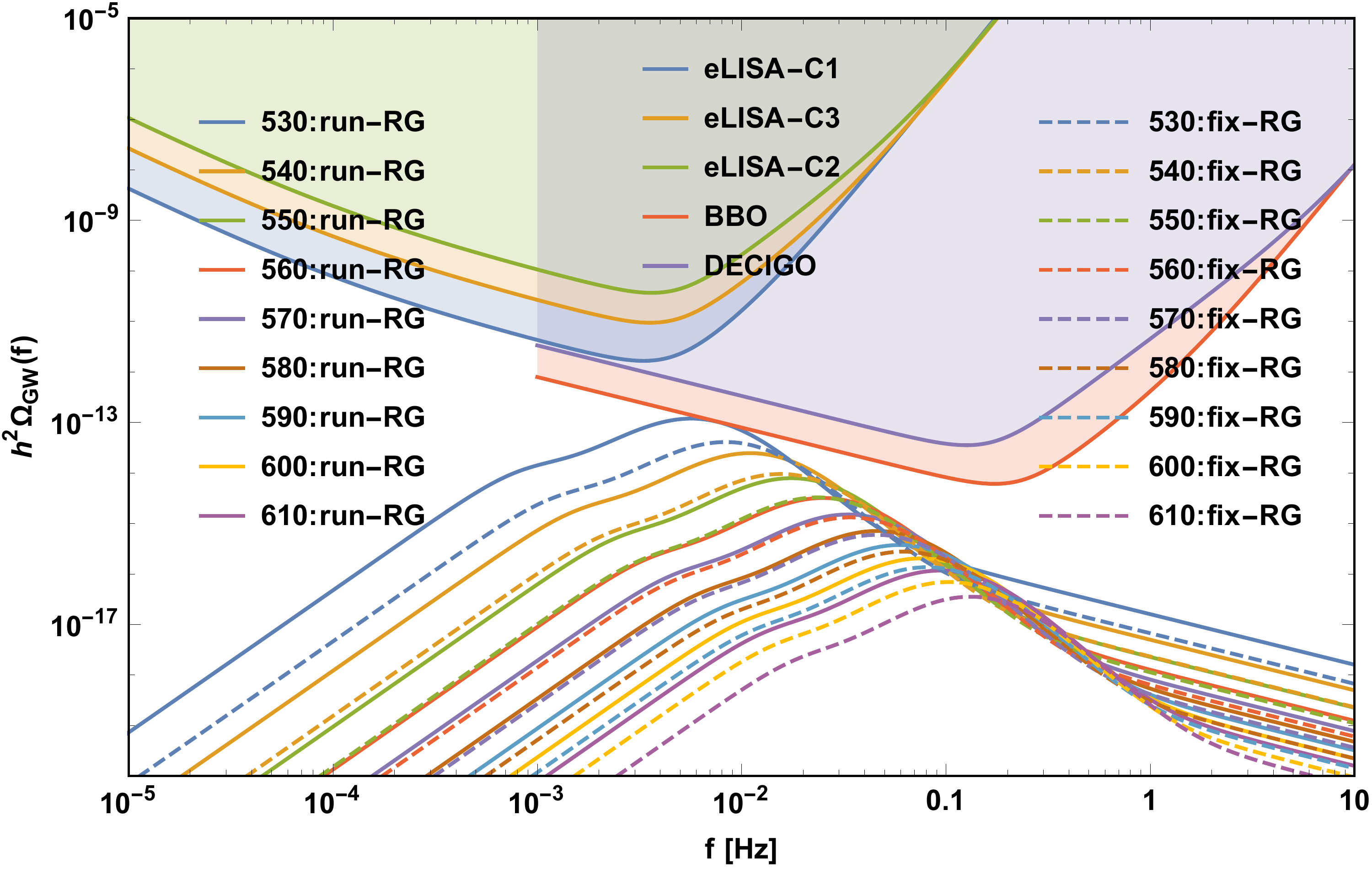}\\
  \caption{The energy density spectrum of the relic GWs from the first-order PT for the full one-loop effective potential with (solid) and without (dashed) RG improvement, where the cut-off scale ranges from $530\,\mathrm{Gev}$ to $610\,\mathrm{GeV}$ with interval $10\,\mathrm{GeV}$.}\label{fig:RGGWTot}
\end{figure}

Having obtained the phenomenological parameters with fixed and running RG scales, we will use the prescription~\eqref{eq:FFPT} of fast first-order PT in~\ref{subsubsec:treeresult} to carry out the predictions of GWs. In Fig.~\ref{fig:RGGW} and Fig.~\ref{fig:RGGWTot}, the energy density spectrums are presented with (solid) or without (dashed) the RG improvement for the different values of the cut-off scale $\Lambda$ ranging from $530\,\mathrm{Gev}$ to $610\,\mathrm{GeV}$ with interval $10\,\mathrm{GeV}$. It turns out as a surprise that, the presence of running RG scale could amplify the peak amplitude by amount of one order of magnitude, and the peak frequency has shifted to the lower frequency regime. Therefore, the effect from RG improvement cannot be simply neglected in future for the precise prediction of energy density spectrum of GWs from first-order PT.

However, compared with previous expectation~\cite{Leitao:2015fmj}, the perspective regime of detecting the relic GWs from the model with a dimension-six term within the sensitivity regions of space-based detectors has shrunken down to the regime with cut-off scale of $\Lambda<530\,\mathrm{GeV}$ at least, probably into the regime of slow first-order PT, which would require lattice computations to fully appreciate the non-perturbative running of strong coupling at low temperature. We hope to report the relevant results in the future.

The difference between~\cite{Leitao:2015fmj} and our results with fixed RG scale might be caused mainly by the different choices of the renormalized radiative corrections~\eqref{eq:CWpotential} and~\eqref{eq:fixRGconSim}, where the contributions from Higgs and Goldstone bosons have been neglected in~\eqref{eq:fixRGconSim}. Furthermore, a simplified treatment of thermal corrections is also invoked in~\cite{Leitao:2015fmj} instead of full ring resummation~\eqref{eq:ringpotential}. We also differ from~\cite{Leitao:2015fmj} in the estimations of those phenomenological parameters involved in the energy density spectrum, although it is not significant for the difference since only the regime of fast first-order PT is considered.

\section{Conclusions}\label{sec:conclusion}

In this paper, we have investigated in some details the GWs from first-order PTs in an example of the SM extended with a dimension-six operator. An unified prescription is proposed for both slow and fast first-order PTs: the reference temperature is chosen at the percolation temperature, the strength factor comes from the total released vacuum energy density from all kinds of bubbles at percolation with different sizes and number densities, and the characteristic length scale of bubble collisions is defined by the mean size of bubbles at percolation. We then investigated the effect of the RG improvement on the predictions of the relic GWs from the first-order PT. When compared to the case with a fixed RG scale, the presence of a running RG scale could amplify the peak amplitude of the energy density spectrum by amount of one order of magnitude, while shift the peak frequency to a lower frequency regime at the same time. Therefore, the effect from RG improvement cannot be simply neglected in future for the precise prediction of energy density spectrum of GWs from first-order PT. However, the promising regime of detections within the sensitivity ranges of some space-based GW detectors has shrunken down to a lower regime of cut-off scale associated with the sextic term than the previous expectation.

\acknowledgments
This work was supported in part by the National Natural Science Foundation of China Grants No.11690022, No.11375247, No.11435006, No.11447601 and No.11647601, and by the Strategic Priority Research Program of CAS Grant No.XDB23030100 and by the Key Research Program of Frontier Sciences of CAS. This work was also supported in part by MEXT KAKENHI Nos. 15H05888 and 15K21733. This work was initiated during a visit of RGC and SJW to Yukawa Institute for Theoretical Physics in the fall of 2016, the warm hospitality extended to them is greatly appreciated. SJW would like to thank David Weir for the very useful discussions on the manuscript during the Spring School on Numerical Relativity and Gravitational-Wave Physics at the Institute of Theoretical Physics, Chinese Academy of Sciences. SJW would also like to thank Cyril Lagger for the stimulating correspondences on the topic of the slow first-order phase transitions and Zhong-Zhi Xianyu for the insightful discussions on the renormalization group improvement. We would also like to thank Yi-Fu Cai, Fa-Peng Huang, Dong-Gang Wang, Xin-Ming Zhang for the discussions at the early stage of this work. We acknowledge the use of supercomputer cluster at the Institute of Theoretical Physics, Chinese Academy of Sciences.

\bibliographystyle{JHEP}
\bibliography{ref}

\end{document}